\RequirePackage{ifpdf}
\documentclass[letterpaper]{article}
\pdfoutput=1

\usepackage[usenames,dvipsnames]{pstricks}
\usepackage{./auxi/jheppub}
\usepackage[T1]{fontenc}
\usepackage[utf8]{inputenc}
\usepackage[english]{babel}
\usepackage[babel]{csquotes}
\usepackage{amsmath}
\usepackage{amsfonts}
\usepackage{amssymb}
\usepackage{mathtools,colonequals}
\usepackage{mathrsfs}
\usepackage{bm}
\usepackage{bbold}
\usepackage{braket}
\usepackage{slashed}
\usepackage{graphicx}
\usepackage{graphics,multicol}
\usepackage{tikz}
\usetikzlibrary{decorations.pathmorphing}
\usetikzlibrary{decorations.markings}
\usepgflibrary{shapes.geometric}
\usepackage{booktabs}
\usepackage{array}
\usepackage{color}
\usepackage{subcaption}
\usepackage[percent]{overpic}
\usepackage{multirow}
\usepackage[normalem]{ulem}

\usetikzlibrary{math}

\newcommand{\al}{\alpha}

\newcommand{\ga}{\gamma}

\newcommand{\D}{\Delta}
\newcommand{\ep}{\epsilon}

\newcommand{\Th}{\Theta}
\newcommand{\la}{\lambda}
\newcommand{\La}{\Lambda}
\newcommand{\Ga}{\Gamma}

\newcommand{\si}{\sigma}

\newcommand{\p}{\phi}

\newcommand{\pa}{\partial}
\DeclarePairedDelimiter{\abs}{\lvert}{\rvert}

\newcommand{\beq}{\begin{equation}}
\newcommand{\eeq}{\end{equation}}
\newcommand{\mc}{\mathcal}

\newcommand{\wh}[1]{\widehat{#1}}

\newcommand{\ii}{\mathrm{i}}

\def\eps{\epsilon}

\def\eps{\epsilon}

\def\D{\Delta}

\newenvironment{sistema}%
  {\left\lbrace\begin{array}{@{}l@{}}}%
  {\end{array}\right.}

\newmuskip\pFqmuskip

\newcommand*\pFq[6][8]{%
  \begingroup 
  \pFqmuskip=#1mu\relax
  \mathchardef\normalcomma=\mathcode`,
  \mathcode`\,=\string"8000
  \begingroup\lccode`\~=`\,
  \lowercase{\endgroup\let~}\pFqcomma
  {}_{#2}F_{#3}{\left(\genfrac..{0pt}{}{#4}{#5};#6\right)}%
  \endgroup
}
\newcommand{\pFqcomma}{{\normalcomma}\mskip\pFqmuskip}

\newmuskip\tFomuskip
\newcommand*\tFo[4][3]{%
  \begingroup 
  \tFomuskip=#1mu\relax
  \mathchardef\normalcomma=\mathcode`,
  \mathcode`\,=\string"8000
  \begingroup\lccode`\~=`\,
  \lowercase{\endgroup\let~}\tFocomma
  {}_{2} F_1 {\left({#2};{#3};#4\right)}%
  \endgroup
}
\newcommand{\tFocomma}{{\normalcomma}\mskip\tFomuskip}

\newcommand{\cross}{\zeta}

\author{Marco Meineri,$^1$}
\author{Joao Penedones,$^2$}
\author{Taro Spirig$^2$}

\affiliation{$^1$ Department of Theoretical Physics, University of Geneva, 24 quai Ernest-Ansermet, 1211 Gen\`eve 4, Switzerland}
\affiliation{$^2$ Fields and Strings Laboratory, Institute of Physics, Ecole Polytechnique Fédéral de Lausanne (EPFL),\\ Route de la Sorge, CH-1015 Lausanne, Switzerland}

\emailAdd{marco.meineri@gmail.com, joao.penedones@epfl.ch, taro.spg@gmail.com}

\bigskip
\abstract{
We study correlation functions of the bulk stress tensor and  boundary operators in Quantum Field Theories (QFT) in Anti-de Sitter (AdS) space. In particular, we derive new sum rules from the two-point function of the stress tensor and its three-point function with two boundary operators. In AdS$_2$, this leads to a bootstrap setup that 
involves the central charge of the UV limit of the bulk QFT 
and may allow to follow a Renormalization Group (RG) flow non-perturbatively by continuously varying the AdS radius.
Along the way, we establish the convergence properties of the newly discovered local block decomposition of the three-point function.
}

\title{Renormalization group flows in AdS 
 and the bootstrap program}

\begin{document}

\maketitle
\newpage

\section{Introduction}
\label{sec:intro}
The renormalization group (RG) flow is one of the most remarkable aspects of quantum field theory (QFT). It makes physical predictions possible, by hiding the effect of fluctuations at the smallest scales inside a manageable number of low energy coefficients. At the same time, it makes computing these coefficients generically extremely hard, even when the high energy limit of the theory is exactly solved. For instance, knowing that QCD is a theory of free quarks in the UV does not bring us close to computing the mass of the pions. Yet, the 
 spectrum is definitely predicted by the QCD Lagrangian, and for instance it can be numerically extracted from the lattice. 

An alternative strategy consists in following the RG flow from the UV, and eventually extrapolate to the IR. This is achieved by introducing an additional tunable scale, which could be the energy of a scattering process, or the size of a sample. The latter option is especially useful in the Hamiltonian truncation framework \cite{Yurov:1989yu,Hogervorst:2014rta}: here one compactifies space, for instance on a sphere of radius $R$, and then diagonalizes the (truncated) Hamiltonian in this subspace. If, for simplicity, we consider a flow set by a unique scale $M$, the observables of the theory are now functions of the parameter $\lambda=MR$, with $\lambda \to \infty$ yielding the flat space theory. 

(Euclidean)\footnote{This work will be concerned with the Euclidean version Anti de Sitter, a.k.a. hyperbolic space, throughout. For simplicity, we shall simply call this space AdS from now on. We shall also assume that the radius of AdS has been set to 1, unless stated otherwise.} Anti de Sitter space (AdS) provides another useful IR regulator \cite{Callan:1989em}. In this case, the space is non-compact, but the spectrum is discrete. More importantly, AdS is maximally symmetric, and its large isometry group is conveniently rephrased as the conformal symmetry of boundary correlators. One can then extract flat space observables, like the S-Matrix, from the $R \to \infty$ limit of the correlations functions on the boundary \cite{Polchinski:1999ry, Gary:2009ae, Penedones:2010ue, Fitzpatrick:2011hu, Paulos:2016fap, Dubovsky:2017cnj, Hijano:2019qmi, Komatsu:2020sag, Li:2021snj, Cordova:2022pbl, Gadde:2022ghy, vanRees:2023air}. 
While this approach uses the full power of the numerical conformal bootstrap \cite{Rattazzi:2008pe}, it is not suited, as is, to follow an RG flow. Indeed, the theory at short scales is never specified, and there is no guarantee that by maximizing a coupling, as it was done in \cite{Paulos:2016fap}, one obtains a one-parameter family of correlators which corresponds to a unique QFT in AdS.

In \cite{Hogervorst:2021spa}, QFT in AdS was studied via Hamiltonian truncation. In this case, the UV Hamiltonian is part of the definition of the setup. The spectrum of the theory is obtained as a function of the parameter $\lambda$, and one can follow the flow well within the strongly coupled region. On the other hand, the truncation of the Hamiltonian breaks the isometries of AdS, and it is numerically challenging to reach large values of $\lambda$ while keeping the truncation errors small.

In this work, we advocate for a third option, where the conformal bootstrap approach is supplemented by conditions which specify the theory under consideration. To this end, our analysis will include not only the four-point function of conformal boundary operators, but correlation functions involving the bulk stress tensor in AdS as well. Our strategy works especially well in AdS$_2$, where we show how to extract the central charge of the UV conformal field theory (CFT) from the two-point function of the stress tensor computed along the flow. The upshot is a set of positive semi-definite constraints on the spectrum, which implement both crossing symmetry and the central charge constraint. This problem exploits the full isometry group of AdS, and at the same time it is suitable to derive rigorous bounds on IR observables as a function of the UV data. The present construction is conceptually similar to the one in \cite{Karateev:2019ymz,Karateev:2020axc}:  the latter can in fact be viewed as the flat space limit of the former (see section \ref{sec:flat}).

While the final result does not extend trivially to higher dimensions, we perform most of the intermediate steps leaving $d$ generic, and we derive along the way various formulas which are interesting on their own. 

Let us spell out the ingredients in more detail, while presenting the outline of the paper. A local QFT in  AdS possesses a stress-tensor $T_{\mu \nu}$, whose trace $\Theta \equiv T_\mu^\mu$ does not vanish in general. In section \ref{sec:twopStress} we discuss its two-point function: we determine the tensor structures and the interesting Euclidean limits. Then, we answer the following question: what is the most general linear function of $\braket{\Theta \Theta}$ which equals a total derivative? In two dimensions, we derive from three positive sum rules, one of which relates the UV central charge to an integral over the two-point function of $\Theta$. In appendix \ref{app:flatStress}, we also review the flat space case, answering the same question. 

Since our aim is to express the sum rule in terms of Operator Product Expansion (OPE) data, in section \ref{sec:spectral} we compute the spectral representation of the two-point function of the stress tensor, which extends an existing result for scalar operators \cite{PhysRevD.33.389}. Along the way, we explain a recipe to obtain the spectral representation for the two-point function of any spin, exploiting the conformal map to the upper half plane and the boundary CFT results of \cite{Lauria:2018klo}.

\tikzmath{\x1 = -.9; 
} 

\definecolor{aquamarine}{rgb}{0.5, 1.0, 0.83}
\begin{figure}
\begin{tikzpicture}[scale=0.75]
\draw[thick,fill=aquamarine] (0,0) circle (2);
\draw[thick] (2.5,0.1) -- (3,0.1);
\draw[thick] (2.5,-0.1) -- (3,-0.1);
\filldraw[black] (1.4,-1.42) circle (1.3pt);
\node[black] at (1.7,-1.7) {$\phi$};
\filldraw[black] (-1.4,-1.42) circle (1.3pt);
\node[black] at (-1.7,-1.7) {$\phi$};
\filldraw[black] (1.4,1.42) circle (1.3pt);
\node[black] at (1.7,1.7) {$\phi$};
\filldraw[black] (-1.4,1.42) circle (1.3pt);
\node[black] at (-1.7,1.7) {$\phi$};
\end{tikzpicture}
\hspace{0.01cm}
\begin{tikzpicture}[scale=0.75]
\node[thick] at (-3.3,-.1) {$\sum\limits_{\Delta} c_{\phi\phi\Delta}^2$};
\draw[thick,white] (0,0) circle (2);
\draw[thick] (2.5,0.1) -- (3,0.1);
\draw[thick] (2.5,-0.1) -- (3,-0.1);
\filldraw[black] (1.4,-1.42) circle (1.3pt);
\node[black] at (1.7,-1.7) {$\phi$};
\filldraw[black] (-1.4,-1.42) circle (1.3pt);
\node[black] at (-1.7,-1.7) {$\phi$};
\filldraw[black] (1.4,1.42) circle (1.3pt);
\node[black] at (1.7,1.7) {$\phi$};
\filldraw[black] (-1.4,1.42) circle (1.3pt);
\node[black] at (-1.7,1.7) {$\phi$};
\draw[thick] (1.4,-1.42) -- (0,-0.7);
\draw[thick] (-1.4,-1.42) -- (0,-0.7);
\draw[thick] (1.4,1.42) -- (0,0.7);
\draw[thick] (-1.4,1.42) -- (0,0.7);
\draw[thick] (0,-0.7) -- (0,0.7);
\node[black] at (0.4,0) {$\Delta$};
\end{tikzpicture}
\hspace{0.01cm}
\begin{tikzpicture}[scale=0.75]
\node[thick] at (-3.3,-0.1) {$\sum\limits_{\Delta} c_{\phi\phi\Delta}^2$};
\draw[thick,white] (0,0) circle (2);
\filldraw[black] (1.4,-1.42) circle (1.3pt);
\node[black] at (1.7,-1.7) {$\phi$};
\filldraw[black] (-1.4,-1.42) circle (1.3pt);
\node[black] at (-1.7,-1.7) {$\phi$};
\filldraw[black] (1.4,1.42) circle (1.3pt);
\node[black] at (1.7,1.7) {$\phi$};
\filldraw[black] (-1.4,1.42) circle (1.3pt);
\node[black] at (-1.7,1.7) {$\phi$};
\draw[thick] (1.4,-1.42) -- (0.7,0);
\draw[thick] (-1.4,-1.42) -- (-0.7,0);
\draw[thick] (1.4,1.42) -- (0.7,0);
\draw[thick] (-1.4,1.42) -- (-0.7,0);
\draw[thick] (-0.7,0) -- (0.7,0);
\node[black] at (0,0.4) {$\Delta$};
\end{tikzpicture}
\hspace{0.7cm}
\begin{tikzpicture}[scale=0.75]
\draw[thick,fill=aquamarine] (0,0) circle (2);
\draw[thick] (2.5,0.1) -- (3,0.1);
\draw[thick] (2.5,-0.1) -- (3,-0.1);
\filldraw[black] (1.4,-1.42) circle (1.3pt);
\node[black] at (1.7,-1.7) {$\phi$};
\filldraw[black] (-1.4,-1.42) circle (1.3pt);
\node[black] at (-1.7,-1.7) {$\phi$};
\node[black] at (0.1,0.9) {$T^{\mu\nu}$};
\filldraw[black] (0,0.4) circle (1.3pt);
\end{tikzpicture}
\hspace{0.01cm}
\begin{tikzpicture}[scale=0.75]
\node[black] at (-4,-.1) {$\sum\limits_{\Delta} c_{\phi\phi\Delta} b_{T\Delta}$};
\draw[thick,white] (0,0) circle (2);
\node[black] at (0+\x1,3) {};
\filldraw[black] (1.4+\x1,-1.42) circle (1.3pt);
\node[black] at (1.7+\x1,-1.7) {$\phi$};
\filldraw[black] (-1.4+\x1,-1.42) circle (1.3pt);
\node[black] at (-1.7+\x1,-1.7) {$\phi$};
\draw[thick] (1.4+\x1,-1.42) -- (0+\x1,-0.7);
\draw[thick] (-1.4+\x1 ,-1.42) -- (0+\x1,-0.7);
\node[black] at (0.1+\x1 ,0.9) {$T^{\mu\nu}$};
\filldraw[black] (0 +\x1,0.4) circle (1.3pt);
\draw[thick] (0+\x1 ,-0.7) -- (0+\x1,0.4);
\node[black] at (0.4+\x1,0) {$\Delta$};
\end{tikzpicture}
\hspace{0.01cm}
\begin{tikzpicture}
\node at (2,0.2) {$\implies \qquad \Delta_\phi = \sum\limits_\Delta \kappa(\Delta,\alpha) c_{\phi\phi\Delta}b_{T\Delta}$};
\node at (0,-1) {};
\node at (2.6,0) {};
\end{tikzpicture}
\hspace{0.6cm}
\begin{tikzpicture}[scale=0.75]
\draw[thick,fill=aquamarine] (0,0) circle (2);
\draw[thick] (2.5,0.1) -- (3,0.1);
\draw[thick] (2.5,-0.1) -- (3,-0.1);
\node[black] at (0.1,-1.1) {$T^{\alpha\beta}$};
\filldraw[black] (0,-0.6) circle (1.3pt);
\node[black] at (0.1,1.1) {$T^{\mu\nu}$};
\filldraw[black] (0,0.6) circle (1.3pt);
\end{tikzpicture}
\hspace{0.01cm}
\begin{tikzpicture}[scale=0.75]
\node[white] at (-4,-.1) {$ole$};
\node[black] at (-3,-.1) {$\sum\limits_{\Delta} b_{T\Delta}^2$};
\draw[thick,white] (0,0) circle (2);
\node[black] at (0.1,-1.1) {$T^{\alpha\beta}$};
\filldraw[black] (0,-0.6) circle (1.3pt);
\node[black] at (0.1,1.1) {$T^{\mu\nu}$};
\filldraw[black] (0,0.6) circle (1.3pt);
\draw[thick] (0,-0.6) -- (0,0.6);
\node[black] at (0.4,0) {$\Delta$};
\end{tikzpicture}
\hspace{0.01cm}
\begin{tikzpicture}
\node at (2.2,0.3) {$\implies \qquad C_T = \sum\limits_\Delta \frac{24 \Gamma(2\Delta)}{4^\Delta \Gamma^2(\Delta+2)}b_{T\Delta}^2$};
\node at (-1,-1) {};
\node at (0,2.6) {};
\node at (3.4,0) {};
\end{tikzpicture}
\caption{The first column depicts the correlation functions involved in our bootstrap setup for QFT on AdS$_2$.  The first row is just the conformal block decomposition and crossing equation for the four-point function of boundary operators. The second row shows the block decomposition of the three-point function of the bulk stress tensor and two boundary operators. On the right, we also show the sum rules \eqref{sumRulesFinal2} involving a free parameter $\alpha >\D_\phi + \frac{1}{2}\D_\mathcal{V}$, with $\D_\mathcal{V}$ the dimension of the bulk relevant operator that seeds the RG flow.
The third row depicts the Källén-Lehmann decomposition of the stress tensor two point function in AdS and a related sum rule that gives the central charge $c_{UV} = 2\pi^2 C_T$ of the bulk UV CFT.
This bootstrap setup is described in more detail in section \ref{sec:problem}.
}
\label{fig:summary}
\end{figure}

To set up the bootstrap problem, local operators on the boundary of AdS need to make an appearance. This happens in section \ref{sec:formFact}, where the three-point functions of the bulk stress tensor with two boundary operators is analyzed. These are the analog of the form factors in flat space. In two dimensions, 
we use and develop the local block decomposition recently introduced in \cite{Levine:2023ywq}\footnote{The authors learned about the local block decomposition from a talk given by Miguel Paulos during the ``Bootstrapping nature'' conference at the Galileo Galilei Institute on 20/11/2022. Section \ref{sec:formFact} contains a review of its derivation, for which we claim no originality. On the other hand, the bound on the growth of the OPE coefficents \eqref{bcBound}, and its consequences, like the value for $\al_\textup{min}$ in eq. \eqref{alphaMin} and the convergence properties of the local block decomposition, are new to the best of our knowledge. The closed form of the local blocks in AdS$_2$, eqs. \eqref{LocalBlockClosed0} and \eqref{LocalBlockClosed2}, and the sum rule \eqref{offDiagSumRule}, are also new. }
to extract a second sum rule, which ties together the coefficients of the spectral decomposition of $\braket{\Theta \Theta}$ with the OPE coefficients of the conformal theory on the boundary of AdS. An important set of results in section \ref{sec:formFact} concern the convergence of the local block decomposition and of our sum rule. We compute bounds on the form factor by means of the Cauchy-Schwarz inequality, and derive from them a region in cross ratio space where the local block decomposition converges uniformly, together with a bound on its rate.  

In section \ref{sec:flat}, we conjecture formulas that implement the flat space limit to give two-particle form factors and the full spectral density in flat space.
While we were concluding this article, the paper \cite{Levine:2023ywq} appeared, which contains an equivalent formula for the two-particle form factor.
We check  our formulas in the cases of a free boson and a free fermion in AdS$_2$ in section \ref{sec:examples}.
Finally, we spell out the constrained system in section \ref{sec:problem}, see in particular eq. \eqref{TotalCrossing}. 
For the reader's convenience, we summarise our main results in figure \ref{fig:summary}.

A number of appendices collect various technical details and some novel results, which we would like to highlight here. In appendix \ref{app:flatStress}, we carefully rederive sum rules involving the spectral density of the stress tensor in flat space and find a new algebraic constraint \eqref{UglyConstraint}
on the two-point function of the stress tensor in two dimensions. 
In appendix \ref{subsec:twopBlocks}, we give a recipe to compute the Källén-Lehmann blocks for two-point functions of integer spin bulk operators in AdS$_{d+1}$. 
In appendix \ref{app:ward}, we consider the problem of isometry preserving boundary conditions for QFT in AdS.
In particular, we provide a definition of the Hamiltonian when the boundary supports relevant operators, thus extending the work of \cite{Breitenlohner:1982jf,Klebanov:1999tb}  to the interacting case.

\section{The two-point function of the stress tensor in AdS$_{d+1}$}
\label{sec:twopStress}
\begin{figure}[t]
\centering
\begin{tikzpicture}[scale=1.2]
\draw[thick] (0,0) circle (2);
\draw[thick,dashed] (-2,0) -- (2,0);
\filldraw[black] (-1,0) circle (1.3pt);
\node [black] at (-1,0.4) {$\mathcal{O}$};
\filldraw[black] (1,0) circle (1.3pt) ;
\node [black] at (1,0.4) {$\mathcal{O}$};
\node[black] at (2,1.6) {$\partial$AdS};
\draw[thin] (-0.1,-0.1) -- (0.1,0.1);
\draw[thin] (-0.1,0.1) -- (0.1,-0.1);
\end{tikzpicture}
\caption{Using the isometries of AdS$_{d+1}$, two points can be brought to the position illustrated in this picture. Here is one way to see it. Start by bringing one point to the center of the ball: this is possible because AdS is homogeneous. Then move on to global coordinates, chosen such that the two points lie on the line at the center of the cylinder. Finally, translate the points along the cylinder appropriately and go back to the ball.}
\label{fig:ball2p}
\end{figure}
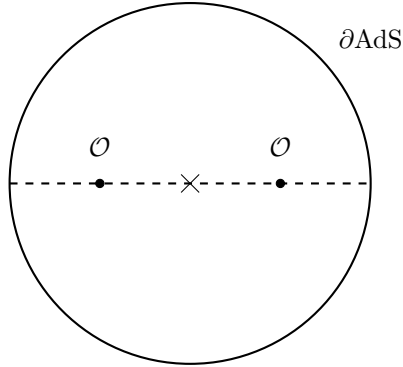

The two-point function $\braket{T^{\mu\nu} T^{\rho \sigma}}$ 
is constrained by the isometry group of AdS$_{d+1}$, $SO(d+1,1)$. This means that we can write it as a linear combination of tensor structures, whose coefficients are invariant under the isometries. To count the tensor structures \cite{Kravchuk:2016qvl}, it is convenient to choose coordinates where AdS$_{d+1}$ looks like a ball. It is   not hard to see that one can place the insertions on a diameter, at equal distance from the center -- see fig. \ref{fig:ball2p}.
This also confirms that there is a unique invariant function of the coordinates of the two points, their distance. 
Now, the correlator must be invariant under the remaining $SO(d)$ subgroup of rotations which stabilizes this configuration. We conclude that the number of tensor structures equals the number of $SO(d)$ singlets in the product of the two stress tensors. It is a simple exercise to show that this number equals 5 for $d>1$ and 4 for $d=1$.\footnote{Group theoretically, this is the number of  $SO(d)$ singlets in the symmetrized tensor products of two spin two $SO(d+1)$ irreducible representations (irreps), or a spin two and a spin zero (the trace), or two spin zero irreps. A quick way to do it is as follows: let us split the index $\mu=(1,i)$, where 1 labels the diameter where the operators lie, and $i=2\dots d+1$. Then there are 5 $SO(d)$ invariant contractions between two identical tensors: $ s_1=T_{11}T_{11},\, s_2=T_i^{i}T_j^{j},\, s_3=T_i^{i}T_{11}, \,
    s_4=T_{ij}T^{ij}, \, s_5=T_{i1}T^{i1}.$ 
    In two dimensions, where $i,j=2$ and the symmetry is just $\mathbb{Z}_2$, $s_2 \sim s_4$.
\label{foot:groupCounting2}    }
It is useful to notice that, since AdS is conformally equivalent to flat space with a plane boundary, the tensor structures are in one-to-one correspondence with the ones found in a boundary CFT, and indeed the counting matches the one of \cite{McAvity:1993ue}. The relation between QFT in AdS and boundary CFT will be further explained and exploited in section \ref{sec:spectral}.

In order to write the tensor structures down, we will work in embedding space \cite{Costa:2014kfa}. This is simply $(d+2)$-dimensional Minkowski space, which we parametrize by Cartesian coordinates $X^A$, $A=0,\dots,d+1$. Euclidean AdS with unit radius is embedded as the upper branch of the hyperboloid $X^2=-1,$ $X^0>0$. Its isometries act as Lorentz transformations on the coordinates $X^A$. From the equation
\beq
dX^A X_A=0~,
\eeq  
we deduce that the timelike vector $X^A$ is normal to AdS. Hence, the induced metric on AdS with unit radius is 
\beq
G_{AB}=\eta_{AB}+X_AX_B~.
\label{Gind}
\eeq
The union of the vector $X^A$ with a basis of the space tangent to a point in AdS forms a basis of $(d+2)$-dimensional Minkowski space. Therefore, 
given a tensor $H$ in embedding space which vanishes if any index is contracted with $X^A$, then $H$ is a tensor on AdS when evaluated on the manifold $X^2=-1$. If $H$ is symmetric, as in all cases in this work, then the condition
\beq
X^{A_1} H_{A_1 \dots A_\ell}=0
\label{Htransv}
\eeq
is sufficient. Any $H$ obeying eq. \eqref{Htransv} is called transverse, and since  we are interested in the physics of AdS we can restrict our considerations to transverse tensors. A tensor $H$ in embedding space can be pulled back to AdS, yielding the map
\beq
h_{\mu_1 \dots \mu_\ell} = \frac{\partial X^{A_1}}{\partial x^{\mu_1}} \dots
\frac{\partial X^{A_\ell}}{\partial x^{\mu_\ell}} H_{A_1\dots A_\ell}~,
\label{pullBack}
\eeq
where $x^\mu$ parametrizes AdS.

Invariance under the isometries and transversality imply that the two-point function $\braket{H_1(X_1)H_2(X_2)}$ can only depend on the following building blocks:
\begin{equation}\label{VandGDef}
    \begin{split}
        V_{1}^A& = X_{2}^A + (X_1 \cdot X_2)\, X_{1}^A~,\\
        V_{2}^A& = X_{1}^A + (X_1\cdot X_2)\, X_{2}^A~,\\
        G_{1}^{AB} & =\eta^{AB} + X_{1}^A X_{1}^B~,\\
        G_{2}^{AB} & = \eta^{AB} + X_{2}^AX_{2}^B~,\\
        G_{12}^{AB}& = \eta^{AB}-\frac{ X_{2}^AX_{1}^B}{X_1\cdot X_2}~.
    \end{split}
\end{equation}
Here, we assumed for simplicity that the theory is parity invariant and the tensors $H_i$ are parity even. Notice that $V_i$ and $G_i$ are transverse to $X_i$, and $G_{12}^{AB}X_{1,A}=G_{12}^{AB}X_{2,B}=0$.
The \emph{connected}  two-point function of the stress tensor is then parametrized as
 \begin{equation}\label{2ptFct}
\mathbb{T}^{A_1B_1A_2B_2} \equiv
 \langle  T^{A_1B_1}(X_1)T^{A_2B_2}(X_2)\rangle_\textup{connected}=\sum_{i=1}^5 h_i(X_1\cdot X_2)\mathbb{T}_i^{A_1B_1A_2B_2}~,
\end{equation}
where the tensor structures are 
\begin{equation}\label{TensorStruct}
\begin{split}
    \mathbb{T}_1^{A_1B_1A_2B_2} &\equiv V_{1}^{A_1} V_{1}^{B_1} V_{2}^{A_2} V_{2}^{B_2},\\
    \mathbb{T}_2^{A_1B_1A_2B_2} &\equiv 
     V_{1}^{A_1} V_{1}^{B_1} G_{2}^{A_2B_2} + G_{1}^{A_1B_1} V_{2}^{A_2} V_{2}^{B_2},\\
    \mathbb{T}_3^{A_1B_1A_2B_2} &\equiv -V_{1}^{A_1} V_{2}^{A_2} G_{12}^{B_1B_2} - V_{1}^{B_1} V_{2}^{B_2} G_{12}^{A_1A_2} - 
    V_{1}^{B_1} V_{2}^{A_2} G_{12}^{A_1B_2} - V_{1}^{A_1} V_{2}^{B_2} G_{12}^{B_1A_2},\\
    \mathbb{T}_4^{A_1B_1A_2B_2} &\equiv G_{1}^{A_1B_1} G_{2}^{A_2B_2},\\
    \mathbb{T}_5^{A_1B_1A_2B_2} &\equiv 
    G_{12}^{B_1B_2} G_{12}^{A_1A_2} + G_{12}^{A_1B_2} G_{12}^{B_1A_2}.
\end{split}
\end{equation}
The overall sign in $\mathbb{T}_3$ is chosen for later convenience. 
From now on  connectedness of all correlators  will   be understood, unless stated otherwise.  
In the following, we will often denote the isometry invariant of two points as 
\beq
\cross=X_1 \cdot X_2~.
\label{2pInvariant}
\eeq
It will be convenient to give a name to the connected two-point function of the trace:
\begin{equation}\label{2ptFctTraceStressTensAdS}
\begin{split}
    A(\cross)&= \langle \Theta(X_1)\Theta(X_2)\rangle = \mathbb{T}_{A_1B_1A_2B_2}G_{1}^{A_1B_1}G_{2}^{A_2B_2}\\&= (\cross^2-1)^2 h_1(\cross)+2(1+d)(\cross^2-1)h_2(\cross)+4\left(\frac{1}{\cross}-\cross\right)h_3(\cross)+(1+d)^2h_4(\cross)\\&+ 2\left(d+\frac{1}{\cross^2}\right)h_5(\cross),    
\end{split}
\end{equation}

The stress tensor is covariantly conserved, which yields three first order differential constraints on the functions $h_i$ in eq. \eqref{2ptFct}.\footnote{The number of constraints equals the number of tensor structures in the correlator of a vector and a symmetric tensor, which is easily found to be three, for any $d\geq 1$, via the same counting argument as before.} To find them, it is convenient to write the covariant derivative directly in embedding space \cite{Costa:2014kfa}. This is easily done by extending the definition of a tensor from AdS to the full embedding space. The way this extension is performed is immaterial, as long as the tensor is transverse when evaluated at $X^2=-1$. In particular, the tensor structures \eqref{TensorStruct} are defined in $(d+2)$ dimensions. Then, the covariant derivative is obtained as the projection onto AdS of the partial derivative:
\beq
\nabla_B H_{A_1 \dots A_\ell} = G_B^C\, G_{A_1}^{C_1} \dots G_{A_\ell}^{C_\ell}
\frac{\pa}{\pa X^C}  H_{C_1 \dots C_\ell} ~,
\label{NablaEmb}
\eeq
where $G$ is the induced metric \eqref{Gind}, whose index is raised with the embedding space metric $\eta$. Eq. \eqref{NablaEmb} is easy to prove: it defines a derivation, it yields a tensor in AdS, and it makes the induced metric $G_{AB}$ covariantly constant. The three constraints from conservation can be isolated by projecting onto three tensor structures: 
\beq\label{consConstrEq}
e_1 = \nabla^A \mathbb{T}_{AB_1 A_2 B_2} V_1^{B_1} G_2^{A_2 B_2}~, \quad
e_2 = \nabla^A \mathbb{T}_{AB_1 A_2 B_2} V_1^{B_1} V_2^{A_2} V_2^{B_2}~,
\quad
e_3 = \nabla^A \mathbb{T}_{AB_1 A_2 B_2} G_{12}^{B_1A_2} V_2^{B_2}~,
\eeq
which are explicitly given in appendix \ref{app:consConstr}.
We shall collect the equations in a vector: 
\beq
E=\frac{\cross^3}{\cross^2-1}\left(e_1,\frac{e_2}{\cross^2-1},\cross\, e_3\right)=0~.
\label{ConsVec}
\eeq

There is one last piece of available information about the functions $h_i$: their short and large distance asymptotics. Let us start from the former. Henceforth, we shall define a QFT in AdS as a CFT deformed by a relevant operator $\mathcal{V}$ (for simplicity we consider a one-parameter flow):\footnote{See \cite{Hogervorst:2021spa} for a detailed discussion of UV complete QFT on an AdS background.}
\beq
S=S_\textup{CFT}+\lambda \int_{AdS} \mathcal{V}~,
\qquad \Delta_\mathcal{V}<d+1~. 
\label{Sdeformed}
\eeq
 When the points get close together, the leading behavior of the correlator \eqref{2ptFct} approximates the two-point function of the stress tensor in flat space. It is convenient to parametrize AdS as
\beq
X^A=\left(\sqrt{1+x^2},x^\mu\right)~, \qquad \mu =1, \dots d+1~,
\label{AdScoordx}
\eeq
where $x^2=\sum_\mu (x^\mu)^2$. 
We can then place one operator at  $x_1^\mu=0$ and the other at a point $x^\mu$. 
In these coordinates, $\zeta=-\sqrt{1+x^2}$. When $x^\mu \to 0$, after pulling the correlator back with eq. \eqref{pullBack}, we get,\footnote{In this paper, we use the symbols $\approx,\, \lessapprox,\, \gtrapprox$ when, in the appropriate limit, the ratio of the l.h.s. over the r.h.s tends to 1, to a number smaller or larger than 1, respectively. Instead, we use $\sim,\,\lesssim,\,\gtrsim$ when the same ratio tends to a finite number different from 0, to a number smaller than infinity, or to a number greater than 0, respectively. In other words, the latter set of symbols is used for estimates and asymptotic inequalities which are correct up to a finite positive factor. Furthermore, we sometimes use the same symbols to bound oscillating quantities: in these cases, the appropriate limits superior and inferior should be understood. }
\begin{equation}
\begin{split}
    \mathbb{T}_1^{\mu\nu\lambda\sigma}& \approx x^\mu x^\nu x^\lambda x^\sigma\\
    \mathbb{T}_2^{\mu\nu\lambda\sigma}& \approx \delta^{\mu\nu}x^\lambda x^\sigma + \delta^{\lambda \sigma} x^\mu x^\nu\\
    \mathbb{T}_3^{\mu\nu\lambda\sigma}& \approx x^\mu x^\lambda \delta^{\nu\sigma} + x^\sigma x^\mu \delta^{\lambda\nu} + x^\nu x^\sigma \delta^{\mu\lambda} + x^\lambda x^\nu \delta^{\sigma\mu}\\
    \mathbb{T}_4^{\mu\nu\lambda\sigma}& \approx \delta^{\mu\nu} \delta^{\lambda \sigma}\\
    \mathbb{T}_5^{\mu\nu\lambda\sigma}&\approx \delta^{\mu \sigma}\delta^{\lambda \nu} + \delta^{\nu \sigma}\delta^{\lambda \mu}.
\end{split}
\end{equation}
The standard renormalization group in flat space allows us to conclude that the correlator must coincide, in the short-distance limit, with the two-point function of the stress tensor for the CFT in eq. \eqref{Sdeformed}. We then obtain \begin{equation}\label{hiUVAdS}
    h_1 \approx\frac{4C_T}{x^{2d+6}}~, \quad h_2\sim o(x^{-2d-4})~, \quad h_3\approx-\frac{C_T}{x^{2d+4}}~, \quad h_4\approx-\frac{C_T}{d+1}\frac{1}{x^{2d+2}}~,\quad h_5\approx\frac{C_T}{2}\frac{1}{x^{2d+2}}~.
\end{equation}
When $d=1$, these leading order behaviors will be sufficient to derive a sum rule which expresses the UV central charge\footnote{We use improperly the term central charge in general dimensions.} $C_T$ in terms of the OPE data along the flow. 
We discuss the short distance limit in more detail in appendix   \ref{SD-AdS2}, for the case $d=1$.
For the moment, let us simply emphasize that a certain combination of the corrections to eq. \eqref{hiUVAdS} is fixed by the leading behavior of the two-point function of the trace, eq. \eqref{2ptFctTraceStressTensAdS}. Indeed, on general grounds, $\Theta$ can be expressed in terms of local operators in the UV CFT. In the UV limit, the stress tensor can be computed in conformal perturbation theory from the action \eqref{Sdeformed}, and we conclude
\beq
\Theta(x)
= (d+1-\D_\mathcal{V}) \lambda\,\mathcal{V} (x) +\dots \,.
\label{ThetaUV}
\eeq
Therefore, the short distance limit of eq. \eqref{2ptFctTraceStressTensAdS} is
\beq
A \approx\frac{(d+1-\D_\mathcal{V})^2\lambda^2}{x^{2\Delta_\mathcal{V}}}~.
\label{Ashort}
\eeq

At large distances, the decay of the two-point function is fixed by the boundary OPE, \emph{i.e.} the expansion of the stress tensor in terms of local operators placed at the conformal boundary of AdS. We will describe this OPE in detail in section \ref{sec:spectral}. For the moment, let us say that in this section we will assume the conformal theory on the boundary of AdS to have a gap 
\beq
\D>d+1~.
\label{BoundaryGap}
\eeq
The rationale for eq. \eqref{BoundaryGap} is that it allows the sum rules discussed in the next subsections to converge. However, in section \ref{sec:spectral}, we shall re-derive the sum rules in the special case $d=1$ and express them directly in terms of OPE data---for instance, contrast eq. \eqref{sumRuleAdS2} with eq. \eqref{sumRuleAdS2FromTT}. The new form of the sum rule will hold also when eq. \eqref{BoundaryGap} is not satisfied, so one should consider it as a temporary technical assumption. 

\subsection{A sum rule in any dimension}
\label{subsec:hdsumrules}

We are ready to look for positive sum rules relating the UV CFT data to integrals of observables along the flow. These sum rules can then be used, in a bootstrap approach, to input information about the starting point of the RG flow.  Our strategy is to write kernels $r(\cross)$, which turn the two-point function of the trace of the stress tensor into a total derivative:\footnote{This is not the most general possibility. Perhaps one can obtain other positive sum rules by replacing $A(\zeta)$ in eq. \eqref{rACputative} with another reflection positive component of the tensor \eqref{2ptFct}.
We shall not explore this possibility here.
}
\beq
r(\cross)A(\cross)=\frac{d}{d\cross} C(\cross)~,
\label{rACputative}
\eeq
where $A$ was defined in eq. \eqref{2ptFctTraceStressTensAdS}, while $r$ and $C$ must be found. In particular, $C$ can but be linear in the functions $h_i$ defined in eq. \eqref{2ptFct}:
\beq
C(\cross)= \sum_{i=1}^5 f_i(\cross) h_i(\cross)~.
\label{Cfh}
\eeq 
Integrating eq. \eqref{rACputative}, the right hand side gives a boundary term and the left hand side gives a positive quantity.
It is important that the connected correlator of $\Theta$ is reflection positive: it can be written as the squared norm of the vector $(\Theta-\braket{\Theta})\ket{0}$. 

Since eq. \eqref{rACputative} is a first order differential equation, it is natural for it to arise from the conservation equations \eqref{ConsVec}:
\beq
 \frac{d }{d\cross}C(\cross)-r(\cross) A(\cross) = E \cdot (q_1(\cross),q_2(\cross),q_3(\cross))^T,
 \label{rACE}
\eeq
where there are 9 unknown functions: the $q_j$'s, $r$ and the $f_i$'s in eq. \eqref{Cfh}. 
Then, imposing that eq. \eqref{rACE} should hold for every choice of the $h_i$'s, yields ten first order differential equations for these unknown functions.
For $d>1$, one  finds only one non-trivial solution.
It is useful to express it as a function of the coordinate
\beq
\xi = -\frac{1+\cross}{2}~,
\label{xiofzeta}
\eeq
which is positive, vanishes when the two insertions coincide and diverges as one point approaches the boundary of AdS.   
The solution reads
\beq
r(\xi) = \left[4\xi (\xi+1) \right]^{\frac{d-1}{2}} (2\xi+1)~,
\label{rAdSd}
\eeq
\begin{multline}
C(\xi) =-\left[4\xi (\xi+1) \right]^{\frac{d+1}{2}}\left( 
\big(4\xi (\xi+1) \big)^2 h_1(\xi)+(d+2)\big(4\xi (\xi+1) \big) h_2(\xi)
+\frac{16\xi (\xi+1)}{2\xi+1} h_3(\xi) \right. \\
\left.+(d+1)h_4(\xi)+\frac{2}{(2\xi+1)^2}h_5(\xi)\right)~.
\label{CAdSd}
\end{multline}
In the rest of the paper, we shall abuse notation writing both $C(\cross)$ and $C(\xi)$, and similarly for other expressions. 

Presented this way, this sum rule may seem mysterious. However, it is a simple generalization of something well known in flat space.
Indeed, one can construct a sum rule in any conformally flat space by using the existence of a conformal Killing vector. The idea is analogous to the $\D$-theorem in flat space \cite{Delfino:1996nf}, which we review in appendix \ref{subsec:sumRules2dFlat}---see eq. \eqref{sumBad}. A conformal Killing vector $v$ obeys the equation
\beq
\nabla_{(\mu} v_{\nu)} = \frac{1}{d+1}\nabla \cdot v\, g_{\mu\nu}~,
\eeq
where the r.h.s. does not vanish unless $v$ is an isometry. One can then construct a vector operator whose divergence is proportional to the trace of the stress tensor:
\beq
j_\mu = v^\nu T_{\mu\nu}~, \qquad  
\nabla_\mu j^\mu = \frac{1}{d+1}\nabla \cdot v \,\Theta~.
\label{jConfKill}
\eeq
The last equation can be used to turn the integral of the two point function of $\Theta$, with the appropriate kernel, into a surface term, only sensitive to the short distance physics. There, $j^\mu$ becomes conserved and the surface integral computes a charge, whose insertion can be evaluated using the conformal Ward identities. 

To realize this idea, we need $v$ not to be an isometry of AdS. AdS is one Weyl transformation away from flat space with a flat or spherical boundary. Under the Weyl transformation, the isometries of AdS are mapped to the conformal transformations which fix the boundary. Hence, we can find $v$ among the conformal Killing vectors of flat space which do not belong to this class. Let us consider a ball with unit radius, centered at the origin of a Cartesian coordinate system $y^\mu$ in flat space. 
The Killing vector corresponding to a dilatation is $v^\mu = y^\mu$, and of course it does not preserve the boundary of the ball. The AdS metric expressed in this coordinate system reads
\beq
ds^2_\textup{AdS} = \left(\frac{2}{1-y^2}\right)^2 dy^\mu\, \delta_{\mu\nu}\, dy^\nu~,
\label{AdSmetricy}
\eeq
and one can check that $v^\mu$ obeys eq. \eqref{jConfKill} with
\beq
\frac{1}{d+1} \nabla \cdot v = \frac{1+y^2}{1-y^2}~.
\label{nablaXiAdS}
\eeq

Using the divergence theorem, we can write
\beq
\lim_{\substack{\ep \to 0 \\R\to 1 }} \int_{\eps<|y|<R}\!d^{d+1}y
\sqrt{g}\, \nabla_\mu \braket{j^\mu(y) \Theta(0)}_\textup{c}=
\lim_{\substack{\ep \to 0 \\R\to 1}} 
\left(\int_{\Sigma(R)}- \int_{\Sigma(\ep)}\right) d\Sigma_\mu \braket{j^\mu(y) \Theta(0)}_\textup{c}~.
\eeq
Here, $g$ is the determinant of the metric \eqref{AdSmetricy} and $\Sigma$ denotes a sphere centered at the origin. The contribution at short distance can be computed in the CFT in flat space, where the Ward identities for dilatation, together with eq. \eqref{ThetaUV}, dictate
\beq
\lim_{\ep\to0}\int_{\Sigma(\ep)}d\Sigma_\mu \braket{j^\mu(y) \Theta(0)}_\textup{c}~ = -\Delta_\mathcal{V} \braket{\Theta}~.
\eeq
On the other hand, the large distance surface term can be expressed more explicitly as follows:
\beq
\lim_{R\to 1}\int_{\Sigma(R)} d\Sigma_\mu \braket{j^\mu(y) \Theta(0)}_\textup{c}=\lim_{R\to 1}\left(\frac{2}{1-R^2}\right)^{d-1}\int d^d\Omega  \braket{T_{\rho\rho}(y=R\Omega)\, \Theta(0)}_\textup{c}~,
\label{LargeBallAdS}
\eeq
where we changed coordinates to $y^\mu = \rho \Omega^\mu$, $\Omega^\mu$ parametrizing the unit $d$-dimensional sphere. We also used the limit to drop a power of $R$. To understand the behavior of the stress tensor close to the boundary of the unit ball, we can approximate the boundary with a plane. More precisely, we replace $\rho=1-z$, and notice that as $z\to 0$, when all other coordinates are fixed, eq. \eqref{AdSmetricy} approaches the Poincaré metric $dz^2/z^2$. The approximate symmetry under rescaling of $z$ allows to write the OPE
\beq
T_{\rho \rho} (\rho) = T_{zz} (1-z) \sim z^{\D-2} O_\D~, \qquad z\to 0~,
\label{TbopeLoose}
\eeq 
where $O_\D$ is the local boundary primary with the lowest dimension above the identity. Replacing eq. \eqref{TbopeLoose} in eq.  
\eqref{LargeBallAdS}, we see that the infrared surface term drops out if the gap in the boundary spectrum obeys eq. \eqref{BoundaryGap}. Hence,
\beq
 \int_{|y|<1} d^{d+1}y \left(\frac{2}{1-y^2}\right)^{d+1} \frac{1+y^2}{1-y^2} 
\braket{\Theta(y) \Theta(0)}_\textup{c}=\D_\mathcal{V} \braket{\Theta}~.
\label{SsumRule}
\eeq

To make contact with eq. \eqref{rAdSd}, one simply needs to express $\xi$ in the coordinate system $y^\mu$. For instance, one can change coordinates to match the $x^\mu$ in eq. \eqref{AdScoordx}:
\beq
x^2=
x^\mu \delta_{\mu\nu} x^\nu = \frac{4y^2}{(1-y^2)^2} ~.
\eeq
Recalling that $\cross=-\sqrt{1+x^2}$, one obtains that the sum rule \eqref{SsumRule} coincides with the one previously found:
\beq
\frac{\D_\mathcal{V}}{S_{d+1}} \braket{\Theta} = \int_{-\infty}^{-1}\!d\cross\, r(\cross)\braket{\Theta(\zeta) \Theta(0)}_\textup{c}~, 
\qquad S_{d+1}=\frac{2\pi^\frac{d+1}{2}}{\Gamma\left(\frac{d+1}{2}\right)}~.
\label{UglyAdSsumRule}
\eeq  
This fixes the short distance limit of the $C$ function \eqref{CAdSd}. 

\subsection{Three sum rules in two dimensions}
\label{subsec:2dsumrules}

As previously mentioned, when $d=1$ there are only four independent structures in the two-point function of the stress tensor. Therefore, the five tensors in eq. \eqref{TensorStruct} obey a constraint:\footnote{Consider the tensor 
$W^{ABCD}_{EF} = X_1^{\left[A\right.} X_2^B \delta^C_E \delta^{\left.D\right]}_{F}$,
 which vanishes identically for $d=1$ because the corresponding  embedding space is 3 dimensional.
The identity is given by $W^{A_1 A_2 CD}_{EF} W^{B_1B_2EF}_{CD}$ symmetrized over $(A_1 B_1)$ and $(A_2 B_2)$. 
}
\beq
-\frac{2}{\cross^4} \mathbb{T}_1+\frac{2}{\cross^2} \mathbb{T}_2+\frac{1}{\cross^3} \mathbb{T}_3-2\frac{\cross^2-1}{\cross^2} \mathbb{T}_4
+\frac{\cross^2-1}{\cross^2} \mathbb{T}_5=0~, \qquad d=1~.
\label{Tdep2d}
\eeq
Correspondingly, the coefficient functions $h_i$ are defined up to a shift by an arbitrary function $g(\cross)$:
\begin{align}
&h_1(\cross) \sim h_1(\cross)-\frac{2}{\cross^4} g(\cross) \notag\\
&h_2(\cross) \sim h_2(\cross)+\frac{2}{\cross^2} g(\cross)\notag\\
&h_3(\cross) \sim h_3(\cross)+\frac{1}{\cross^3} g(\cross) \label{hred2d}\\
&h_4(\cross) \sim h_4(\cross)-2\frac{\cross^2-1}{\cross^2} g(\cross)\notag\\
&h_5(\cross) \sim h_5(\cross)+\frac{\cross^2-1}{\cross^2} g(\cross)~. \notag
\end{align}
Consistently, the vector $E$ and the function $A$ in eq. \eqref{rACE} are invariant under the gauge symmetry \eqref{hred2d}, and so must be $C(\cross)$. One way to proceed is then to trade the $h_i$'s for four linearly related gauge invariant functions, and express $C(\cross)$ as a combination of those, rather than through eq. \eqref{Cfh}.\footnote{In fact, solving eq. \eqref{rACE} treating the $h_i$'s as independent yields the same result as the rigorous procedure. Alternatively, one can work in physical space, and organize the components of the stress tensor according to their eigenvalue under the (abelian) rotational symmetry. In that basis, the tensor structures trivialize and there is no redundancy. We use this strategy in subsection \ref{subsec:twopBlocksStress}, while the relation between this basis and the $h_i$ basis can be found in eq. \eqref{FGHIofh}.} Then, eq. \eqref{rACE}, which we want satisfied for every choice of the $h_i$'s, yields eight first order differential equations, for the coefficients of the new functions and their first derivatives. The unknowns are $r(\cross)$, $q_j(\cross)$ for $i=1,\,2,\,3$ and the four coefficient functions in $C(\cross)$. There are three independent solutions. The kernels of the independent solutions read
\begin{subequations}
\begin{align}
r_1(\xi) &=2\xi+1~, \\
r_2(\xi) &=8\xi +4(2\xi+1) \log (\xi+1)~, \\
r_3(\xi) &=(1+2\xi) \log \left(1+\frac{1}{\xi}\right)-2 ~. \label{r3AdS2}
\end{align}
\label{riAdS2}
\end{subequations}
The kernels $r_i$ are positive for $\xi>0$.
The corresponding $C$ functions are  reported in appendix \ref{app:details}, eq. \eqref{CAdS2} and eq. \eqref{CAdS2ofT}. 
The first sum rule is simply \eqref{SsumRule} for $d=1$.
Let us now consider the convergence of the integral of the left hand side of eq. \eqref{rACE}, for the three choices:
\beq
\int_0^\infty\! d\xi\, r_i(\xi) A(\xi)~, \qquad i=1,\,2,\,3~.
\label{IntLhsAdS2}
\eeq 
In the UV, using the short-distance limit \eqref{Ashort} for $A(\cross)$ and the leading behavior $\xi \sim x^2/4$, we find that the choice $r_2(\xi)$ makes the integral converge for any relevant perturbation, while the other two choices require $\Delta_\mathcal{V} < 1$. In the IR, the condition \eqref{BoundaryGap} insures that the integral \eqref{IntLhsAdS2} converges for any choice of kernel among the \eqref{riAdS2}.\footnote{In fact, for $r_3$ the integral converges at large $\xi$ as long as $\Delta>0$, which follows from unitarity.} We now evaluate the $C$ function corresponding to $r_2$---eq. \eqref{C2AdS2}--- in the UV, using eq. \eqref{hiUVAdS}, and get $C_2(\xi=0)=C_T.$ On the other hand, a result established in appendix \ref{app:details}---see eq. \eqref{CAdS2IR}---shows that all three $C$-functions vanish in the IR if eq. \eqref{BoundaryGap} is obeyed. Hence
\beq
C_T = 2\int_0^\infty\! d\xi\, \left[8\xi +4(2\xi+1) \log (\xi+1)\right] A(\xi)~.
\label{sumRuleAdS2}
\eeq
This is one of the main results of our paper. In section \ref{sec:spectral}, we shall develop the spectral decomposition of the two-point function of the stress tensor. Plugging it in eq. \eqref{sumRuleAdS2}, we shall express the central charge in terms of boundary CFT data. 

Let us now discuss the third sum rule. 
Since its convergence in the UV depends on the scaling dimension of the perturbing operator, the short distance limit of the corresponding $C$ function cannot be fixed by the correlator of the CFT stress tensor, which is blind to $\D_\mathcal{V}$. One can quickly confirm this by plugging eq. \eqref{hiUVAdS} in eqs. (\ref{C3AdS2}), or simply staring at eqs. (\ref{C3AdS2ofT}). 
In appendix \ref{SD-AdS2}, we relate  $C_3(0)$ to the short distance behavior of the two point function of the stress tensor in AdS$_2$. It would be useful to relate its value to other observables, 
but we leave that for the future.

\section{The spectral representation in AdS$_{d+1}$ and boundary CFT}
\label{sec:spectral}

Our next task is to express the two-point function of the trace of the stress tensor in terms of boundary data. This way, the sum rule \eqref{sumRuleAdS2} will be turned into a positive constraint on a set of OPE coefficients, which can be later exploited via the bootstrap. The boundary OPE decomposition of the two-point function is the AdS analogue of the K\"allén-Lehman spectral decomposition in flat space: it expresses the correlator as a sum over contributions labeled by the eigenvalue of the quadratic Casimir of the spacetime symmetry algebra. Each contribution corresponds to all the states in the spectrum of the theory which lie in a given representation of the isometries. In AdS, the isometry group is $SO(d+1,1)$, and states are labeled by their eigenvalue $\Delta$ under global time translations, and their representation $\rho$ under rotations in a global time slice. Via the bulk state/boundary operator correspondence \cite{Paulos:2016fap}, one can reinterpret these states as local scaling operators on the boundary of AdS. 

The spectral representation of the two-point function of $\Theta$, or of any scalar for that matter, can be written as
\beq
\braket{\Theta(X_1)\Theta(X_2)} = \sum_{\Delta} b^2_{\Theta\Delta}  f_\Delta(\xi)~.
\label{specDecScalar}
\eeq
The \emph{spectral blocks} $f_\Delta$ are kinematical, while the real OPE coefficients $b_{\Theta\Delta}$ contain the dynamical information. For scalar external operators, the sum only runs over scalar primaries, for reasons to be clarified shortly, therefore we omitted the spin label $\rho$. The spectral blocks can be obtained for instance as the solution to a Casimir equation. Instead of writing the equation explicitly, it is useful to notice that the functional form of the blocks only depends on the representations of the exchanged and external operators. Hence, it is the same whether the QFT in AdS is massive or conformal. This observation is useful, because, as mentioned in subsection \ref{subsec:hdsumrules}, a conformal field theory in AdS is the same as a boundary CFT in flat space.\footnote{This fact is often employed to take advantage of the AdS/CFT technology when performing perturbative computations in boundary CFT, see for instance \cite{Kapustin:2005py, Giombi:2020rmc, Giombi:2021uae, Cuomo:2021kfm}.} Correspondingly, eq. \eqref{specDecScalar} expresses the OPE decomposition of the two-point function in the boundary channel, which has been discussed extensively in the literature \cite{Cardy:1984bb,McAvity:1995zd,Liendo:2012hy,Billo:2016cpy,Lauria:2017wav,Lauria:2018klo,Isachenkov:2018pef,Buric:2020zea}. Of course, this argument works for any choice of external operators, so that the spectral representation for two-point functions of bulk operators of any spin in AdS can be obtained from the conformal block decomposition in boundary CFT. In particular, one can check that the spin of the exchanged operator is bounded by the spin of the bulk operator \cite{Lauria:2018klo}, so that only scalars appear in eq. \eqref{specDecScalar}. In \cite{Lauria:2018klo} a recipe was given to compute the blocks in closed form for the two-point function of symmetric traceless operators of any rank.
We will review and take advantage of this in appendix \ref{subsec:twopBlocks}.  

For the moment, let us make the map to boundary CFT explicit. It is convenient to write AdS in Poincaré coordinates:
\beq
X = \frac{1}{z}\left(\frac{1+x_a x^a+z^2}{2}, x^a, \frac{1-x_a x^a-z^2}{2}\right)~, \qquad ds^2 = \frac{dx^a dx_a +dz^2}{z^2}~, \quad a=1, \dots d~. 
\label{PoincarePatch}
\eeq
In a CFT, traceless symmetric flat space bulk operators of dimension $\Delta$ and spin $\ell$ are related to operators in AdS by
\beq
\Phi_\textup{AdS}(x,z) = z^{\Delta-|\ell|} \Phi_\textup{flat}(x,z)~,
\label{AdStoFlat}
\eeq
while boundary operators are unchanged.\footnote{The absolute value in eq. \eqref{AdStoFlat} is meant to avoid confusion in the two dimensional case, where we will denote with $\ell$ the eigenvalue under $so(2)$, which can be negative.} The prescription for boundary operators can be verified, for instance, comparing the boundary OPE in the two frames, as dictated by the symmetries. For scalar bulk operators:\footnote{In the boundary CFT literature, it is customary to normalize the OPE coefficient $b_{\Delta}$ such that boundary operators are weighted by powers of $(2z)$. We find that for the AdS sum rules the convention chosen here is more convenient instead.}
\begin{align}
\Phi_\textup{flat}(x,z) &\approx b_{\Phi\Delta}\,z^{-\Delta_\Phi+\Delta}\,\mc{O}(x)+\dots~, \label{OPEFlat}\\
\Phi_\textup{AdS}(x,z) &\approx b_{\Phi\Delta}\,z^{\Delta}\,\mc{O}(x)+\dots~. \label{OPEAdS}
\end{align}
The second line---eq. \eqref{OPEAdS}---is valid for a massive QFT as well, and is the OPE responsible for the spectral decomposition \eqref{specDecScalar}. The spectral blocks in eq. \eqref{specDecScalar} are easily obtained from the two-point function of a boundary CFT---see \emph{e.g.} \cite{McAvity:1995zd}---up to the transformation \eqref{AdStoFlat}: 
\beq
 f_\Delta(\xi) = (4\xi)^{-\Delta}
  {}_2 F_1 \left(\Delta,\Delta-\frac{d-1}{2};2\Delta+1-d;-\frac{1}{\xi}\right)~.
 \label{scalarSpecBlocks}
\eeq
This result was originally found in AdS$_4$ directly \cite{PhysRevD.33.389}.  In view of the connection with boundary CFT, we shall use the terms spectral blocks and boundary blocks interchangeably hereafter. The agreement with the OPE limit \eqref{OPEAdS} can easily be checked using that in the Poincaré patch
\beq
\xi = \frac{(x_1-x_2)^2+(z_1-z_2)^2}{4z_1z_2}~.
\eeq

In two dimensions, we can now replace the spectral decomposition \eqref{specDecScalar} in the sum rule for the central charge, eq. \eqref{sumRuleAdS2}, integrate block by block, and express the central charge as a positive sum over the squared OPE coefficients:
\beq
C_T = \sum_{\Delta>2} 
\frac{12\, \Gamma\!\left(\D+\frac{1}{2}\right)}{\sqrt{\pi}(\D-2)^2(\D+1)^2 \Gamma(\D)} b^2_{\Theta\Delta}~.
\label{specSumRuleAdS2}
\eeq
In eq. \eqref{specSumRuleAdS2}, the sum excludes the identity operator. This follows from the fact that $A(\xi)$ in eq. \eqref{sumRuleAdS2} is the connected correlator. Notice also that the integral over the spectral block \eqref{scalarSpecBlocks} converges if $\Delta>2$, which is the gap announced in eq. \eqref{BoundaryGap}. Correspondingly, a pole appears in eq. \eqref{specSumRuleAdS2} at $\D=2$. Nevertheless, in the rest of the section we will show how to extend eq. \eqref{specSumRuleAdS2} to theories with a completely general boundary spectrum.

The main tool will be the generalization of eq. \eqref{specDecScalar} to the case of a correlation function of operators with spin. Although we only need the two dimensional case for our purposes, in appendix \ref{subsec:twopBlocks} we derive from \cite{Lauria:2018klo} a recipe valid for traceless symmetric representations in general dimensions. In subsection \ref{subsec:twopBlocksStress}, we specialize   to two dimensions, and we re-derive the sum rule \eqref{specSumRuleAdS2} in its most general form.

\paragraph{The other sum rules} can also be expressed in terms of the BOPE $b_{\Theta \D}$ using \eqref{specDecScalar}. Firstly consider the sum rule \eqref{UglyAdSsumRule} which is valid in any dimension.
This gives
\beq
 \D_\mathcal{V}  \braket{\Theta} = \sum_{\Delta >d+1 } 
\frac{ 2\pi^{\frac{d}{2}} \Gamma\!\left(\D-\frac{d}{2}+1\right)}{ (\D-d-1)(\D+1) \Gamma(\D)} b^2_{\Theta\Delta}~,
\label{s1bbb}
\eeq
where we assumed the gap \eqref{BoundaryGap} in the spectrum.
Finally, there is one more sum rule valid for $d=1$ (namely \eqref{IntLhsAdS2} with $i=3$). This leads to
\beq
C_3(0)= \sum_{\Delta>2} 
\frac{2\D \Gamma\!\left(\D+\frac{1}{2}\right)
\left(
 H_{\D-1} - 1
\right)
}{ \sqrt{\pi} (\D-2)  \Gamma(\D+2)} b^2_{\Theta\Delta}~,
\label{s3bbb}
\eeq
 where $H_w\equiv \int_0^1dx \frac{1-x^w}{1-x}$ is the Harmonic Number function.

\subsection{The spectral representation in AdS$_2$, and another derivation of the sum rules}
\label{subsec:twopBlocksStress}

In two dimensions, the kinematics simplifies dramatically, since the group of rotations is abelian and its representations are one-dimensional. Therefore, the tensor structures trivialize in the appropriate basis: correlators of operators with definite spin are proportional to a unique structure. The spectral blocks for any spin can most simply be found by using the method of images. In flat space, the method of images allows to relate conformal blocks of an $n$-point function with a boundary to the holomorphic conformal blocks of a $2n$-point function without the boundary \cite{Cardy:1984bb}.  In order to describe the procedure, it is convenient to label operators by their holomorphic dimensions $(h,\bar{h})$.
 Since in the map from flat space to AdS all dependence on the scaling dimension of the external operators must drop, we will focus on the two-point function of operators with vanishing scaling dimension. In this case, $(h,\bar{h})=(\ell/2,-\ell/2)$, $\ell$ being the spin. Consider one such operator placed at position $(x,z)$ on the upper half plane, where the coordinates are defined as in \eqref{PoincarePatch}. Then, we are instructed to replace it with a pair of operators, placed at $(x,z)$ and $(x,-z)$, whose holomorphic dimensions are $(\ell/2,0)$ and $(-\ell/2,0)$ respectively. The boundary block associated to the primary of scaling dimension $\D$ equals the four-point function block where an operator of holomorphic dimensions $(\D,0)$ is exchanged between each external operator and its mirror image. All such blocks can be written in closed form \cite{Osborn:2012vt}. It is convenient to use complex coordinates $w=x+\ii z$ and $\bar{w}=x-\ii z$ to capture the two dimensional tensor structures. Denoting with $\phi^{\ell}$ a bulk operator of spin $\ell$, and with a subscript the restriction of the correlator to a single conformal block, one gets, for a boundary CFT:
 \begin{multline}
 \braket{\phi^{\ell_1}_\textup{flat}(w_1,\bar{w}_1) 
 \phi^{\ell_2}_\textup{flat}(w_2,\bar{w}_2)}_\D  \\
 = b_{\ell_1\Delta } b_{\ell_2\Delta }
 \left(\frac{\bar{w}_1-\bar{w}_2}{w_1-\bar{w}_2}\right)^{\ell_1} \left(\frac{w_1-\bar{w}_2}{w_1-w_2}\right)^{\ell_2} 
 (4\xi)^{-\D} {}_2 F_1 \left(\D-\ell_1,\D+\ell_2;2\D;-\frac{1}{\xi}\right)~.
 \label{blocks2PtFlat2}
\end{multline}
Eq. \eqref{blocks2PtFlat2} also implies a choice of normalization for the OPE coefficients. The spectral blocks are obtained by applying the Weyl transformation \eqref{AdStoFlat}:
 \begin{multline}
 \braket{\phi^{\ell_1}_\textup{AdS}(w_1,\bar{w}_1) 
 \phi^{\ell_2}_\textup{AdS}(w_2,\bar{w}_2)}_\D \\
 =  b_{\ell_1\Delta } b_{\ell_2\Delta }z_1^{-|\ell_1|} z_2^{-|\ell_2|}
 \left(\frac{\bar{w}_1-\bar{w}_2}{w_1-\bar{w}_2}\right)^{\ell_1} \left(\frac{w_1-\bar{w}_2}{w_1-w_2}\right)^{\ell_2} 
  (4\xi)^{-\D} {}_2 F_1 \left(\D-\ell_1,\D+\ell_2;2\D;-\frac{1}{\xi}\right)~.
 \label{blocks2PtAdS2}
\end{multline}
For scalar external operators, eq. \eqref{blocks2PtAdS2} correctly reduces to eq. \eqref{scalarSpecBlocks}. In a parity invariant theory, which we consider in this work, the symmetry $x \leftrightarrow -x$ implies
\beq
b_{-\ell\D } = (-1)^\ell b_{\ell\D}~.
\label{parityOnb}
\eeq
This can be checked from eq. \eqref{blocks2PtAdS2} with the help of a hypergeometric identity, or directly from the bulk-to-boundary correlator.

Since our focus is on the correlation functions of the stress tensor in AdS, let us restrict to that case and derive the constraints imposed by conservation on the boundary data. The components of definite spin are $T_{ww}$ ($\ell=2$), $T_{\bar{w}\bar{w}}$ ($\ell=-2$) and the trace, which is proportional to $T_{w\bar{w}}$. Once parity invariance is taken into account, there are four independent pairings among these operators, matching the number of tensor structures computed in section \ref{sec:twopStress}. The spectral blocks corresponding to any two-point function involving these operators can be obtained specifying eq. \eqref{blocks2PtAdS2}. Together with $b_{\Theta \D}$ defined in eq. \eqref{specDecScalar}, we shall denote the other OPE coefficient as $b_{\ell=\pm2,\D} = -\frac{1}{4} b_{T\D}$. In AdS, conservation of the stress tensor reads
\beq
\frac{(w-\bar{w})^4}{4} \partial_{\bar{w}}T_{ww}+\left(\pa_w-\frac{2}{w-\bar{w}}\right)\frac{(w-\bar{w})^4}{4} T_{w\bar{w}}  = 0~,
\eeq
together with the equation obtained swapping $w$ and $\bar{w}$, which we shall not need since we already assumed parity invariance.
 By replacing this equation in the two-point function of the stress tensor, one gets three constraints, which simply are eq. \eqref{ConsVec} pulled down to physical space. Requiring the identity to hold block by block, one obtains
\beq
(\D-2)\, b_{T\D}+\D\, b_{\Theta \D}=0~.
\label{bOPECons}
\eeq 
It is interesting to notice that the trace of the stress tensor does not exchange a state of dimension $\Delta=2$. Hence, the pole in eq. \eqref{specSumRuleAdS2} is actually absent: this is the first hint that the sum rule might be valid without the restriction \eqref{BoundaryGap}.

In fact, eqs. \eqref{blocks2PtAdS2} and \eqref{bOPECons} allow to give another derivation of the sum rule \eqref{specSumRuleAdS2}. It is not hard to see that the spectral blocks in eq. \eqref{blocks2PtAdS2} have individually the correct small $\xi$ behavior to reproduce the two-point function of the stress tensor in the UV. More precisely, the blocks in the two point function of $T_{ww}$ behave as:
\beq
\braket{T_{ww}(w_1,\bar{w_1})T_{ww}(w_2,\bar{w_2})}_\D \overset{\xi\to 0}{\approx}
 \frac{6\,\Gamma(2\D)}{4^\D \Gamma(\D+2)^2}  \frac{b_{T\D}^2}{(w_1-w_2)^4}~,
 \label{smallXiTblocks}
\eeq
while the blocks in the correlators of the other components are less singular. This is precisely the behavior of a $2d$ CFT in flat space, where the stress tensor is holomorphic, and it is consistent with eq. \eqref{hiUVAdS}, up to the appropriate change of basis, if we set 
\beq
C_T=\sum_{\D>0} \frac{24\, \Gamma(2\D)}{4^\D \Gamma(\D+2)^2} b_{T\D}^2~.
\label{sumRuleAdS2FromTT}
\eeq
Notice that the sum now runs over all the operators exchanged by the $\ell=2$ component of the stress tensor, which excludes the identity operator ($\D=0$). This can be quickly seen, for instance, in embedding space: the one-point function of the stress tensor is proportional to the induced metric \eqref{Gind}, which does not have a $(ww)$ component. The result is also consistent with the conservation equation \eqref{bOPECons}.
Trading $b_{T\D}$ for $b_{\Theta\D}$ via eq. \eqref{bOPECons}, one sees that eq. \eqref{sumRuleAdS2FromTT} coincides with eq. \eqref{specSumRuleAdS2}, except now the condition \eqref{BoundaryGap} is not necessary anymore. This new derivation is not quite rigorous yet, because we did not prove that the sum on the right hand side of eq. \eqref{sumRuleAdS2FromTT} converges. The problem is that the approximation \eqref{smallXiTblocks} is obtained at fixed $\D$, and one might wonder if it is legal to exchange the small $\xi$ limit and the sum over spectral blocks. There are various ways to prove that eq. \eqref{sumRuleAdS2FromTT} is correct. In the next section, we provide asymptotics for the spectral density which in particular imply the convergence of the sum. 

\paragraph{The other sum rules}-- namely eqs. (\ref{s1bbb}, \ref{s3bbb}) -- can also be derived from the short distance behavior of the conformal block decomposition of the two-point function of the stress tensor (see appendix \ref{app:details}). 
This leads to 
\beq
 \D_\mathcal{V}  \braket{\Theta} = \sum_{\Delta >0} 
\frac{ 2\sqrt{\pi} (\Delta-2) \Gamma\!\left(\D+\frac{1}{2}\right)}{  \D  \Gamma(\D+2)} b^2_{T\Delta}~,
\label{s1better}
\eeq
\beq
C_3(0)= \sum_{\Delta>0 } 
\frac{2(\D -2)\Gamma\!\left(\D+\frac{1}{2}\right)
\left(
 H_{\D-1} - 1
\right)
}{ \sqrt{\pi}  \D   \Gamma(\D+2)} b^2_{T\Delta}~,
\label{s3better}
\eeq
which are equivalent to (\ref{s1bbb}, \ref{s3bbb}) using \eqref{bOPECons}.
This derivation shows that  the condition \eqref{BoundaryGap} is actually not necessary.

\subsection{Asymptotics of the spectral coefficients and convergence of the sum rules}
\label{subsec:taubSpectral}

While the short distance singularity in the two-point function of $T_{ww}$ is accounted for block by block, the same cannot be said for the two-point function of the trace $\Theta$, eq. \eqref{Ashort}. In this case, only the infinite sum in the spectral decomposition \eqref{specDecScalar} can reproduce the non-analytic ultraviolet behavior. Hence, the strength of the singularity is sensitive to the large $\D$ behavior of the coefficients $b_{\Theta\D}^2$. The solution to this problem falls into the class of Tauberian theorems that has played an important role in the conformal bootstrap literature \cite{Pappadopulo:2012jk}. A simple bound on the growth of the OPE coefficients can be found by moving from $\xi$ to a $\rho$-coordinate, defined in such a way that all the descendants in eq. \eqref{specDecScalar} contribute with positive OPE coefficients\footnote{On the other hand, the power series in $\xi$ does not have positive coefficients, as one immediately sees from the minus sign in the last argument of the Hypergeometric function in eq. \eqref{scalarSpecBlocks}.}. Then one estimates the averaged growth of all OPE coefficients, primaries and descendants alike, which bounds the asymptotics of the $b_{\Theta\D}^2$ \cite{Lauria:2017wav}. 

However, the precise asymptotics of the $b_{\Theta\D}^2$ alone can also be found in the literature \cite{Fitzpatrick:2012yx,Komargodski:2012ek,Qiao:2017xif}, owing to the coincidence of the $d=1$ spectral blocks \eqref{scalarSpecBlocks} with the collinear conformal blocks, see eq. \eqref{2ptblockTo1d}. Let us define the following spectral density for the connected correlator:
\beq
\rho(x)=\sum_{\D>0} b_{\Theta\D}^2 \delta(x-\D)~.
\eeq
Then, as reviewed in appednix \eqref{subsec:TaubRQ}, \cite{Qiao:2017xif} proved that, given the singularity \eqref{Ashort},
\beq
\int_{0}^\D\! dx\, \sqrt{x}\,\rho(x) \overset{\D \to \infty}{\approx}
\frac{2\sqrt{\pi}\, (2-\D_\mc{V})^2\la^2}{\D_\mathcal{V}\Gamma(\D_\mathcal{V})^2}  \left(\frac{\D}{2}\right)^{2\D_\mathcal{V}}~.
\label{tauberian}
\eeq
The translation of the result in \cite{Qiao:2017xif} to eq. \eqref{tauberian} is transparent if we choose the coordinate
\beq
u=\frac{1}{\xi+1}~,
\eeq
in terms of which the spectral block reads coincide with the collinear conformal blocks,
and the short distance singularity \eqref{Ashort} becomes 
\beq
A(z) \approx (2-\D_\mc{V})^2 \la^2 [4(1-u)]^{-\D_\mathcal{V}}~.
\eeq

Armed with the result \eqref{tauberian}, we can go back to eqs. \eqref{sumRuleAdS2FromTT}, \eqref{s1better} and \eqref{s3better} and prove the convergence of the three sums. One simply writes the sums in terms of an integral over $\rho(x)$, using eq. \eqref{bOPECons}, then integrates by parts to make the integral on the l.h.s. of eq. \eqref{tauberian} appear. Finally, the asymptotics on the r.h.s. of eq. \eqref{tauberian} tell us that
\beq
\D_\mathcal{V}<2 \quad \implies \quad \textup{\eqref{sumRuleAdS2FromTT} converges.}
\eeq
On the other hand, the sum rules \eqref{s1better} and \eqref{s3better} require a more relevant flow:
\beq
\D_\mathcal{V}<1 \quad \implies \quad \textup{\eqref{s1better} and \eqref{s3better} converge.}
\eeq
Both conditions agree with the results from section \ref{sec:twopStress}---see comments after eq. \eqref{IntLhsAdS2}.

\section{Form factors in AdS}
\label{sec:formFact}

In the quest for a positive semi-definite set of conditions involving bulk and boundary OPE data, we need a sum rule which contains both the $b_{T\D}$ coefficients and the three-point function coefficients $c_{\p\p\D}$. The appropriate observable to study is then the correlator of the stress tensor with two boundary operators, whose OPE decomposition is easily seen to contain the product $b_{T\D}c_{\p\p\D}$. Like with flat space physics, we will refer to these three-point functions as form factors of the stress tensor: as we shall see in section \ref{sec:flat}, this is more than an analogy. 

In subsections \ref{subsec:threePointd} and \ref{subsec:threePoint2}, we study the OPE decomposition of the form factors in $d+1$ and in $2$ dimensions respectively. The rest of the section is about two dimensional physics. In \ref{subsec:naive}, we derive a sum rule for the product $b_{T\D}c_{\p\p\D}$ by integrating the stress tensor against a Killing vector. However, the sum, while valid when the scaling dimension of the boundary operators is small, does not converge in general. We solve this problem in subsection \ref{subsec:localBlocks}, thanks to an alternative OPE decomposition of the form factor, the local block decomposition \cite{Levine:2023ywq}. The new sum rule is absolutely convergent for any dimension of the boundary operators.
 Along the way, we study the convergence properties of the local block decomposition, which follow from the Cauchy-Schwarz inequality after a number of technical steps. In particular, in subsection \ref{sec:convergenceFF}, we prove that the convergence is uniform in a large region in cross ratio space, and we derive bounds on the rate of convergence.

\subsection{Three-point functions in AdS$_{d+1}$}
\label{subsec:threePointd}

In this section, we discuss the kinematics of the three-point function of the stress tensor with two boundary operators. We only consider a scalar boundary primary operator $\phi$ with scaling dimension $\Delta_\phi$, inserted at the boundary points $P_1$ and $P_2$. The stress tensor is inserted at the bulk point $X$. 
We consider the trace of the stress tensor $\Theta$ and the traceless symmetric part of the stress tensor $T_{\mu\nu}$ separately.

The three-point function involving the trace of the stress tensor can be written as follows:
\begin{equation}
    \langle \phi(P_1) \phi(P_2) \Theta(X) \rangle= \frac{1}{(-2P_1\cdot P_2)^{\Delta_\phi}}\sum_\Delta c_{\phi\phi\Delta}b_{\Theta\Delta} g_\Delta(\chi) ~.
    \label{3pGdelta}
\end{equation}
Each function $g_\Delta(\chi)$ sums up the contribution of a single conformal family, and
 $\chi$ is invariant under the isometries:
\begin{equation}\label{ChiEmbedding}
    \chi=\frac{-P_1 \cdot P_2}{2(X \cdot P_1)(X \cdot P_2)}~.
\end{equation}
In Poincaré coordinates, eq. \eqref{PoincarePatch}, the cross ratio reads
\begin{equation}
\chi = \frac{z^2(x_1-x_2)^2 }{\left(z^2+(x-x_1)^2\right)\left(z^2+(x-x_2)^2\right)}~.
\label{ChiPoincare}
\end{equation}
Here, $x^a_1,\,x^a_2$ and $(z,x^a)$ are the coordinates of the boundary operators and of the bulk operator respectively. In eq. \eqref{3pGdelta}, the bulk-to-boundary OPE coefficients $b_{\Theta\Delta}$ are the same as in eq. \eqref{specDecScalar}, defined by eq. \eqref{OPEAdS}. On the other hand, $c_{\phi\phi\Delta}$ is the coefficient of the three-point function of two $\phi$'s with the primary operator of dimension $\Delta$.
The conformal block $g_\Delta(\chi)$ solves a hypergeometric differential equation, which is derived from the action of the quadratic Casimir operator of $SO(d+1,1)$ on the two boundary operators:
\begin{equation}
    4\chi^2(\chi - 1)\partial_\chi^2 g_\Delta(\chi) + 4\chi\left(\frac{d}{2}-1+\chi\right) \partial_\chi g_\Delta(\chi)=-C_{\Delta}g_\Delta(\chi),
\end{equation}
where $C_\Delta=\Delta(\Delta-d)$ is the Casimir eigenvalue.
 The correct solution is selected by the boundary condition $g_\Delta(\chi)\sim \chi^{\Delta/2}$, as $\chi \rightarrow 0$, dictated by the OPE.  One obtains
\begin{equation}
g_\Delta(\chi) =  \chi^{\Delta/2} {}_2F_1 \left(\frac{\Delta}{2},\frac{\Delta}{2}; \Delta+1-\frac{d}{2};\chi \right)~.
\label{block3p0}
\end{equation}
The correlator can then be decomposed as follows: 
\begin{equation}\label{3ptFctAdS2Trace}
    \langle \phi(P_1) \phi(P_2) \Theta(X)\rangle=\frac{1}{(-2P_1\cdot P_2)^{\Delta_{\phi}}}\sum_{\Delta} c_{\phi\phi\Delta}b_{\Theta\Delta} \chi^{\Delta/2} {}_2F_1 \left(\frac{\Delta}{2},\frac{\Delta}{2}; \Delta+1-\frac{d}{2};\chi \right)~.
\end{equation}

The three-point function involving the traceless symmetric part of the stress tensor can be written, as usual, as a linear combination of tensor structures. There are two structures for $d>1$, and only one for $d=1$.\footnote{This follows from similar considerations to the ones in footnote \ref{foot:groupCounting2}. The three insertions can be brought to a plane, in the configuration of figure \ref{fig:interface}. The residual symmetry is $SO(d-1)\times \mathbb{Z}_2$. Counting the singlets of the spin two representation of $SO(d+1)$ under this subgroup, one gets the quoted result.} The tensor structures are conveniently written by uplifting $T_{\mu\nu}$ to embedding space \cite{Costa:2014kfa} and contracting the resulting tensor with polarization vectors 
\begin{equation}
    W^M W^N\langle \phi(P_1) \phi(P_2) T_{MN}(X)\rangle = \frac{h_1(\chi) T_1 + h_2(\chi) T_2}{(-2P_1 \cdot P_2)^{\Delta_{\phi}}}~,
    \label{3ptFctAdSPolarizations}
\end{equation}
where the polarization vector $W$ satisfies $W\cdot X=W^2=0$   and $T_{MN}$ relates to $T_{\mu \nu}$ as 
\begin{equation}
    T_{\mu \nu}-\frac{1}{d+1}g_{\mu\nu} \Theta=\frac{\partial X^M}{\partial x^\mu}\frac{\partial X^N}{\partial x^\nu} T_{MN}~,
\end{equation}
where $g_{\mu\nu}$ is the AdS metric.
The tensor structures must be polynomials of second order in $W$, Lorentz invariant in embedding space, and homogeneous of degree zero in $P_1$ and $P_2$: 
\begin{equation}
    T_1=\frac{(W \cdot P_1)(W \cdot P_2)}{(X \cdot P_1)(X \cdot P_2)}~, \qquad T_2=\frac{(W\cdot P_1)^2}{(X \cdot P_1)^2} + \frac{(W \cdot P_2)^2}{(X \cdot P_2)^2}~.
    \label{TensorStructures}
\end{equation}
The two functions $h_1$ and $h_2$ in eq. \eqref{3ptFctAdSPolarizations} can be expanded in conformal blocks:
\begin{equation}
h_1(\chi) = \sum_{\Delta,\ell} h_{1,(\Delta,\ell)}(\chi)~, \qquad
h_2(\chi) = \sum_{\Delta,\ell}  h_{2,(\Delta,\ell)}(\chi)~, \qquad
\end{equation}
where the spin $\ell$ takes the values 0 and 2 (spin 1 primaries would be exchanged in a correlator involving two non-identical boundary operators).
In this case, the Casimir equation turns into a system of  coupled differential equations:
\begin{multline}
4 (\chi-1) \chi^2 \partial_\chi^2 h_{1,(\Delta,\ell)}(\chi)+4\chi\left(\frac{d}{2}-2+3\chi\right)\partial_\chi h_{1,(\Delta,\ell)}(\chi)+ 8\chi \partial_\chi h_{2,(\Delta,\ell)}(\chi)+4(\chi-1)h_{1,(\Delta,\ell)}(\chi)
    \\=-C_{\Delta,\ell} h_{1,(\Delta,\ell)}(\chi) ~, \notag
\end{multline}
\begin{multline}
4(\chi-1)\chi^2 \partial_\chi^2 h_{2,(\Delta,\ell)}(\chi)+2\chi \partial_\chi h_{1,(\Delta,\ell)}(\chi)+4\chi\left(\frac{d}{2}-2+3\chi\right) \partial_\chi h_{2,(\Delta,\ell)}(\chi) +2h_{1,(\Delta,\ell)}(\chi)\\=-C_{\Delta,\ell} h_{2,(\Delta,\ell)}(\chi)~,
\label{CasimirEqAdS}
\end{multline}
with $C_{\Delta,\ell}=\Delta(\Delta-d) + \ell(\ell+d-2)$. We will find the blocks in closed form in two dimensions. In the general case, efficient techniques are available to compute them order by order in powers of the cross ratio \cite{Penedones:2015aga, Costa:2016xah, Lauria:2018klo}.

\subsection{Form factor of the traceless part of $T_{\mu\nu}$ in $\mathrm{AdS}_2$}
\label{subsec:threePoint2}

As mentioned, in two dimensions the tensor structures are not independent:
\begin{equation}
	2(2\chi - 1)T_1 + T_2=0~.
\end{equation}
The three-point function can thus be written in terms of a single tensor structure, say $T_1$, and its coefficient can be expanded in a sum of conformal blocks $h_{\Delta}(\chi)$. Notice that there is no spin label, since the rotation group reduces to $\mathbb{Z}_2$.
The Casimir equation simplifies to
\begin{equation}
4\chi^2(\chi-1)\partial_\chi^2 h_{\Delta}(\chi) + 2\chi(2\chi-1)\partial_\chi h_{\Delta}(\chi) - 4\chi h_{\Delta}(\chi) = -\Delta(\Delta-1)h_{\Delta}(\chi).
\end{equation}
The solution with the appropriate boundary condition is
\begin{equation}
h_{\Delta}(\chi)=\chi^{\frac{\Delta}{2}} {}_2F_1 \left(\frac{\Delta}{2}+1,\frac{\Delta}{2}-1; \Delta + \frac{1}{2};\chi\right)~.
\label{blocks3p2}
\end{equation}
The same solution for the block can be found applying the method of images, as described in subsection \ref{subsec:twopBlocksStress}---see also eq. \eqref{sl2blockwtoz2}.
The three-point function of two boundary primaries and a bulk traceless symmetric tensor in AdS$_{2}$ is thus given by
\begin{equation}\label{3ptFctAdS2TracelessSymm}
        W^M W^N\langle \phi(P_1) \phi(P_2) T_{MN}(X)\rangle 
        =\frac{T_1}{(-2P_1 \cdot P_2)^{\Delta_{\phi}}}\sum_{\Delta}  c_{\phi\phi \Delta}b_{T\Delta}\, \chi^{\frac{\Delta}{2}} {}_2F_1\left(\frac{\Delta}{2}+1,\frac{\Delta}{2}-1; \Delta + \frac{1}{2};\chi\right)~,
\end{equation}
where we incorporated the OPE coefficients, with a normalization compatible with the one fixed in eq. \eqref{blocks2PtAdS2} for $b_{T\Delta }$. 

\subsection{Ward identities and a (naive) sum rule}
\label{subsec:naive}

\begin{figure}[t]
\centering
\begin{tikzpicture}[scale=1.2]
\draw[thick] (0,0) circle (2);
\draw[thick,blue] (-2,0).. controls (0,0.8) .. (2,0);
\draw[thick, blue, ->] (0,0.6) -- (0,1);
\node[blue] at (0.4,1) {$d\Sigma^\mu$}; 
\node[blue] at (-1,0) {$\Sigma$};
\filldraw[black] (0,2) circle (1.3pt);
\node [black] at (0,2.4) {$\phi(P_1)$};
\filldraw[black] (1,-1.73) circle (1.3pt) ;
\node [black] at (1,-2.2) {$\phi(P_2)$};
\node[black] at (2,1.6) {$\partial$AdS};
\end{tikzpicture}
\caption{Integrating along the arbitrary surface $\Sigma$ against the normal vector $d\Sigma^\mu$, as depicted in the figure, yields a Ward identity.}
\label{fig:ward}
\end{figure}
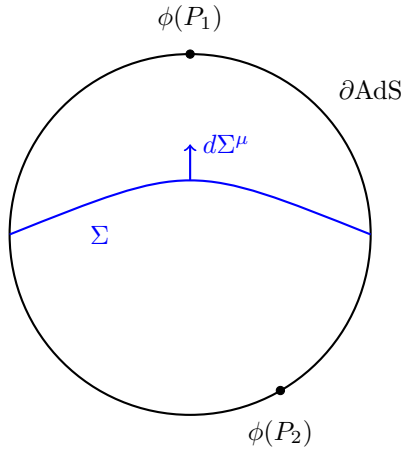

Given a Killing vector $\xi^\mu$, the following integral expression of the Ward identities holds
\begin{equation}\label{WardIdAdS}
    \left\langle \phi(P_1) \left(\int_\Sigma d\Sigma^\mu \xi^\nu T_{\mu\nu} \right) \phi(P_2) \right\rangle_\textup{connected} =\left\langle \phi(P_1) \delta_{\xi} \phi(P_2)\right\rangle,
\end{equation}
where $d\Sigma^\mu$ is normal to the codimension-one surface $\Sigma$, oriented as in figure \ref{fig:ward}. The right hand side contains the variation of one of the boundary operators with respect to the isometry, which acts as a conformal transformation. Notice that eq. \eqref{WardIdAdS} can be derived without extra assumptions on the contact terms of the bulk stress tensor in the presence of boundary operators. We show this explicitly in appendix \ref{app:ward}.

Our aim is to turn eq. \eqref{WardIdAdS} into a sum rule for the product of OPE coefficients $c_{\phi\phi \Delta}b_{T \Delta}$. An obvious way of doing this is replacing the conformal block decompositions \eqref{3ptFctAdS2Trace} and \eqref{3ptFctAdS2TracelessSymm} in eq. \eqref{WardIdAdS}, recalling the relation \eqref{bOPECons} between $b_{\Theta \Delta}$ and $b_{T\Delta}$. This strategy was pursued in the context of a boundary CFT, and in general number of dimensions, in \cite{Herzog:2021spv}. However, performing the integral in eq. \eqref{WardIdAdS} term by term might be dangerous, because as we will see the integration touches the boundary of the region of convergence of the OPE. Let us push through nonetheless: we will later discuss the issue and find a more sophisticated solution.

\begin{figure}[t]
\centering
\begin{tikzpicture}[scale=1.2]
\begin{scope}
    \clip (-2.6,0) rectangle (2.6,2.6);
    \draw[dashed] (0,0) circle (2);
\end{scope}
    \draw[thick,->] (-2.5,0) -- (2.6,0);
    \node[black] at (2.6,-0.2) {$x$};
    \filldraw[black] (-2,0) circle (1.3pt);
    \node[black] at (-2.2,0.2) {$\phi$};
    \node[black] at (-2.2,-0.2) {$-1$};
    \filldraw[black] (2,0) circle (1.3pt);
    \node[black] at (2.2,0.2) {$\phi$};
    \node[black] at (2.1,-0.2) {$1$};
    \draw[thick,blue] (0,0) -- (0,2.8);
    \node[blue] at (-0.4,2.5) {$\Sigma$};
    \filldraw[black] (0,1.5) circle (1.3pt);
    \node[black] at (0.4,1.5) {$T^{\mu\nu}$};
    \draw[very thick,black] (0,-0.1) -- (0,0.1);
\end{tikzpicture}
\caption{In Poincaré coordinates, we choose the surface of integration $\Sigma$ to be on the $z$-axis. As we integrate the spectral sum term-by-term, a convergence issue may arise when we cross the dashed surface.}
\label{fig:interface}
\end{figure}
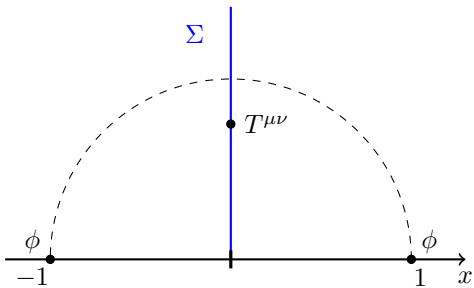
Let us work in Poincaré coordinates, eq. \eqref{PoincarePatch}, and choose for the isometry a translation along the boundary: $\xi^\mu$ is constant and directed along the $x$ coordinate. Without loss of generality, we pick the surface $\Sigma$ to be a straight line at $x=0$, separating the two insertions---see figure \ref{fig:interface}. We get the following equation: 
\begin{equation}
-\frac{1}{2}(x_1-x_2)^{2\Delta_\phi+1} \int_0^\infty\! dz\,\langle \phi(x_1) \phi(x_2)T_{xx}(x=0,z) \rangle
= \Delta_\phi~.
\label{sum_rule_interm}
\end{equation}
It is convenient to use the isometries to further specify $x_1=1$, $x_2=-1$, see figure \ref{fig:interface}. Replacing eqs. \eqref{3ptFctAdS2Trace} and \eqref{3ptFctAdS2TracelessSymm}, appropriately projected to physical space, into eq. \eqref{sum_rule_interm}, we get
\beq
\frac{1}{2}\int_0^1\! dz \left(1+\frac{1}{z^2}\right)
\sum_{\Delta>0}c_{\phi \phi \D} b_{T \D}\left[h_\Delta\big(\chi(z)\big) +
\frac{\Delta-2}{\Delta}g_\Delta\big(\chi(z)\big)\right]=\Delta_\phi~,
\label{sumRuleBlocks}
\eeq
where the relation between $\chi$ and $z$ is obtained from eq. \eqref{ChiPoincare}:
\begin{equation}
\chi = \frac{4z^2}{(1+z^2)^2}~.
\label{chiofz}
\end{equation}
In eq. \eqref{sumRuleBlocks}, we dropped the identity operator, which is not exchanged in the connected correlator.
Integrating the block expansion term by term, one gets
\beq
\sum_{\D>0} 
\frac{\pi^{\frac{1}{2}} \D\, \Gamma\left(\D+\frac{1}{2}\right)}{(\D+1) \Gamma\left(\frac{\D}{2}+1\right)^2}
c_{\phi \phi \D} b_{T \D}=\D_\phi ~, \qquad \textup{naive.}
\label{NaiveSumRule}
\eeq

It should be noted that, if the boundary has a relevant operator $\D$ which couples to the stress tensor, the corresponding integral in eq. \eqref{sumRuleBlocks} does not converge. Eq. \eqref{NaiveSumRule} still holds in this case as written, so that the relevant operator simply contributes a coefficient equal to the analytic continuation in $\D$ of the expression for $\D>1$. The reason for this is explained in appendix \ref{subsec:improved}, where we point out that the correct Ward identity is always obtained by simply dropping the divergent part of the integral: this is an extension of the works \cite{Breitenlohner:1982jf,Klebanov:1999tb} to the generic interacting situation. The correctness of eq. \eqref{NaiveSumRule} in the presence of a relevant boundary operator can be checked in the free scalar case with $m^2<0$, with boundary conditions such that $\D_\phi<1/2$.

While eq. \eqref{sum_rule_interm} is a true formula, eq. \eqref{NaiveSumRule} is unfortunately not guaranteed to converge. For instance, it diverges for a free scalar with sufficiently large mass, as we shall see in subsection \ref{subsec:free_scalar}. The illegal step in the computation above is swapping the integral with the sum over conformal blocks in eq. \eqref{sumRuleBlocks}. To understand the issue, we must assess the region of convergence of the OPE. Let us refer to figure \ref{fig:interface}.
The OPE of the stress tensor with the boundary of AdS converges absolutely for $z<1$ \cite{Pappadopulo:2012jk,Bianchi:2022ulu}. An inversion $x^\mu \to x^\mu/(x_a x^a+z^2)$ where $x^\mu=(x^a,z)$ is    an isometry of AdS, and 
sends $z \to 1/z$ in the configuration of figure \ref{fig:interface}.
Therefore the OPE converges absolutely for $z>1$ as well. In fact, one can prove that the OPE converges in the complex region $|z|<1$, $z \in \mathbb{C},$ and in all configurations related to it by isometries \cite{Hogervorst:2013sma,Bianchi:2022ulu}. In the Euclidean kinematics, this leaves the point $z=1$ outside. Were the sum \eqref{NaiveSumRule} positive, eq. \eqref{sum_rule_interm} would still guarantee convergence, but the product $c_{\phi\phi\Delta}b_{T\Delta}$ does not have fixed sign. In fact, in the case of a free theory it oscillates---see eqs. \eqref{CoeffbTScalar} and \eqref{bTCFermion}. 

Luckily, it is possible to derive from eq. \eqref{sum_rule_interm} a sum rule which converges absolutely for any value of $\Delta_\phi$. The key is to use a different representation of the correlator of two boundary and one bulk operators, which involves the same CFT data but different functions of the cross ratio. This is the \emph{local block} decomposition, introduced in \cite{Levine:2023ywq}.

\subsection{The local block decomposition and a better sum rule}
\label{subsec:localBlocks}

\begin{figure}[t]
\begin{subfigure}{0.5\textwidth}
\centering
\begin{tikzpicture}[scale=0.85]
    \draw (2.7,2.1) -- (2.7,1.7);
    \draw (2.7,1.7) -- (3.1,1.7);
    \node at (3,2.03) {$\chi'$};
    \draw [decorate,decoration={zigzag,amplitude=1.2pt,segment length =2 pt}] (-3,0) -- (-0.7,0);
    \draw[dashed] (0.7,0) -- (3.1,0);
    \filldraw[black] (-0.7,0) circle (1.3pt);
    \node[black] at (-0.7,-0.3) {0}; 
    \filldraw[black] (0.7,0) circle (1.3pt);
    \node[black] at (0.7,-0.3) {1};
    \filldraw[black] (0.6,0.8) circle (1.3pt);
    \node[black] at (0.8,0.8) {$\chi$};
    \draw[black,thick, ->] (0.46,1.05) arc (135:495:0.36); 
\end{tikzpicture}
\end{subfigure}
\begin{subfigure}{0.5\textwidth}
\centering
\begin{tikzpicture}[scale=0.85]
    \draw (2.7,2.1) -- (2.7,1.7);
    \draw (2.7,1.7) -- (3.1,1.7);
    \node at (3,2.03) {$\chi'$};
    \draw [decorate,decoration={zigzag,amplitude=1.2pt,segment length =2 pt}] (-3,0) -- (-0.7,0);
    \draw[dashed] (0.7,0) -- (3.1,0);
    \filldraw[black] (-0.7,0) circle (1.3pt);
    \node[black] at (-0.7,-0.3) {0}; 
    \filldraw[black] (0.7,0) circle (1.3pt);
    \node[black] at (0.7,-0.3) {1};
    \filldraw[black] (0.6,0.8) circle (1.3pt);
    \node[black] at (0.8,0.8) {$\chi$};
    \draw[thick, ->] (-3,0.25) -- (-1.7,0.25);
    \draw[thick] (-1.7,0.25) -- (-0.7,0.25);
    \draw[thick] (-0.7,0.25) arc (91:-91:0.39);
    \draw[thick,->] (-0.7,-0.53) -- (-1.8,-0.53);
    \draw[thick] (-1.8,-0.53) -- (-3,-0.53);
\end{tikzpicture}
\end{subfigure}
\caption{Analytic structure of the conformal blocks and of the form factor, with the integration contour used to define the dispersion relation \eqref{dispersiveG}. The point $\chi=1$ is a branch point for each conformal block, but it is regular for the full correlator.}
\label{fig:chiPath}
\end{figure}

This subsection is dedicated to the derivation of the local block decomposition defined in \cite{Levine:2023ywq} and, through that, of a new version of the sum rule.
 We postpone to subsection \ref{sec:convergenceFF} the discussion of the convergence of both the former and the latter, which is technical, albeit important and new.

In order to define the local blocks, we start by mapping our considerations on the convergence of the OPE onto the $\chi$ complex plane. From eq. \eqref{chiofz}, we see that the $|z|=1,$ Re$z>0$ semi-circle is mapped to the line $\chi\geq 1$, the map being two-to-one. Correspondingly, the analytic structure of the conformal blocks \eqref{block3p0} and \eqref{blocks3p2} is the one depicted in figure \ref{fig:chiPath}, where both the wiggly and the dashed lines are cuts. However, while $\chi=0$ is a branch point dictated by the OPE, $\chi=1$ is not a special point for the correlator, which is analytic there---see figure \ref{fig:interface}, where  it corresponds to the bulk operator being inserted at $z=1$. Hence, the full correlator is analytic along the dashed line in figure \ref{fig:chiPath}, contrary to the conformal blocks \cite{Hamilton:2005ju,Kabat:2016zzr,Behan:2020nsf,Bianchi:2022ulu,Levine:2023ywq}. This justifies the following dispersive representation for any function $\mathcal{G}(\chi)$ with the same analytic structure as a form factor:
\begin{equation}
\mathcal{G}(\chi) = \chi^\alpha \int_{-\infty}^0 \frac{d\chi'}{2\pi \ii} \frac{1}{\chi'-\chi} \textup{Disc} \left(\frac{\mathcal{G}(\chi')}{\chi'^\alpha} \right)~, \qquad 
\alpha_\textup{min}<\alpha<\alpha_\textup{max}
\label{dispersiveG}
\end{equation}
In the previous equation, $\alpha_\textup{min}$ is fixed by the requirement that, in deforming the integration contour from the first to the second panel in figure \ref{fig:chiPath}, the arc at infinity can be dropped. In subsection \ref{sec:convergenceFF}, we will find that in general\footnote{To be precise, in the case of the traceless part of $T_{\mu\nu}$, this statement is valid for the modified form factor defined in eq. \eqref{TildeGT}. As explained in subsection \ref{subsec:spin2LBConv}, this leads to the correct local block decomposition \eqref{localBlockT}.} 
\beq
\al_\textup{min}=\D_\p+\frac{\D_\mc{V}}{2}~,
\label{alphaMin}
\eeq
where $\D_\mc{V}$ is the dimension of the relevant operator defined in eq. \eqref{Sdeformed}  which starts the RG flow.
Similarly, the upper bound on $\alpha$ is required to drop the arc around $\chi=0$. When $\mathcal{G}(\chi)$ is a connected form factor, this is a simple condition coming from the OPE: 
\begin{equation}
\alpha_\textup{max}=\frac{\Delta_\textup{gap}}{2}+1~,
\label{alphaMax}
\end{equation} 
with $\Delta_\textup{gap}$ the gap above the identity in the spectrum.  In fact, as we shall point out at the end of this subsection, keeping the arc around $\chi=0$ is easy, and the final form of the local block decomposition will be free of the limitation \eqref{alphaMax}. Notice that restricting to the connected component of the form factor does not change the analyticity properties of $\mathcal{G}(\chi)$, nor the validity of eq. \eqref{alphaMin}. Indeed, the identity block has no branch point at $\chi=1$, and in particular only $\Theta$ exchanges it, due to the isometries---see eq. \eqref{bOPECons}. Hence, the disconnected component is a constant, and dropping the arc at infinity only requires the extra condition $\al>0$, which is weaker than eq. \eqref{alphaMin}.


The local block decomposition is obtained by replacing the conformal block decomposition in the right hand side of eq. \eqref{dispersiveG}, and swapping the OPE sum with the integration over $\chi'$. We will prove the swapping property in subection \ref{sec:convergenceFF}. This leads to the definition of the local blocks, which for spin 0 and spin 2 bulk operators read respectively
\begin{multline}
G^\alpha_\Delta(\chi) = \chi^\alpha \int_{-\infty}^0 \frac{d\chi'}{2\pi \ii} \frac{1}{\chi'-\chi} \textup{Disc} \left(\frac{g_\Delta(\chi')}{\chi'^\alpha} \right) \\
=\chi^\alpha \sin\left[\frac{\pi}{2} \left(\Delta-2\alpha\right)\right]
\int_{-\infty}^0 \frac{d\chi'}{\pi} \frac{1}{\chi'-\chi} |\chi'|^{\Delta/2-\alpha}
{}_2 F_1 \left(\frac{\Delta}{2},\frac{\Delta}{2}; \Delta+\frac{1}{2};\chi' \right)
 ~,
 \label{localBlock0}
 \end{multline}
 \begin{multline}
H^\alpha_\Delta(\chi) = \chi^\alpha \int_{-\infty}^0 \frac{d\chi'}{2\pi \ii} \frac{1}{\chi'-\chi} \textup{Disc} \left(\frac{h_\Delta(\chi')}{\chi'^\alpha} \right) \\
=\chi^\alpha \sin\left[\frac{\pi}{2} \left(\Delta-2\alpha\right)\right]
\int_{-\infty}^0 \frac{d\chi'}{\pi} \frac{1}{\chi'-\chi} |\chi'|^{\Delta/2-\alpha}
{}_2F_1 \left(\frac{\Delta}{2}+1,\frac{\Delta}{2}-1; \Delta + \frac{1}{2};\chi'\right)~.
\label{localBlock2}
\end{multline}
The conformal blocks $g_\Delta(\chi)$ and $h_\Delta(\chi)$ were defined in eqs. \eqref{block3p0} and \eqref{blocks3p2} respectively.  There are closed form expressions for the local blocks, which look especially simple in Mellin space. We derive them in appendix \ref{app:local_blocks}---eqs. \eqref{MellinLocalBlock}, \eqref{LocalBlockClosed0} and \eqref{LocalBlockClosed2}. Furthermore, a variety of useful asymptotics for the local blocks are derived in appendix \ref{app:asymptBlocksFF}. Convergence of the integrals require $0<\alpha<\Delta/2+1$ and  $1<\alpha<\Delta/2+1$ respectively for $G_\D^\al$ and $H_\D^\al$, the lower bound coming, again, from large $\chi$ and the upper bound from $\chi \sim 0$. The upper bound in particular explains eq. \eqref{alphaMax}.

As explained, eq. \eqref{dispersiveG} translates into alternative decompositions to eqs. \eqref{3ptFctAdS2Trace} and \eqref{3ptFctAdS2TracelessSymm}, which involve the same OPE data:
\begin{align}
 \langle \phi(P_1) \phi(P_2) \Theta(X)\rangle &=\frac{1}{(-2P_1\cdot P_2)^{\Delta_{\phi}}}\sum_{\Delta>0} c_{\phi\phi\Delta}b_{\Theta\Delta} G_\D^\al(\chi)  ~,
 \label{localBlockTheta}\\
 W^M W^N\langle \phi(P_1) \phi(P_2) T_{MN}(X)\rangle 
        &=\frac{T_1}{(-2P_1 \cdot P_2)^{\Delta_{\phi}}}\sum_{\Delta>0}  c_{\phi\phi \Delta}b_{T\Delta}\, H_\D^\al(\chi)~.
        \label{localBlockT}
\end{align}
Recall that, as in the rest of the paper, the disconnected component---\emph{i.e.} the identity operator in the OPE---has been subtracted out.
 As explained in \cite{Levine:2023ywq}, the existence of two inequivalent decompositions is related to the existence of infinitely many sum rules, which the OPE data must satisfy. For us, the crucial difference between the two expansions is that
\begin{center} 
  \emph{the local block decomposition converges uniformly in $\chi \in [0,1]$, for 
  \beq
  \alpha>\Delta_\phi+\frac{\Delta_\mathcal{V}}{2}~,
  \label{alphaBoundLocalBlock}
  \eeq
  where $\Delta_\phi$ is the dimension of the boundary operator and $\Delta_\mathcal{V}$ is the dimension of the deformation which triggers the RG flow---see eq. \eqref{Sdeformed}.} 
  \end{center}
  
We will prove (a generalization of) this fact in the next subsection.

To obtain the new sum rule, it is sufficient to replace $g_\Delta$ and $h_\Delta$ with $G_\Delta^\alpha$ and $H_\Delta^\alpha$ in eq. \eqref{sumRuleBlocks}.
Since the local block decomposition is uniformly convergent for $\alpha$ obeying \eqref{alphaBoundLocalBlock}, we are allowed, this time, to swap the integration and the sum. Denoting by a bar the integral on each local block:
\begin{subequations}
\begin{align}
\Bar{G}_\Delta^\alpha = \int_0^1 dz \left(1+\frac{1}{z^2}\right) G_\Delta^\alpha\left(\left(\frac{2z}{1+z^2}\right)^2\right)~, \label{Gbar}\\
\Bar{H}_\Delta^\alpha = \int_0^1 dz \left(1+\frac{1}{z^2}\right) H_\Delta^\alpha\left(\left(\frac{2z}{1+z^2}\right)^2\right)~, \label{Hbar}
\end{align}
\label{GHbar}
\end{subequations}
we define
\beq
	\kappa(\Delta,\alpha)=\frac{1}{2}\left[\Bar{H}_\Delta^\alpha + \frac{\Delta-2}{\Delta} \Bar{G}_\Delta^\alpha \right]~,
	\label{kappaGHbar}
\eeq
and we land on our final sum rule:
\beq
\sum_{\D>0} \kappa(\D,\al)\,
c_{\phi \phi \D} b_{T \D}=\D_\phi ~.
\label{offDiagSumRule}
\eeq
The coefficients $\kappa(\D,\al)$ have the following closed form expression, which can be derived either in Mellin space or by simply expanding the local blocks in powers of $\chi$ and integrating term by term:
\begin{multline}
\kappa(\D,\al)=
\frac{\pi^{\frac{1}{2}} \D\, \Gamma\left(\D+\frac{1}{2}\right)}{(\D+1) \Gamma\left(\frac{\D}{2}+1\right)^2} \\
- \frac{\pi^\frac{1}{2}}{2}\frac{\D-2}{\D} \frac{\Ga\left(\alpha-\frac{1}{2}\right) \Ga\left(\D+\frac{1}{2}\right) \Ga\left(\alpha\right)}{\Ga\left(\alpha+\frac{\D}{2}+\frac{1}{2}\right) \Ga\left(\alpha-\frac{\D}{2}+1\right) \Ga\left(\frac{\D}{2}\right)^2} \left[ \pFq{3}{2}{1,\al-\frac{1}{2},\al}{\al-\frac{\D}{2}+1,\al+\frac{\D}{2}+\frac{1}{2}}{1}\right.  \\
\left.+\frac{\alpha}{\al-1}
 \pFq{4}{3}{1,\al-1,\al+1;\al-\frac{1}{2}}{\al,\al-\frac{\D}{2}+1,\al+\frac{\D}{2}+\frac{1}{2}}{1} \right]~.
\label{kappaSumRule}
\end{multline}
Like for the naive sum rule \eqref{NaiveSumRule}, the contribution of operators with $\D<1$ is obtained by analytic continuation. Furthermore, the integrals \eqref{Gbar} and \eqref{Hbar} also fail to converge when $\al<1/2$, since the local blocks \eqref{LocalBlockClosed0}-\eqref{LocalBlockClosed2} behave as
\begin{align}
G_\D^\al(\chi) &\overset{\chi \to 0}{\approx}  \chi^{\D/2}-
\frac{\Gamma(\alpha)^2 \Gamma\left(\Delta +\frac{1}{2}\right)}{\Gamma\left(\frac{\Delta}{2}\right)^2\Gamma\left(\alpha-\frac{\Delta}{2}+1\right) \Gamma\left(\alpha+\frac{\Delta}{2}+\frac{1}{2}\right)} \chi^\al~, \\
H_\D^\al(\chi) &\overset{\chi \to 0}{\approx}  \chi^{\D/2}-\frac{\Gamma(\alpha+1)\Gamma(\alpha-1) \Gamma\left(\Delta +\frac{1}{2}\right)}{\Gamma\left(\frac{\Delta}{2}+1\right)\Gamma\left(\frac{\Delta}{2}-1\right)\Gamma\left(\alpha-\frac{\Delta}{2}+1\right) \Gamma\left(\alpha+\frac{\Delta}{2}+1\right)}\chi^\al ~.
\end{align}
However, notice that the power $\chi^\al$ is incompatible with the OPE, generically: in fact, this discrepancy is precisely the source of the sum rules derived in \cite{Levine:2023ywq}. In particular, the cancellation of this term in the sum over local blocks leads to the following sum rule:
\beq
\sum_{\D>0} \frac{\D-2}{\D} \frac{\Gamma \left(\D+\frac{1}{2}\right)}{\Gamma \left(\frac{\D}{2}\right)^2}
\frac{1}{\Gamma\left(\al+\frac{\D}{2}+\frac{1}{2}\right)\Gamma\left(\al-\frac{\D}{2}+1\right)}
c_{\p\p\D}b_{T\D}=0~.
\label{zeroSumRule}
\eeq
The convergence of this sum rule can be tested by using a bound on the growth of the OPE coefficients, \eqref{bcBound}, which is derived in appendix \ref{app:OPEasymptotics}. Precisely when eq. \eqref{alphaBoundLocalBlock} is obeyed, eq. \eqref{zeroSumRule} converges absolutely. In terms of our main sum rule \eqref{offDiagSumRule}, this means that we can safely evaluate it for $\D_\p+\D_\mc{V}/2<\al<1/2$, defining the $\alpha$ dependence by analytic continuation. Eq. \eqref{zeroSumRule} can be verified for a free boson and a free fermion. 

Eqs. \eqref{offDiagSumRule} and \eqref{kappaSumRule} are the last missing ingredients for building our bootstrap problem. In section \ref{sec:flat},  where we analyse the flat space limit of our sum rules, we will see that eq. \eqref{offDiagSumRule} reduces to a family of sum rules for the flat space form factors. 
 Then, in section \ref{sec:examples}, we check the sum rule in a few examples, and confirm its convergence. Finally, in section \ref{sec:problem}, we put everything together and formulate a positive semi-definite bootstrap problem.

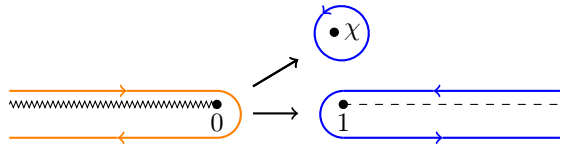
\begin{figure}[t]
\centering
\begin{tikzpicture}[scale=1.2]
    \draw (2.8,2.1) -- (2.8,1.8);
    \draw (2.8,1.8) -- (3.1,1.8);
    \node at (3,2.03) {$\chi'$};
    \draw [decorate,decoration={zigzag,amplitude=1.2pt,segment length =2 pt}] (-3,0) -- (-0.7,0);
    \draw[dashed] (0.7,0) -- (3.1,0);
    \filldraw[black] (-0.7,0) circle (1.3pt);
    \node[black] at (-0.7,-0.2) {0}; 
    \filldraw[black] (0.7,0) circle (1.3pt);
    \node[black] at (0.7,-0.2) {1};
    \filldraw[black] (0.6,0.8) circle (1.3pt);
    \node[black] at (0.8,0.8) {$\chi$};
    \draw[thick,orange, ->] (-3,0.15) -- (-1.7,0.15);
    \draw[thick,orange] (-1.7,0.15) -- (-0.7,0.15);
    \draw[thick,orange] (-0.7,0.15) arc (91:-91:0.26);
    \draw[thick,orange,->] (-0.7,-0.37) -- (-1.8,-0.37);
    \draw[thick,orange] (-1.8,-0.37) -- (-3,-0.37);
    \draw[thick,->] (-0.3,0.2) -- (0.2,0.5);
    \draw[black,thick,blue, ->] (0.47,1.) arc (135:495:0.3);
    \draw[thick,blue, ->] (3.1,0.15) -- (1.7,0.15);
    \draw[thick,blue] (1.7,0.15) -- (0.7,0.15);
    \draw[thick,blue] (0.7,0.15) arc (91:269:0.26);
    \draw[thick,blue,->] (0.7,-0.37) -- (1.8,-0.37);
    \draw[thick,blue] (1.8,-0.37) -- (3.1,-0.37);
    \draw[thick,->] (-0.3,0.2) -- (0.2,0.5);
    \draw[thick,->] (-0.3,-0.1) -- (0.2,-0.1);
\end{tikzpicture}
\caption{Pulling the contour back in the computation of the local blocks shows that there is no condition on the analyticity in $\al$ from the OPE singularity at $\chi'=0$. However, the representation \eqref{dispersiveG} is incorrect for the blocks with dimension $\D<2\al-2$. This means in particular that these local blocks don't vanish when $\D=2\al-2n$, $n$ integer, a fact which has interesting consequences in free (and weakly coupled) theories, see section \ref{sec:examples}.}
\label{fig:contourLocalBlock}
\end{figure}

Before all of this, let us tie one loose end. The integrals in \eqref{localBlock0} and \eqref{localBlock2} diverge at $\chi=0$ for $\alpha$ large enough, which led to the condition \eqref{alphaMax}. Yet, it is easy to see that \emph{the local block decompositions \eqref{localBlockTheta}-\eqref{localBlockT} do not require any upper bound on $\alpha$}. The local blocks associated to low dimensional primaries, with $\Delta<2(\alpha-1)$, should be evaluated by integrating along a keyhole contour, which does not drop the arc around $\chi=0$. In fact, the exact expressions of the local blocks---eqs. \eqref{LocalBlockClosed0} and \eqref{LocalBlockClosed2}---have no singularity at the threshold marked by eq. \eqref{alphaMax}, and are finite for larger $\alpha$. As it is clear from the computation in appendix \ref{app:local_blocks}, the closed form expression of the local blocks $G_\Delta^\alpha$ and $H_\Delta^\alpha$ is obtained by pulling back the contour towards the right, as in figure \ref{fig:contourLocalBlock}. Therefore it consists of the sum of the original conformal block and a contribution from the cut along $\chi>1$. This is why, in eq. \eqref{kappaSumRule}, the first line matches the naive sum rule \eqref{NaiveSumRule}, while the remaining part is the effect of subtracting from each block the contribution of the right cut in figure \ref{fig:chiPath}.

As announced, the next subsection is dedicated to the proof of the convergence properties of the local block decomposition and of the sum rule.

\subsection{Convergence of the local block decomposition and  the sum rule}
\label{sec:convergenceFF}

We first discuss in detail the convergence properties of the local block decomposition for the trace of the stress tensor. Afterwards, we explain the additional complications arising in the case of the traceless part of the form factor.

\subsubsection{Spin zero bulk operator}

The steps necessary to establish eq. \eqref{offDiagSumRule} are the following:
\begin{enumerate}
\item Proving that eq. \eqref{alphaMin} allows to drop the arc around $\chi=\infty$ in the contour deformation of figure \ref{fig:chiPath}; 
\item Proving that, after replacing the conformal block decomposition in the r.h.s. of eq. \eqref{dispersiveG}, the sum over primaries and the integral can be swapped;
\item Proving that the local block decomposition can be integrated term by term in the Ward identity \eqref{sumRuleBlocks}.
\end{enumerate}
The first two items establish that, for some region in the $\chi$ complex plane, the local block decomposition converges pointwise to the form factor. The last property will follow from uniform convergence in the region of integration, as stated around eq. \eqref{alphaBoundLocalBlock}. 

The main tool to prove the statements above is a bound on the asymptotic growth of the form factor, as $|\chi|\to \infty$. 
To this end, it is useful to use the variable $z^2$, defined in eq. \eqref{chiofz}.
The limit $|\chi|\to \infty$ corresponds to $z^2 \to -1$, while the phase of $\chi$ maps to the direction in which this limit is approached, with the constraint $|z^2|\leq 1$.\footnote{Notice that the relation between $\chi$ and $z^2$ is the same as the relation which defines the $\rho$ coordinate for the four-point function of a CFT in flat space \cite{Pappadopulo:2012jk,Hogervorst:2013sma}, $z^2$ playing the role of $\rho$. The $\chi$ complex plane is mapped to the unit $z^2$ disk, with the two sides of the cut $\chi>1$ being mapped to the arcs of the unit circle above and below the real axis.} In particular, $\chi \to +\infty$ corresponds to approaching $z^2=-1$ along the unit circle, while $\chi \to -\infty$ is obtained by keeping $z^2$ real.
Using eq. \eqref{FFblockZ} to express the blocks as a function of $z$, and setting $x=-z^2$, with $\nu=$Arg$(x)$, we write the expansion of the form factor of $\Theta$ as follows:
\begin{equation}
(-2P_1\cdot P_2)^{\Delta_{\phi}}   \langle \phi(P_1) \phi(P_2) \Theta(X)\rangle=\sum_{\Delta} c_{\phi\phi\Delta}b_{\Theta\Delta} e^{\ii \frac{\D}{2}(\nu-\pi)}
     \big(2\sqrt{|x|}\big)^\D {}_2F_1 \left(\frac{1}{2},\Delta; \Delta+\frac{1}{2};x \right)~.
     \label{FFDec0x}
\end{equation}
In the ,  $|\chi|\to \infty$ limit, $\nu \to 0$ or $2\pi$ for Arg$(\chi)<0$ or $>0$ respectively. Since the expansion in powers of $x$ of the Hypergeometric function in \eqref{FFDec0x} has positive coefficients, we can bound the r.h.s as follows:
\begin{equation}
\left|(-2P_1\cdot P_2)^{\Delta_{\phi}}   \langle \phi(P_1) \phi(P_2) \Theta(X)\rangle \right| \leq \sum_{\Delta} \left|c_{\phi\phi\Delta}b_{\Theta\Delta} \right|
     \big(2\sqrt{|x|}\big)^\D {}_2F_1 \left(\frac{1}{2},\Delta; \Delta+\frac{1}{2};|x| \right) 
     \label{FFbound0x} 
\end{equation}
For later convenience, we give a name to the modified block appearing on the r.h.s. of eq. \eqref{FFbound0x}:
\beq
g^+_\D(z)=\big(2z\big)^\D {}_2F_1 \left(\frac{1}{2},\Delta; \Delta+\frac{1}{2};z^2 \right)~.
\label{posBlock0}
\eeq 
Notice that the bound \eqref{FFbound0x} is saturated, as $x\to 1$, by a free boson and a free fermion---see section \ref{sec:examples}. In turn, the right hand side of eq. \eqref{FFbound0x} can be bounded via the Cauchy-Schwarz inequality: this is done in appendix \ref{app:Cauchy-Schwarz}. In the limit $x\to 1$, from eq. \eqref{asymptBoundCauchy}, we get
\beq
\left|(-2P_1\cdot P_2)^{\Delta_{\phi}}   \langle \phi(P_1) \phi(P_2) \Theta(X)\rangle  \right|\overset{z^2\to -1}{\lesssim}  \left(1-\abs{z^2}\right)^{-2\D_\p-\D_\mathcal{V}}~.
\label{CauchySchwarzSection4}
\eeq
Going back to the $\chi$ cross ratio, and choosing $\mathcal{G}$ in eq. \eqref{dispersiveG} to be the form factor of the trace of the stress tensor, we get the bound
\beq
\abs{\mathcal{G}_\Theta(\chi)} \overset{\chi\to \infty}{\lesssim} \left(\frac{|\chi|}{1-\cos \arg \chi}\right)^{\Delta_\phi+\Delta_{\mc{V}}/2}~.
\label{boundOnGspin0}
\eeq
This bound is not uniform, reflecting the fact that the contribution of large dimensional operators stops being exponentially suppressed when $\chi>0$, \emph{i.e.} $|z|=1$.

Armed with eq. \eqref{boundOnGspin0}, let us tackle the first item in the list. We want to bound the contribution from the arc:
\beq
\mathcal{A}(\al)=\lim_{r\to \infty} \frac{1}{r^\al}\int_{-\pi}^\pi\! \frac{d\theta}{2\pi}\, \mathcal{G}_\Theta\left(r e^{\ii \theta}\right)
e^{-\ii \al \theta}~,
\label{AofAlpha}
\eeq
where we set $\chi = r \exp \ii \theta$.
Firstly, it is easy to see that $\mathcal{A}(\al)$ is well defined (\emph{e.g.} the limit exists) and it is never divergent. Indeed, the bound \eqref{boundOnGspin0} implies that the integral of the discontinuity, eq. \eqref{dispersiveG}, converges at infinity for $\al> \D_\p+\D_\mathcal{V}/2$, and since $\mathcal{G}(\chi)$ is obviously finite, the arc can at most contribute a constant. Furthermore, a non-vanishing contribution as the arc is deformed to infinity must localize to $\theta=\arg \chi =0$. We would like to use the dependence on $\al$ to show that this is impossible: intuitively, $\mathcal{G}(\chi)/\chi^\al$ might provide a representation of a localized distribution for a fixed value of $\al$, but not for any $\al$. To show this, let us fix an integer
\beq
n>2\alpha_\textup{min}-1~,
\label{nInteger}
\eeq
and rewrite the integral in eq. \eqref{AofAlpha} as follows:
\beq
\int_{-\pi}^\pi\! \frac{d\theta}{2\pi}\, \mathcal{G}_\Theta\left(r e^{\ii \theta}\right)e^{-\ii \al \theta}=\sum_{m=0}^{n-1} \frac{(-\ii\al)^m}{m!}
\int_{-\pi}^\pi\! \frac{d\theta}{2\pi}\,\theta^m\,\mathcal{G}_\Theta\left(r e^{\ii \theta}\right) +
\int_{-\pi}^\pi\! \frac{d\theta}{2\pi} \left(e^{-\ii \al \theta}-\sum_{m=0}^{n-1} 
\frac{(-\ii\al \theta)^m}{m!}\right)\mathcal{G}_\Theta\left(r e^{\ii \theta}\right)~.
\eeq
The advantage of this reshuffling is that the second integral can easily be bounded in the $r\to \infty$ limit:
\begin{equation}
\left|\int_{-\pi}^\pi\! \frac{d\theta}{2\pi} \left(e^{-\ii \al \theta}-\sum_{m=0}^{n-1} 
\frac{(-\ii\al \theta)^m}{m!}\right)\mathcal{G}_\Theta\left(r e^{\ii \theta}\right)\right|
\leq \int_{-\pi}^\pi\! \frac{d\theta}{2\pi} \left|e^{-\ii \al \theta}-\sum_{m=0}^{n-1} 
\frac{(-\ii\al \theta)^m}{m!}\right|\left| \mathcal{G}_\Theta\left(r e^{\ii \theta}\right)\right|
\overset{r\to\infty}{\lesssim} r^{\al_\textup{min}}~.
\end{equation}
Crucially, the angular dependence of the inequality \eqref{boundOnGspin0}, together with eq. \eqref{nInteger}, guarantee the convergence of the integral at $\theta=0$. Therefore, when $\al>\al_\textup{min}$, we deduce that $\mathcal{A}$ is a polynomial:
\beq
\mathcal{A}(\al)=\lim_{r\to \infty} \frac{1}{r^\al} \mathcal{P}_r(\al)~, \qquad
\mathcal{P}_r(\al)=\sum_{m=0}^{n-1} \frac{(-\ii\al)^m}{m!} 
\int_{-\pi}^\pi\! \frac{d\theta}{2\pi}\,\theta^m\,\mathcal{G}_\Theta\left(r e^{\ii \theta}\right) ~.
\label{Apolynomial}
\eeq
While we cannot bound the $r$ dependence of the integrals in eq. \eqref{Apolynomial}, we see that they are independent of $\al$. This makes it obvious that $A(\al)$ must vanish, since the limit must exist and be finite for all $\al>\al_\textup{min}$. To make the result explicit, one can for instance use the fact that the coefficient of a polynomial of degree $n-1$ can be expressed as a linear combination of values of the polynomial at $n$ distinct points:
\beq
\int_{-\pi}^\pi\! \frac{d\theta}{2\pi}\,\theta^m\,\mathcal{G}_\Theta\left(r e^{\ii \theta}\right) = \sum_{k=1}^n \beta_{k\,m} \mathcal{P}_r(\al_k)~,
\eeq
where $\beta$ is simply related to the inverse of the Vandermonde matrix, and we can choose all the $\al_k>\al_\textup{min}$. Since $\mathcal{P}_r(\al) \approx \mathcal{A}(\al) r^\al$, while the l.h.s. is independent of $\al,$ we conclude that
\beq
\mathcal{A}(\al)=0~, \qquad \al>\al_\textup{min}~,
\eeq
which proves eq. \eqref{alphaMin}.

Let us move on to the proof of statement number 2 in the list above. The region of integration of eq. \eqref{dispersiveG} touches the boundary of convergence of the OPE, which is not a sum of positive terms, hence the swapping property is non-trivial. We shall give a  direct $\epsilon-\delta$ proof. We want to bound the difference
\begin{multline}
\left|\mathcal{G}_\Theta(\chi)-\sum_\Delta^{\Delta_M}c_{\p\p\D}b_{\Theta \D}G_\D^\al(\chi)\right|=
\left|\chi^\al \int_{-\infty}^0\! \frac{d\chi'}{2\pi} \frac{1}{\chi-\chi'}\sum_{\Delta_M}^\infty c_{\p\p\D}b_{\Theta \D} \textup{Disc}
\left(\frac{g_\D(\chi')}{\chi'^\al}\right)\right| \\
\leq 2\left|\chi\right|^\al \int_{-\infty}^0\! \frac{d\chi'}{2\pi} 
\frac{\left|\chi'\right|^{-\al}}{\left|\chi-\chi'\right|} \sum_{\Delta_M}^\infty \left|c_{\p\p\D}b_{\Theta \D}\right|
g^+_\D(|z(\chi')|)~.
\end{multline}
In the second line, we used the definition \eqref{posBlock0}. Let us take $\chi$ away from the negative real axis, so that the integrand is finite along the integration contour. For the sum rule, we will eventually be interested in the range $\chi \in [0,1]$.\footnote{Notice that $\chi=0$ is obviously fine: both the local block decomposition and the (connected) form factor vanish at that point.} Now, we choose $\eps>0$, and we define $\chi^*(\eps)<0$ such that
\beq
2\left|\chi\right|^\al \int_{-\infty}^{\chi^*}\! \frac{d\chi'}{2\pi} 
\frac{\left|\chi'\right|^{-\al}}{\left|\chi-\chi'\right|} \sum_{\Delta_M}^\infty \left|c_{\p\p\D}b_{\Theta \D}\right| g^+_\D(|z(\chi')|) < \frac{\epsilon}{2}~,
\label{boundTail1}
\eeq
for any $\D_M>0$. This is possible, provided $\alpha>\Delta_\phi+\D_{\mc{V}}/2$, because the tail of the sum on the l.h.s. of the inequality has the same asymptotics as the whole sum, and is bounded by the r.h.s. of eq. \eqref{boundOnGspin0}. As for the complement of the integration region, $\chi' \in [\chi^*,0]$, there the OPE converges absolutely and uniformly---a fact that follows from uniform convergence of the $z^2$ power series \cite{Pappadopulo:2012jk}. Hence, it can be integrated term by term, and there is a value $\D_M^*(\eps)>0$ such that 
\beq
2\left|\chi\right|^\al \sum_{\Delta_M}^\infty \int_{\chi^*}^0\! \frac{d\chi'}{2\pi} 
\frac{\left|\chi'\right|^{-\al}}{\left|\chi-\chi'\right|}  \left|c_{\p\p\D}b_{\Theta \D}\right| g^+_\D(|z(\chi')|) < \frac{\epsilon}{2}~, \quad \forall\, \D_M>\D_M^*(\eps)~.
\label{boundTail2}
\eeq
Eqs. \eqref{boundTail1} and \eqref{boundTail2} imply that, for any $\eps>0$, there is a $\D_M^*$, such that
\beq
\left|\mathcal{G}_\Theta(\chi)-\sum_\Delta^{\Delta_M}c_{\p\p\D}b_{\Theta \D}G_\D^\al(\chi)\right| < \eps~, \quad \forall\, \D_M>\D_M^*(\eps)~,
\eeq
which is what we wanted to prove. 

This establishes that the local block decomposition converges pointwise to the form factor, at least away from the negative real axis. It is now not hard to prove that
\begin{center}
\emph{the local block decomposition converges uniformly in any bounded region of the $\chi$ complex plane which excludes the negative real axis, if
\beq
\al> \D_\phi+\frac{\D_{\mc{V}}}{2}~.
\label{alphaBoundUniform0}
\eeq}
\end{center}
In particular, any such region can be enclosed in the following pie missing a slice:
\beq
\mathcal{R}_{R,\omega} = \left\{\chi \ \textup{s.t.} \ |\chi|<R<\infty, \ |\pi-\arg \chi|>\omega>0 \right\}~.
\eeq
There, the following inequality holds, for any fixed $\chi' \in (-\infty,0]$:
\beq
\left|\chi-\chi'\right| \geq \left| \chi'  \right|\sin \omega~.
\eeq 
Therefore, for $\al>0$, we can bound each local block with scaling dimension $\D>2\al$ as follows:
\beq
\left|G_\D^\al(\chi)\right| \leq \frac{R^\al}{\sin \omega} 
\int_{-\infty}^0 \! \frac{d\chi'}{2\pi} \left|\chi'\right|^{-\al-1} g^+_\D(|z(\chi')|)
\equiv M_\D^\al~, \qquad \chi \in \mathcal{R}_{R,\omega}~.
\label{MDeltaDef}
\eeq
In turn, the series
\beq
\sum_{\D>2\al}^\infty \left|c_{\p\p\D}b_{\Theta \D}\right| M_\D^\al < \infty~, \qquad \textup{if } \al> \D_\phi+\frac{\D_{\mc{V}}}{2}~,
\eeq
because in defining $M_\D^\al$ we did not modify the $\chi' \to -\infty$ behavior of the integrand, with respect to the original local block. Uniform convergence of the local block decomposition then follows from Weierstrass M-test. 
This implies the third point in the wishlist at the beginning of this subsection.
\vspace{0.5cm}

Once established the validity of the local block decomposition, one can ask about its rate of convergence. 
The starting point is the following bound on the growth of the OPE coefficients:
\beq
R(\D_M)\equiv \sum_\D^{\D_M} 2^{\D} \sqrt{\D} |c_{\p\p\D}b_{\Theta \D} |
\lesssim  \D_M^{2\D_\p+\D_{\mc{V}}}~, \qquad \D_M \to \infty~.
  \label{bcBound}
\eeq  
Eq. \eqref{bcBound} follows from the Cauchy-Schwarz inequality, and is proven in appendix \ref{app:OPEasymptotics}.
A bound on the rate of absolute convergence of the local block decomposition can now be derived using the asymptotic formula for the local blocks, valid at large $\D$ and fixed $\chi$ and $\al$, derived in appendix \ref{app:asymptBlocksFF}: 
\beq
G_\D^\al(\chi) \overset{\D\to\infty}{\approx} 
-\chi^\al \sin\left[\frac{\pi}{2}(\D-2\al)\right] 
\frac{2^{\D+2\al-1}}{\pi^{3/2}}  \Gamma(\al)^2 \D^{\frac{1}{2}-2\al}~,
\label{LocBlockLargeD0}
\eeq
A simple way to proceed is 
to write the sum $\sum_\D |c_{\p\p\D}b_{\Theta \D} G_\D^\al|$ as an integral, and integrate by parts in $\D$ to be able to use eq. \eqref{bcBound}. One gets:
\beq
\sum_{\D_M}^\infty |c_{\p\p\D}b_{\Theta \D} G_\D^\al(\chi)| \overset{\D_M\to\infty}{\lesssim} \chi^\al \D_M^{2\D_\phi+\D_{\mathcal{V}}-2\al}
\label{locBlockRate0}
\eeq
We see that the series converges at least as a power law, whose exponent depends on $\al$, in agreement with eq. \eqref{alphaBoundUniform0}. Notice, however, that depending on the value of $\al$, the local block decomposition might converge must faster, and even reduce to a finite sum, as we shall see in the free theory examples.

\subsubsection{Spin two bulk operator}
\label{subsec:spin2LBConv}

The proofs of convergence of the local block decomposition for the form factor of the traceless part of the stress tensor proceed in the same way as in the previous subsection, but for a few important details. The Cauchy-Schwarz inequality only imposes a bound weaker than \eqref{boundOnGspin0} on the form factor:
\beq
\abs{\mathcal{G}_T(\chi)} \overset{\chi\to \infty}{\lesssim} \left(\frac{|\chi|}{1-\cos \arg \chi}\right)^{\Delta_\phi+1}~.
\label{boundOnGspin2Weak}
\eeq
However, the bound \eqref{boundOnGspin2Weak} can never be saturated. Indeed, the asymptotics of the boundary OPE coefficients are fixed by conservation in terms of the ones of $\Theta$---eq. \eqref{bOPECons}. By using the large $\D$ uniform approximation of the spin two blocks given in eq. \eqref{largeDuniformFF2}, and the asymptotics \eqref{bcBound} one gets a stronger bound. Let us first assume that $\D_\p+\D_\mc{V}/2>1$, for reasons that will soon become clear. Then,
\begin{multline}
\left|\sum_\D c_{\p\p\D}b_{T\D} h_\D(\chi) \right| \leq 
\sum_\D \left| c_{\p\p\D}b_{T\D} \right| h_\D^+(|z|) \\
\overset{|z|\to 1}{\approx} 
\left. \frac{R(\D)}{\sqrt{\pi}} |z|^\D e^{\frac{1-|z|^2}{2}\D} K_2\left(\frac{1-|z|^2}{2}\D\right)\right]^\infty_{\D_0}-\int^\infty_{\D_0}\! d\D \frac{R(\D)}{\sqrt{\pi}}\frac{d}{d\D} 
\left(|z|^\D e^{\frac{1-|z|^2}{2}\D} K_2\left(\frac{1-|z|^2}{2}\D\right)\right) \\
 \overset{|z|\to 1}{\lesssim}
 \int^\infty_{\D_0}\! d\D \D^{2\D_\p+\D_\mc{V}-1} |z|^\D e^{\frac{1-|z|^2}{2}\D} K_2\left(\frac{1-|z|^2}{2}\D\right)~.
 \label{spin2boundStep}
\end{multline}
In the first line we defined 
\beq
h_\D^+(z) = (2z)^\D(1-z^2)^2 {}_2F_1 \left(\frac{5}{2},\Delta+2; \Delta+\frac{1}{2};z^2 \right)~,
\eeq
whose Taylor expansion has positive coefficients. In the second line, we assumed that the leading singularity as $|z| \to 1$ comes from the tail of the sum: we will comment more on this below. Finally, to go to the third line, we used that
\beq
\frac{d}{d\D} \left(|z|^\D e^{\frac{1-|z|^2}{2}\D} K_2\left(\frac{1-|z|^2}{2}\D\right)\right)<0~,
\eeq
to apply the bound \eqref{bcBound} and integrate by parts. Now, using that $\log|z|+(1-|z|^2)/2$ has a maximum at $|z|=1,$ we simplify as follows:
\beq
\left|\sum_\D c_{\p\p\D}b_{T\D} h_\D(\chi) \right| 
 \overset{z^2\to -1}{\lesssim}
 \int^\infty_{\D_0}\! d\D \D^{2\D_\p+\D_\mc{V}-1} K_2\left(\frac{1-|z|^2}{2}\D\right)
 \overset{|z|\to 1}{\sim} (1-|z|^2)^{-2\D_\p-\D_\mc{V}}.
 \label{spin2boundStep2}
 \eeq
In the last step, we used the assumption $\D_\p+\D_\mc{V}/2>1$: in the opposite case, the small argument singularity of the Bessel function, \emph{i.e.}
\beq
K_2(x) \overset{x\to 0}{\approx} \frac{2}{x^2}~,
\label{BesselKsmall}
\eeq  
dominates, and the leading singularity would instead be $(1-|z|^2)^{-2}$. 
We have obtained,
 \beq
\abs{\mathcal{G}_T(\chi)} \overset{\chi\to \infty}{\lesssim} \left(\frac{|\chi|}{1-\cos \arg \chi}\right)^{\Delta_\phi+\Delta_{\mc{V}}/2}~, \qquad \textup{if }\D_\p+\frac{\D_\mc{V}}{2}>1~.
\label{boundOnGspin2}
\eeq
The inequality \eqref{boundOnGspin2} coincides with the bound on the form factor of the trace \eqref{boundOnGspin0}. We see that conservation of the stress tensor imposes a milder growth for the tail of the conformal block decomposition. From here on, all the steps of the previous section can be followed verbatim.

What happens if $\D_\phi+\D_\mc{V}/2<1$? 
Each conformal block has a power law large $\chi$ singularity:
\beq
h_\D(\chi) \overset{\chi\to\infty}{\approx} 
\chi\,  F(\D)\times
\begin{sistema}
e^{\ii \frac{\pi}{2}\D}~, \qquad \arg \chi \in [0,\pi) \\
e^{-\ii \frac{\pi}{2}\D} \qquad \arg \chi \in [\pi,2\pi)~,
\end{sistema}
\qquad F(\D)=-\frac{2^{\D+1}\Gamma\left(\D+\frac{1}{2}\right)}{\sqrt{\pi} \Ga\left(\D+2\right)}~,
\label{hLargeChi}
\eeq
This should be contrasted with the logarithmic limit of $g_\D(\chi)$. Eq. \eqref{hLargeChi} is of course reflected at large $\Delta$ by the asymptotics \eqref{BesselKsmall}. On one hand, this implies that the integral \eqref{localBlock2} which defines the local block only converges for $\al>1$. Correspondingly, as explained below eq. \eqref{spin2boundStep2}, the singularity \eqref{hLargeChi} is leading if $\D_\phi+\D_\mc{V}/2<1$, and the arguments of the previous subsection would only guarantee convergence of the local block decomposition and the sum rule  for $\al>1$. However, this situation can be improved upon.

Firstly, notice that eq. \eqref{hLargeChi} showcases the discontinuity across the positive real axis which, as pointed out above, is incompatible with locality. If such singularity dominates at large $\chi$, the discontinuity must cancel, which implies the following sum rule:
\beq 
\sum_\D c_{\p\p\D}b_{T\D} F(\D) \sin \frac{\pi}{2} \D =0~.
\label{DiscConstraint2}
\eeq
Notice that, precisely when $\D_\phi+\D_\mc{V}/2<1$, this sum is absolutely convergent due to the asymptotic bound \eqref{bcBound}. One is led to consider a modified form factor, where the linear behavior at large $\chi$ is subtracted off:
\beq
\tilde{\mathcal{G}}_T(\chi) = \mathcal{G}_T(\chi) - \chi \sum_\D c_{\p\p\D}b_{T\D} F(\D) \cos \frac{\pi}{2}\D~.
 \label{TildeGT}
\eeq
We expect the function $\tilde{\mathcal{G}}_T(\chi)$ obeys the bound in eq. \eqref{boundOnGspin2}, and so its dispersion relation still converges for $1>\al>\D_\phi+\D_\mc{V}/2$.\footnote{Proving this rigorously requires bounding appropriately $\left|h_\D(\chi)-\chi F(\D) e^{\pm\ii \pi \D/2}\right|$, for $\arg \chi \lessgtr \pi$.
} However, the discontinuity is not expressed purely in terms of local blocks, if we insist using the spectrum of the original correlator:\footnote{In writing eq. \eqref{GtildeDisc}, we used the sum rule \eqref{DiscConstraint2} to first rewrite $\tilde{\mathcal{G}}_T$ as
\beq
\tilde{\mathcal{G}}_T(\chi) = \mathcal{G}_T(\chi) - \chi \sum_\D c_{\p\p\D}b_{T\D} F(\D) 
\times
\begin{sistema}
e^{\ii\frac{\pi}{2} \D} \ \  \qquad  \qquad \arg \chi \in [0,\pi) \\
e^{-\ii\frac{\pi}{2} \D} \qquad  \qquad \arg \chi \in [\pi,2\pi)~.
\end{sistema}
\eeq
This expression emphasizes that the linear term in $\chi$ can be canceled out block by block. This is important for the swapping property, see the comment below eq. \eqref{boundTail1}.
}
\beq
\tilde{\mathcal{G}}_T(\chi) =
\sum_\D c_{\p\p\D}b_{T\D} \chi^\alpha \int_{-\infty}^0 \frac{d\chi'}{2\pi \ii} \frac{1}{\chi'-\chi}  \left[\textup{Disc}\left(\frac{h_\Delta(\chi')}{\chi'^\alpha}\right)-2\ii\chi
\frac{F(\D)\sin \left(\frac{\pi}{2}(\D-2\al)\right)}{\left|\chi'\right|^{\al}}\right]
\label{GtildeDisc}
\eeq 
Our aim now is to use eq. \eqref{GtildeDisc} to show that also $\mathcal{G}_T(\chi)$ admits a local block decomposition in the same range of $\al$, where the local blocks are obtained by analytic continuation of the defining integral \eqref{localBlock2}, \emph{i.e.} they coincide with the exact expression \eqref{LocalBlockClosed2}. The analytic continuation of the local block can be obtained as follows:
\begin{multline}
H^\alpha_\Delta(\chi) = \chi^\alpha \int_{-\infty}^0 \frac{d\chi'}{2\pi \ii} \frac{1}{\chi'-\chi}  \left[\textup{Disc}\left(\frac{h_\Delta(\chi')}{\chi'^\alpha}\right)-2\ii\chi 
\frac{F(\D)\sin \left(\frac{\pi}{2}(\D-2\al)\right)}{\left|\chi'\right|^{\al}}\right] \\
+ \chi F(\D) \left(\cos \frac{\pi}{2}\D-\frac{\sin \frac{\pi}{2}\D}{\tan \pi \al}\right)~.
\end{multline}
The integral in the first line now converges for $\al>0$. This allows to rewrite eq. \eqref{GtildeDisc} as
\beq
\tilde{\mathcal{G}}_T(\chi) =
\sum_\D c_{\p\p\D}b_{T\D} \left(H_\D^\al(\chi)-\chi F(\D)\left(\cos \frac{\pi}{2}\D-\frac{\sin \frac{\pi}{2}\D}{\tan \pi \al}\right)\right).
\label{GtildeLocBlock}
\eeq 
Using eq. \eqref{DiscConstraint2} and the definition of $\tilde{\mathcal{G}}_T$ \eqref{TildeGT}, we conclude that
\beq
\mathcal{G}_T(\chi) =
\sum_\D c_{\p\p\D}b_{T\D} H_\D^\al(\chi)~, \qquad \al>\D_\p+\frac{\D_\mc{V}}{2}~.
\eeq
This explains the validity of the bound in eq. \eqref{alphaBoundLocalBlock}, and the statement that, also for the traceless part of the stress tensor,
\begin{center}
\emph{the local block decomposition converges uniformly in any bounded region of the $\chi$ complex plane which excludes the negative real axis, if
\beq
\al> \D_\phi+\frac{\D_{\mc{V}}}{2}~.
\label{alphaBoundUniform2}
\eeq}
\end{center}

In section \ref{sec:examples}, we will verify the sum rule \eqref{DiscConstraint2} explicitly for a free scalar of mass $m^2<0$, \emph{i.e.} $\D_\p<1$. We also explicitly checked that both the local block decomposition and the sum rule converge to the correct value when $\D_\phi<1$ and $1>\al>\D_\p$. The same can be done for a free Majorana fermion with mass $m<0$. 

Finally, a bound on the rate of convergence of the local block decomposition can be obtained via the fixed $\chi$ large $\D$ asymptotics of $H_\D^\al$ derived in appendix \ref{app:asymptBlocksFF}:\footnote{While the limit was derived there for $\al>1$, it is valid in general, as one can for instance check from the exact expression  \eqref{LocalBlockClosed2}, where for $|\chi|<1$ one can use the Taylor expansion of the ${}_3F_2$.} 
\beq
H_\D^\al(\chi) \overset{\D\to\infty}{\approx} 
-\chi^\al \sin\left[\frac{\pi}{2}(\D-2\al)\right] 
\frac{2^{\D+2\al-1}}{\pi^{3/2}}  \Gamma(\al+1)\Gamma(\al-1)
 \D^{\frac{1}{2}-2\al}~.
\label{LocBlockLargeD2}
\eeq

As in the scalar case, one obtains
\beq
\sum_{\D_M}^\infty |c_{\p\p\D}b_{T \D} H_\D^\al(\chi)| \overset{\D_M\to\infty}{\lesssim} \chi^\al \D_M^{2\D_\phi+\D_{\mathcal{V}}-2\al}~.
\eeq

\subsubsection{Rate of convergence of the sum rule}

In this subsection, we briefly discuss the rate of convergence of eq. \eqref{offDiagSumRule}. In doing so, we will of course also show that the sum rule converges in the first place. Notice that this direct proof does not replace the work done in the previous two subsections: here, we will not show that the series converges \emph{to the correct value}.

The absolute convergence of the sum rule can be easily tested directly, using the bound \eqref{bcBound} and the asymptotics \eqref{LocBlockLargeD0} and \eqref{LocBlockLargeD2} of the local blocks. By replacing the latter estimates in eq. \eqref{GHbar}, one gets the following asymptotics for the coefficient $\kappa(\D,\al)$ of the sum rule: 
\beq
\kappa(\D,\al) \overset{\D\to\infty}{\approx} 
2^\D \D^{\frac{1}{2}-2\al} \frac{\Gamma(2\al)}{\sqrt{\pi}(\al-1)}\sin\left[\frac{\pi}{2}(\D-2\al)\right] ~.
\label{kappaAsymptotics}
\eeq
Eq. \eqref{bcBound} then implies the following bound on the asymptotics of the sum rule:
\beq
\left|\sum_{\D_M}^{\infty} \kappa(\D,\al)\,
c_{\phi \phi \D} b_{T \D} \right| \leq
\sum_{\D_M}^{\infty} \kappa(\D,\al)\,
\left|c_{\phi \phi \D} b_{T \D} \right| 
 \overset{\D_M\to\infty}{\lesssim}  \D_M^{2(\al_\textup{min}-\al)}~,
 \label{kappaSumRuleConvergence}
\eeq
where $\al_\textup{min}$ was defined in eq. \eqref{alphaMin}.
This result is of course in agreement with all the previous discussions. Notice that the actual rate of convergence can be much faster, as we will demonstrate in section \ref{sec:examples}. We will see choices of $\al$ for which the inequality \eqref{kappaSumRuleConvergence} is saturated, and others for which the series reduces to a finite sum. On the other hand, we will not find examples where the sum rule converges for $\al<\D_\p+\D_\mc{V}/2$. 

For comparison, one can also use the bound on the OPE growth \eqref{bcBound} to study the convergence of the naive sum rule \eqref{NaiveSumRule}. In this case, the tail goes as
\beq
\left|\sum_{\D_M}^{\infty} \frac{\pi^{\frac{1}{2}} \D\, \Gamma\left(\D+\frac{1}{2}\right)}{(\D+1) \Gamma\left(\frac{\D}{2}+1\right)^2}\,
c_{\phi \phi \D} b_{T \D} \right| \overset{\D_M\to\infty}{\lesssim}  \D^{2\D_\p+\D_\mc{V}-3/2}~.
\label{absConvNaive}
\eeq
We see that absolute convergence is only guaranteed for small values of $\D_\p$. In section \ref{sec:examples}, we will see that free theories saturate the bound \eqref{absConvNaive}.

\section{The flat space limit}
\label{sec:flat}

Interest for quantum field theory in AdS largely stems from the fact that the infinite radius limit yields flat space physics. In this section, we connect the correlation functions studied so far with their flat space counterparts. In particular, we find how the sum rules we uncovered in AdS transition to well known flat space sum rules. This is especially satisfying for the form factor sum rule \eqref{offDiagSumRule}, where the local block decomposition is seen to naturally provide the correct weight function in the infinite radius limit.

Following the logic of \cite{Paulos:2016fap}, it is tempting to conjecture the following formulas for the scattering amplitude $S(s)$, two-particle form factor $\mathcal{F}_2^\Theta(s)$ and spectral density $\rho_\Theta(s)$:
\begin{align}
S(s) &= \lim \sum_{\Delta>0} e^{-i\pi(\Delta -2\Delta_\phi)} \left[ h_1(\Delta) c_{\phi\phi\Delta} \right]^2 \,\delta\left( \frac{\sqrt{s}}{m}- \frac{\Delta}{\Delta_\phi} \right)\frac{2}{ \Delta_\phi}~,
\label{FSL:S}
\\
\mathcal{F}_2^\Theta(s)&= -\sqrt{2} m [s(s-4m^2)]^\frac{1}{4} \lim \sum_{\Delta>0} e^{-i\frac{\pi}{2}(\Delta -2\Delta_\phi)} h_1(\Delta) h_2(\Delta)   c_{\phi\phi\Delta}  
 b_{T\Delta}  
\, \delta\left( \frac{\sqrt{s}}{m}- \frac{\Delta}{\Delta_\phi} \right)\frac{2}{  \Delta_\phi}~,
\label{FSL:F2}
\\
\rho_\Theta(s)&=\frac{m^2}{2\pi} \lim \sum_{\Delta>0}  \left[ h_2(\Delta) b_{T\Delta} \right]^2 \,\delta\left( \frac{\sqrt{s}}{m}- \frac{\Delta}{\Delta_\phi} \right) \frac{2}{ \Delta_\phi}~,
\label{FSL:rho}
\end{align}
where we follow the conventions of \cite{Karateev:2019ymz} and $\lim$ denotes the flat space limit $\Delta \sim \Delta_\phi \to \infty$. More precisely, the limit of the sum $\sum_\Delta \delta\left( \frac{\sqrt{s}}{m}- \frac{\Delta}{\Delta_\phi} \right) \dots $ should be thought in the averaged sense, so that it gives rise to smooth functions of $s$.
Formula \eqref{FSL:S} was derived in \cite{Paulos:2016fap} in a slightly different form.  The main point is that the scattering amplitude in flat space is an average of pure phases $e^{-i\pi(\Delta -2\Delta_\phi)}$. In other words, 
\begin{align}
1= \lim \sum_{\Delta>0}   \left[ h_1(\Delta) c_{\phi\phi\Delta} \right]^2 \,\delta\left( \frac{\sqrt{s}}{m}- \frac{\Delta}{\Delta_\phi} \right) \frac{2}{ \Delta_\phi}\,.
\end{align}
The universality of the flat space limit behaviour of the distribution of OPE coefficients $c_{\phi\phi\Delta}$ is a prediction that can be used to determine $h_1(\Delta)$.
Indeed, one can use a free theory in AdS to obtain 
\beq
\left[h_1(\Delta) \right]^2=  
\left[c_{\phi\phi\Delta}^{\rm free} \right]^{-2} \approx  
2^{2\D-1}  \D_\p   \sqrt{\frac{\pi }{\D}} \left(\frac{\Delta }{2\Delta_\phi }-1\right)^{\Delta-2 \Delta_\phi
   +\frac{1}{2}} \left(\frac{\Delta
   }{2\Delta_\phi
   }+1\right)^{-\Delta-2 \Delta_\phi
   +\frac{3}{2} }~,
\eeq
where we used that $s\geq 4m^2$ in a physical scattering, so that only operators for which $\D\geq 2\D_\p$ appear in eqs. \eqref{FSL:S}-\eqref{FSL:rho}.

The formulas \eqref{FSL:F2} and  \eqref{FSL:rho} are conjectured based on the  idea that in the flat space limit the discrete spectrum of a QFT in AdS turns into a continuum in flat space. In addition, we think of boundary operators as preparing asymptotic states in the flat space limit and the scattering in Minkowski spacetime corresponds to propagation in global time $\Delta \tau =\pi$ in AdS. Similarly, the form factor should correspond to propagation  in global time $\Delta \tau =\pi/2$. 
A formula equivalent to \eqref{FSL:F2} appeared in \cite{Levine:2023ywq} while this paper was being completed.

The normalization $h_2(\Delta)$ can be fixed as follows.
Replacing \eqref{FSL:rho} in the sum rule
\cite{Karateev:2019ymz},
\beq
C_T=\frac{c_{UV}}{2\pi^2} = \frac{6}{\pi} \int_0^\infty \frac{ds}{s^2} \rho_\Theta(s)~,
\eeq
and comparing with \eqref{sumRuleAdS2FromTT}, one concludes that
\beq
\left[h_2(\Delta)\right]^2= \frac{ 2\pi ^2 \Delta^3}{\Delta_\phi^2} \frac{\Gamma(2\Delta)}{4^\Delta \Gamma^2(\Delta+2)} \approx \frac{\pi^\frac{3}{2} }{\Delta_\phi^2 \sqrt{\Delta}}\,.
\eeq
Formula \eqref{FSL:F2} is built so that the elastic relations \cite{Karateev:2019ymz} 
\beq
\rho_\Theta(s) = \frac{|\mathcal{F}_2^\Theta(s)|^2}{4\pi \sqrt{s(s-4m^2)}} \,,\qquad\qquad
S(s)= \frac{\mathcal{F}_2^\Theta(s)}{[\mathcal{F}_2^\Theta(s)]^*}\,,
\eeq
are automatically satisfied if there are only two-particle states with smooth\footnote{By smooth we mean that $\gamma(n+1)-\gamma(n)\sim 1/\Delta_\phi$ in the flat space limit.} energy shifts $\gamma(n)$, \emph{i.e.} $\Delta = 2\Delta_\phi+2n + \gamma(n)$ for $n=0,1,2,\dots$.
As discussed in \cite{Paulos:2016fap}, inelastic effects in flat space correspond to the existence of other states beyond two-particle in the OPE decomposition of the boundary four-point function.

A stringent consistency check for eq. \eqref{FSL:F2} comes from the flat space sum rule  \cite{Karateev:2019ymz}
\beq
-2m^2 = \mathcal{F}_2^\Theta(0) =\oint_0 \frac{ds}{2\pi i} \frac{\mathcal{F}_2^\Theta(s)}{s\left(1-\frac{s}{4m^2}\right)^q} =
 \int_{4m^2}^\infty \frac{ds}{s\left(\frac{s}{4m^2}-1\right)^q} \frac{ e^{i\pi q} \mathcal{F}_2^\Theta(s) - e^{-i\pi q} \left[\mathcal{F}_2^\Theta(s)\right]^* }{2\pi i}
\,. 
\eeq
where we deformed the contour integral around the origin to wrap the cut along the real $s$-axis from $4m^2$ to infinity and assumed that $q$ is large enough to drop the arc at infinity. We also used real analyticity of the form factor. More precisely, $\mathcal{F}_2^\Theta(s)=\mathcal{F}_2^\Theta(s+i\epsilon)$ and $\left[\mathcal{F}_2^\Theta(s)\right]^*=\mathcal{F}_2^\Theta(s-i\epsilon)$.
Replacing the educated guess \eqref{FSL:F2}, one finds 
\begin{multline}
1=   -\frac{4\sqrt{2}}{\pi}
 \lim \sum_{\Delta>0}   \sin\left[{\frac{\pi}{2}(\Delta -2\Delta_\phi-2q)}\right] \,h_1(\Delta) h_2(\Delta)      (2\Delta\Delta_\phi)^{-\frac{1}{2}} \left( \frac{\Delta^2}{4\Delta_\phi^2 }-1 \right)^{\frac{1}{4}-q} c_{\phi\phi\Delta}  b_{T\Delta} \label{guessSR} \\
 = \lim \sum_{\Delta>0}   \sin\left[{\frac{\pi}{2}(2\Delta_\phi-\Delta +2q)}\right]
 \frac{2^{\D+\frac{3}{2}} }{\D \D_\p} 
 \left(\frac{\Delta }{2\Delta_\phi }-1\right)^{\Delta/2-\Delta_\phi+\frac{1}{2}-q} 
 \left(\frac{\Delta}{2\Delta_\phi   }+1\right)^{-\Delta/2 - \Delta_\phi
   +1-q } c_{\phi\phi\Delta}  
 b_{T\Delta}~.
\end{multline}
We would like to compare this sum rule with eq. \eqref{offDiagSumRule}. It is clear from the oscillating factor in eq. \eqref{guessSR} that we need $\al \approx \D_\p$ in the flat space limit. We shall see that the correct choice is  $\al=\D_\p+q$. We set this value of $\alpha$  in the local blocks, with $q$ a fixed parameter, and approximate them when $\D,\,\D_\p \to\infty$ with ratio fixed and larger than 2. This is worked out in appendix \ref{app:asymptBlocksFF}, and reads
\begin{multline}
G_\D^{\D_\p+q}(\chi) \approx H_\D^{\D_\p+q}(\chi) \approx \\
 \sin \left[\frac{\pi}{2} \left(2\D_\p-\D+2q\right)\right]
\frac{\chi^{\D_\p+q}}{\chi+\left(\frac{\D}{2\D_\p}\right)^2-1} 
\frac{2^{\D-\frac{1}{2}}\D}{\sqrt{\pi} \D_\p^{3/2}} 
\left(\frac{\Delta }{2\Delta_\phi }-1\right)^{\Delta/2-\Delta_\phi+\frac{1}{2}-q} 
 \left(\frac{\Delta}{2\Delta_\phi   }+1\right)^{-\Delta/2 - \Delta_\phi
   +1-q }~.
\label{largeDal02}
\end{multline}
Integrating eq. \eqref{largeDal02} as in eqs. \eqref{Gbar} and \eqref{Hbar}, we obtain the appropriate limit of the coefficients $\kappa(\D,\al)$ in eq. \eqref{kappaSumRule}. The integrals can be performed via a saddle point approximation, and are dominated by $\chi=z=1$. This underlines the importance of the local block decomposition in obtaining the flat space limit: recall that the original conformal block decomposition stops converging at this value of the cross ratio. The result is
\beq
\kappa\left(\D,\al=\D_\p+q\right) \approx 
\sin \left[\frac{\pi}{2} \left(2\D_\p-\D+2q\right)\right]
\frac{2^{\D+\frac{3}{2}}}{\D} 
\left(\frac{\Delta }{2\Delta_\phi }-1\right)^{\Delta/2-\Delta_\phi+\frac{1}{2}-q} 
 \left(\frac{\Delta}{2\Delta_\phi   }+1\right)^{-\Delta/2 - \Delta_\phi
   +1-q }~.
\eeq
Plugging this asymptotics in eq. \eqref{offDiagSumRule}, and comparing with eq. \eqref{guessSR}, we obtain a perfect match.

\section{Examples}
\label{sec:examples}

This section is dedicated to extensive checks of the formulas derived in this paper, in a few solvable examples. We start from a conformal field theory, as a quick warm up, then we move on to the theories of a massive free boson and fermion. The last two examples allow to verify   our sum rules, and provide non-trivial checks of the conjectured flat space formulas \eqref{FSL:S}-\eqref{FSL:rho}. In both theories, the presence of a free parameter---the mass in units of the AdS radius---enriches the physics and allows to test many of the subtleties related  to the convergence of the local block decomposition.

\subsection{Boundary CFT}

The simplest possible example is a conformal field theory in AdS. After a Weyl transformation, this setup is equivalent to a CFT with a boundary in flat space. 

By definition, the trace of the stress tensor vanishes. Then, eq. \eqref{bOPECons} implies that the only quasiprimary appearing in the OPE of the stress tensor has $\D=2$. This is the well-known displacement operator. It follows that both the two-point function of the stress tensor and the form factor are fixed by a single coefficient and a single block. This is a well known consequence of holomorphy in a boundary CFT, see \emph{e.g.} \cite{Billo:2016cpy}.

In particular, plugging $\D=2$ in the spectral block \eqref{blocks2PtAdS2} for the holomorphic components of the two-point function, $\braket{T_{ww} T_{ww}}$, one correctly reproduces the fact that the correlators of holomorphic operators are unaffected by a conformal boundary \cite{Billo:2016cpy}. 
This is nicely verified by the sum rule \eqref{sumRuleAdS2FromTT}, which in this case reduces to
\beq
C_T=\frac{1}{4}b_{T,2}^2~.
\eeq

The other two sum rules involving the two-point function of the stress tensor, eqs. \eqref{s1better} and \eqref{s3better}, are trivially satisfied as $0=0$. For eq. \eqref{s1bbb}, this is obvious. Instead, the value $C_3(0)$ on the l.h.s. of eq. \eqref{s3bbb} can be deduced from eqs. \eqref{TTcomplex} and \eqref{eq:sdFGHI}, which give the short distance limit of the two-point function of the stress tensor for a generic theory in AdS$_2$. Comparing with the results in a boundary CFT, one easily concludes that $C_3(0)=0$.

Moving on to the form factor, one can make use of holomorphy and the method of images, as explained above, to conclude that the three-point function of the stress tensor with two boundary operators is fixed up to a coefficient. The functional coincides with the three-point function of holomorphic operators:
\beq
\braket{\phi(x_1)\phi(x_2) T_{ww}(x,z)} \propto
\frac{1}{x_{12}^{2\D_\p-2}(x-x_1+i z)^2(x-x_2+i z)^2}~.
\label{threePointHolo}
\eeq
This is true both in flat space with a boundary at $z=0$, and in AdS in Poincaré coordinates, due to eq.\eqref{AdStoFlat} which states that the stress tensor is Weyl invariant in two dimensions (up to anomalous contributions that anyway vanish in AdS). One can check that eq. \eqref{3ptFctAdS2TracelessSymm}, once specified to the $\D=2$ contribution and pulled back to physical space, reproduces eq. \eqref{threePointHolo}. On the other hand, the proportionality coefficient in eq. \eqref{threePointHolo} is precisely fixed by the Ward identities explained in subsection \ref{subsec:naive}. Therefore, this case does not allow for an independent check of the ensuing sum rules. Instead, we move on to more interesting cases of massive QFTs.

\subsection{Free scalar in AdS}
\label{subsec:free_scalar}

The action of the free scalar in AdS$_2$ is given by
\beq
    S=\int_{\mathrm{AdS}} \frac{1}{2} \nabla_\mu \Phi \nabla_\nu \Phi + \frac{1}{2}m^2\Phi^2~.
\eeq
The (full, rather than connected) boundary four-point function and its conformal block decomposition are easily obtained---see \emph{e.g.} \cite{Paulos:2019fkw}:
\beq
1+\eta^{-2\Delta_\phi} + (1-\eta)^{-2\Delta_\phi} = \eta^{-2\Delta_\phi}\left[1+  \sum_{n=0}^\infty c_{\phi\phi\Delta_n}^2 G_{\Delta_n}(\eta)\right]~,
\eeq
where $\eta$ is the usual cross ratio,\footnote{This cross ratio is usually denoted by $z$, but so is the radial coordinate in Poincaré AdS. } defined as the position of one operator, while the others are placed at $0$, $1$ and $\infty$. The functions
\beq
G_\D(\eta) = \eta^\D \tFo{\D,\D}{2\D}{\eta}
\label{FourPBlocks}
\eeq
are conformal blocks and
\beq
c_{\phi\phi\Delta_n}^2 = \frac{2 \Gamma^2(\Delta_n) \Gamma(\Delta_n+2\Delta_\phi-1)}{
\Gamma^2(2\Delta_\phi)\Gamma(2\Delta_n-1)\Gamma(\Delta_n-2\Delta_\phi+1)}~.
\label{csqfree}
\eeq
The sum runs over  $\Delta_n=2\Delta_\phi+2n$ with $n=0,1,2,\dots$. The mass and $\D_\phi$ are related by the well-known equation
\beq
m^2=\Delta_\phi(\Delta_\phi-1)~.
\eeq

The trace of the stress tensor is $\Theta = -m^2 \Phi^2$, and its connected two-point function reads as follows:
\beq\label{TwoPtThetaScalar}
\langle \Theta(X_1)\Theta(X_2)\rangle = 2m^4 \left[\Pi_{\Delta_\phi}(X_1,X_2)\right]^2
= 2m^4 \sum_{n=0}^\infty a_n \Pi_{\Delta_n}(X_1,X_2) ~,
\eeq
with \cite{Fitzpatrick:2011dm}
\beq
a_n = \frac{(\frac{1}{2})_n}{2\sqrt{\pi} n!} \frac{(2\Delta_\phi+2n)_\frac{1}{2} (2\Delta_\phi +n)_n}{
\left((\Delta_\phi+n)_\frac{1}{2}\right)^2 (2\Delta_\phi+n-\frac{1}{2})_n}~,
\eeq
and 
\beq
\Pi_{\Delta}(X_1,X_2) =\frac{\Gamma(\Delta)}{2\sqrt{\pi}\Gamma(\Delta+\frac{1}{2}) } f_\Delta(\xi)~.
\eeq
The block $f_\Delta(\xi)$ is defined in eq. \eqref{scalarSpecBlocks}.
Therefore,
\beq
b_{\Theta \Delta_n}^2 = 2\Delta_\phi^2(\Delta_\phi-1)^2  \frac{\Gamma(\Delta_n)}{2\sqrt{\pi}\Gamma(\Delta_n+\frac{1}{2}) } a_n~.
\eeq
Using \eqref{bOPECons}, we conclude that
\beq
b_{T \Delta_n}^2 = \frac{\Delta_\phi^2(\Delta_\phi-1)^2 \Delta_n^2 }{(\Delta_n-2)^2}  \frac{\Gamma(\Delta_n)}{\sqrt{\pi}\Gamma(\Delta_n+\frac{1}{2}) } a_n~.
\label{bTsqfree}
\eeq
This OPE data obeys the sum rule \eqref{sumRuleAdS2FromTT} with the correct value of $C_T=\frac{1}{2\pi^2}$.

Instead, the sum rules \eqref{s1better} and \eqref{s3better} do not directly apply to the free boson example because $\D_\mathcal{V}=0$. 
This is a degenerate case where the perturbing operator $\mathcal{V}$ can mix with the identity.
Nevertheless, 
the sums in \eqref{s1better} and \eqref{s3better} converge  to 
$\frac{1}{2\pi}\Delta_\p(\D_\p-1) $ and  $\frac{1}{\pi}\D_\p(1-\D_\p)  \braket{\p^2}$, respectively. This expectation value is given by $\braket{\p^2}=\frac{1}{\pi}  (\psi(\D_\p)-\psi(1))$, with $\psi$ the digamma function.

For later reference, the short distance limit of the two-point function of the trace \eqref{TwoPtThetaScalar} is 
\beq
	\langle \Theta(X_1)\Theta(X_2)\rangle\approx \frac{\Delta_\phi^2 ( \Delta_\phi-1)^2\log^2 x}{8\pi^2}~,
	\label{ThShortDistFreeScalar}
\eeq
where $x=\sqrt{x^2}$ is defined by eq. \eqref{AdScoordx}, not to be confused with the Poincaré coordinates system. 

Let us now compute the three-point function of the lightest boundary operator $\phi$ with the stress tensor. We start from the trace:
\begin{equation}
\langle \phi(P_1) \phi(P_2)  \Theta(X)\rangle =
-2m^2
 \frac{ \Gamma(\Delta_\phi)}{2\sqrt{\pi}\Gamma(\Delta_\phi+\frac{1}{2}) } \frac{\chi^{\Delta_\p}}{(-2P_1\cdot P_2)^{ {\Delta}_{ {\phi}}}}~. \label{FFThetaScalar}
\end{equation}
One can see that the bound \eqref{boundOnGspin0} is respected, and in particular its exponent is saturated, if we set $\D_\mc{V}=0.$ This is not completely precise, since the two-point function of the trace \eqref{ThShortDistFreeScalar} has a logarithmic behavior, but we will not refine the bound here. In particular, the phase dependence allowed by eq. \eqref{boundOnGspin0} is absent from eq. \eqref{FFThetaScalar}, therefore one can drop the arcs in the dispersion relation without the need for the careful argument that follows eq. \eqref{AofAlpha}.
Expanding eq. \eqref{FFThetaScalar} in conformal blocks:
\beq
-2m^2\frac{ \Gamma(\Delta_\phi)}{2\sqrt{\pi}\Gamma(\Delta_\phi+\frac{1}{2}) } \chi^{\Delta_\p}=\sum_{n=0}^\infty b_{\Theta  \Delta_n} c_{\phi\phi\Delta_n}
\, g_{\Delta_n}(\chi)~,
\eeq
with $g_\Delta(\chi)$ defined in eq. \eqref{block3p0}, we get
\beq \label{CoeffbThScalar}
b_{\Theta  \Delta_n} c_{\phi\phi\Delta_n}=(-1)^{n+1}\,2m^2 \frac{ \Gamma(\Delta_\phi)}{2\sqrt{\pi}\Gamma(\Delta_\phi+\frac{1}{2}) } \frac{ \left[(\Delta_\phi)_n \right]^2}{n! (2\Delta_\phi+n-\frac{1}{2})_n}~.
\eeq
Eq. \eqref{CoeffbThScalar}, together with eq. \eqref{FFThetaScalar}, allow us to directly check the Cauchy-Schwarz inequality \eqref{asymptBoundCauchy}: indeed, taking $\chi \to -\infty$, the oscillating sign in the OPE coefficients cancel out against the blocks, which in turn coincide with the positive functions defined in eq. \eqref{posBlock0}: see the discussion around eq. \eqref{FFDec0x}. Hence, the correlator in that limit matches the l.h.s. of eq. \eqref{asymptBoundCauchy}, and we see that the exponent of the inequality is saturated, again up to logarithmic corrections which would also solve the divergence of the prefactor.

By applying eq. \eqref{bOPECons}, we also get 
\beq
b_{T  \Delta_n} c_{\phi\phi\Delta_n}=(-1)^n\frac{\Delta_\phi(\Delta_\phi-1) \Gamma(\Delta_\phi)}{\sqrt{\pi}\Gamma(\Delta_\phi+\frac{1}{2}) } 
\frac{\Delta_n}{\Delta_n-2}
\frac{ \left[(\Delta_\phi)_n \right]^2}{n! (2\Delta_\phi+n-\frac{1}{2})_n}~,
\label{CoeffbTScalar}
\eeq
which matches the direct computation of the form factor of the traceless part:
\begin{align}
W^M W^N\langle \phi(P_1) \phi(P_2)  T_{MN}\rangle &=
2 \Delta_\p^2
 \frac{ \Gamma(\Delta_\phi)}{2\sqrt{\pi}\Gamma(\Delta_\phi+\frac{1}{2}) } \frac{\chi^{\Delta_\p}}{(-2P_1\cdot P_2)^{ {\Delta}_{ {\phi}}}} T_1 \label{FFTScalar}
\\
&=T_1\sum_{\Delta_n} \frac{b_{T  \Delta_n} c_{\phi\phi\Delta_n} }{(-2P_1\cdot P_2)^{ \Delta_\phi}}
\, h_{\Delta_n}(\chi)~,
\end{align}
with $h_\Delta(\chi)$ defined in eq. \eqref{blocks3p2}.
One can check that the square of \eqref{CoeffbThScalar} gives the product of \eqref{csqfree} and \eqref{bTsqfree}. Interestingly, eq. \eqref{FFTScalar} respects the bound \eqref{boundOnGspin2} also for $\D_\p<1$. We will explain this fact below.

Eqs. \eqref{FFThetaScalar}, \eqref{CoeffbTScalar} and \eqref{FFTScalar} also allow us to put to the test the sum rules defined in section \ref{sec:formFact} and various properties of the local block decomposition. Firstly, one can check the sum rule \eqref{NaiveSumRule}. This is an alternating sum that only converges for $\Delta_\phi <\frac{5}{4}$, and it is absolutely convergent for $\Delta_\phi<\frac{3}{4}$. The latter value saturates the bound \eqref{absConvNaive}, with $\D_\mc{V}=0$. This is no surprise, once we consider the large $\D$ behavior of the product \eqref{CoeffbTScalar}:
\beq
\left| c_{\p \p \D} b_{T\D} \right| \overset{\D\to \infty}{\sim} 2^{-\D} \D^{2\D_\p-3/2}~.
\eeq
The bound \eqref{bcBound} is saturated, if we set $\D_\mc{V}=0$ there (but the comment on logarithmic corrections applies here as well). 

As expected, the sum rule \eqref{offDiagSumRule} works much better. The bound \eqref{alphaBoundLocalBlock} implies that the sum rule must converge for $\al>\D_\phi$ in this case. This can be easily verified numerically for various choices of $\al$.
Since the OPE coefficient saturate the inequality \eqref{bcBound}, we expect that, for generic values of $\al$, the bound on the rate of convergence of the sum rule \eqref{kappaSumRuleConvergence} is also saturated. For instance, if we choose $\al=\Delta_\p+(2k+1)/2$, with integer $k$, the oscillating factor in the local blocks \eqref{localBlock0} and \eqref{localBlock2} precisely cancels the alternating sign in eq. \eqref{CoeffbTScalar} and the sum rule \eqref{offDiagSumRule} has sign definite summands, at least asymptotically---see eq. \eqref{kappaAsymptotics}. In these cases the saturation of the bound on the rate of convergence is guaranteed. In particular, all summands are positive for $k=0$. 

On the contrary, if we pick $\al=\Delta_\p+k$, $k$ a positive integer, the discontinuity across the cut for $\chi<0$ vanishes block by block. In these cases, all the local blocks for which the defining integrals converge, vanish. The only non-vanishing contributions to the sum rule come from the first $k$ operators ($n=0,\dots k-1$)---see eqs. \eqref{dispersiveG} and \eqref{alphaMax}---whose local blocks are fully determined by the arc around $\chi=0$. In other words, the series \eqref{offDiagSumRule} reduces to a finite sum. For instance, setting $k=1$, one correctly gets
\beq
\kappa(2\D_\p,\Delta_\p+1)\, c_{\p \p \D_0} b_{T \D_0} = \D_\p.
\eeq

Now, let us explicitly discuss the case $\D_\p<1$, which was a source of subtleties in the local block decomposition for the spin two part of the stress tensor---see subsection \ref{subsec:spin2LBConv}. The sum rule \eqref{DiscConstraint2} reads:
\beq
\sum_{n=0}^\infty (-1)^n c_{\p\p\D_n}b_{T\D_n} \frac{2^{\D_n+1}\Gamma\left(\D_n+\frac{1}{2}\right)}{\sqrt{\pi} \Ga\left(\D_n+2\right)}=0~, \qquad \D_\p<1~.
\label{DiscConstraintFreeScalar}
\eeq
This is a rather remarkable equation. When $\D_\p \geq 1$ all the summands are non-negative, and the series must diverge, according to the comment below eq. \eqref{DiscConstraint2}. When $\D_\p$ crosses $1$, all the OPE coefficients \eqref{CoeffbTScalar} change sign except the $n=0$ one: the series does not oscillate, rather it turns into a positive sum rule for the OPE coefficients with $n>0$. It is amusing to notice that for this to happen, the factor $\D/(\D-2)$ imposed by conservation---eq. \eqref{bOPECons}---is crucial. It is easy to perform the sum in Mathematica in terms of a linear combination of ${}_3 F_2$'s evaluated at one. We verified numerically that the result vanish, a fact that should follow from hypergeometric identities. This matches all our expectations. What is special to this case is that eq. \eqref{DiscConstraintFreeScalar} also implies
\beq
\sum_{n=0}^\infty c_{\p\p\D_n}b_{T\D_n} F(\D_n) \cos \frac{\pi}{2}\D_n=0~,
\eeq
\emph{i.e.} $\tilde{\mathcal{G}}_T=\mathcal{G}_T$---see eq. \eqref{TildeGT}. In other words, canceling the single-block-discontinuity at $\chi \to \infty$ fully erases the linear term in $\chi$, which is why eq. \eqref{FFTScalar} obeys the bound \eqref{boundOnGspin2} for $\D_\p<1$, as pointed out above.

Of course, all this implies that both the local block decomposition of eq. \eqref{FFTScalar} and the sum rule \eqref{offDiagSumRule} converge for $1>\al>\D_\p$, statements that are easily verified numerically.

Similar observations to the ones made so far apply for the sum rule \eqref{zeroSumRule}, whose validity can be verified analytically, and, depending on the values of $\al$, follows from a nice interplay between the oscillation of the OPE coefficients and the one of the gamma functions in eq. \eqref{zeroSumRule}.

\vspace{0.5cm}
\paragraph{Flat space limit.} The scattering amplitude, the form factor and  the spectral density of $\Theta$ in flat space read \cite{Karateev:2019ymz}
\beq
S(s)=1~,\qquad\qquad
\mathcal{F}_2^\Theta(s)=-2m^2 ~,\qquad\qquad
\rho_\Theta(s) = \frac{m^4}{\pi \sqrt{s(s-4m^2)}}~.
\label{freescalarflat}
\eeq
Remarkably, our conjectured formulas (\ref{FSL:S},\ref{FSL:F2},\ref{FSL:rho})  work beautifully and match  \eqref{freescalarflat}.

\subsection{Free fermion in AdS}

Let us consider the theory of a free Majorana fermion in AdS$_2$ \cite{Doyon:2004fv}. We write Majorana fermions $\Psi$ and their conjugate $\overline{\Psi}$ as 
\begin{align}
 \Psi=\begin{pmatrix}\psi \\ -i\overline{\psi}
    \end{pmatrix}, \qquad
 \overline{\Psi}=\begin{pmatrix} i\overline{\psi}& \psi\end{pmatrix}.
\end{align} 
Here, $\psi$ is a complex Grassman variable and $\bar{\psi}$ is the conjugate of $\psi$. We follow the conventions of \cite{Beccaria:2019dju}.
The action for a free Majorana fermion in AdS$_2$ is given by
\beq
	\int d^2x \sqrt{g}\, \overline{\Psi}\left(\slashed{D}-m\right) \Psi = 2\int \frac{dxdz}{z^2} \left(z \psi \overline{\partial} \psi +z \overline{\psi} \partial \overline{\psi} -im\overline{\psi}\psi \right)~,
	\label{MajoAction}
\eeq
where $ \slashed{D}=\gamma^\mu D_\mu$, $D_\mu$ being the covariant derivative in curved space. On the r.h.s, the action is evaluated in the Poincaré coordinates \eqref{PoincarePatch}. The spin connection could be dropped, as the fermion fields satisfy the Majorana condition $\bar{\Psi} \gamma^\mu \Psi=0$, leaving the partial derivatives $\partial =\frac{1}{2}(\partial_x -i \partial_z)$, $\overline{\partial}=\frac{1}{2}(\partial_x+i\partial_z)$.  The gamma matrices $\gamma^\mu$ are defined via their flat space analog,
\begin{align}
 \Gamma^y=\begin{pmatrix} 0 & 1\\ 1 & 0
    \end{pmatrix}~, \qquad
     \Gamma^z=\begin{pmatrix} 0 & -i \\ i & 0\end{pmatrix}~,
\end{align} 
as $\gamma^\mu = e^{\mu}_a \Gamma^a$. Here, $e^\mu_a=z \delta_\mu^a$ is the zweibein which satisfies $g_{\mu\nu}=\eta_{ab}e^a_\mu e^b_\nu$ with $\eta_{ab}$ the flat space metric. The equations of motions are $(\slashed{D}-m)\Psi=0$. The stress tensor is found by varying the action against the zweibein:
\begin{align}
	T^{\mu \nu}= \frac{1}{2}\overline{\Psi}\left( \gamma^\nu\partial^\mu + \gamma^\mu \partial^\nu \right)\Psi-g^{\mu\nu}\overline{\Psi}(\partial\!\!\!/ -m) \Psi ~.
\end{align}
The trace of the stress tensor is then equal to $\Theta=g_{\mu\nu}T^{\mu\nu}=m\overline{\Psi}\Psi=2im\overline{\psi}\psi$ using the equations of motion. 
  
At the boundary, $\psi$ is identified with $-\overline{\psi}$, and only one real degree of freedom survives. From \cite{Mazac:2018ycv}, we get the boundary four-point function\footnote{We break the convention of the rest of the paper, denoting with $\psi$ the boundary operator and $\D_\psi$ its dimension, instead of $\phi$ and $\D_\p$, to abide to much older conventions on the notation of fermionic fields.}
\beq
-1+\eta^{-2\Delta_\psi} + (1-\eta)^{-2\Delta_\psi} = \eta^{-2\Delta_\psi}\left[1+  \sum_{n=0}^\infty c_{\psi\psi\Delta_n}^2 G_{\Delta_n}(\eta)\right]~,
\eeq
where this time $\Delta_n=2\Delta_\psi+2n+1$ with $n=0,1,2,\dots$. The OPE coefficients read
\beq
c_{\psi\psi\Delta_n}^2 = \frac{2 \Gamma^2(\Delta_n) \Gamma(\Delta_n+2\Delta_\psi-1)}{
\Gamma^2(2\Delta_\psi)\Gamma(2\Delta_n-1)\Gamma(\Delta_n-2\Delta_\psi+1)}~,
\eeq
and $G_\Delta(\eta)$ was defined in eq. \eqref{FourPBlocks}. 

The bulk-to-bulk propagators for the free fermion are given by \cite{Beccaria:2019dju} 
\begin{align}
	\langle \psi(x_1,z_1) \psi(x_2,z_2) \rangle &= C \frac{\overline{w}_1-\overline{w}_2}{\sqrt{z_1z_2}}\frac{F_2(\xi)}{(\xi+1)^{m+1}}~,\label{Bulk2Bulk:1}\\
	\langle \overline{\psi}(x_1,z_1) \overline{\psi}(x_2,z_2) \rangle &= C \frac{w_1-w_2}{\sqrt{z_1z_2}}\frac{F_2(\xi)}{(\xi+1)^{m+1}}~,\label{Bulk2Bulk:2}\\
\langle \psi(x_1,z_1) \overline{\psi}(x_2,z_2) \rangle &=C \frac{w_2-\overline{w}_1}{\sqrt{z_1z_2}}\frac{F_1(\xi)}{(\xi+1)^{m+1}}~,\label{Bulk2Bulk:3}\\
	 \langle \overline{\psi}(x_1,z_1)\psi(x_2,z_2) \rangle &=C \frac{\overline{w}_2-w_1}{\sqrt{z_1z_2}}\frac{F_1(\xi)}{(\xi+1)^{m+1}}~,\label{Bulk2Bulk:4}
\end{align}
where $w_j=x_j+iz_j$, $\Delta_\psi=m+\frac{1}{2}$, 
\beq 
	C = \frac{1}{\sqrt{\pi}4^{2+m}}\frac{\Gamma(m+1)}{\Gamma(m+\frac{1}{2})}~,
\eeq
and 
\beq
	F_1(\xi)=\tFo{m+1,m}{2m+1}{\frac{1}{1+\xi}}, 
	\quad
	F_2(\xi)=\tFo{m+1,m+1}{2m+1}{\frac{1}{1+\xi}}~.
\eeq	
One can readily verify that the boundary condition and the conjugation properties are respected.

We can now compute the bulk two-point function of the trace of the stress tensor by performing Wick contractions: 
\begin{equation}\label{TwoPtThetaFermion}
\begin{split}
 \langle \Theta(X_1) \Theta(X_2)\rangle &= -4m^2\langle \colon \overline{\psi}(X_1) 			\psi(X_1) \colon \, \colon \overline{\psi}(X_2) \psi(X_2) \colon \rangle \\
	&= -16m^2C^2 \left( (\xi + 1)\frac{F_1(\xi)^2}{(\xi+1)^{2m+2}} - \xi \frac{F_2(\xi)^2}{(\xi+1)^{2m+2}} \right)~.
\end{split}
\end{equation}
The normal ordering is defined as in infinite space: it subtracts self contractions, thus yielding the connected correlator.
Comparing eq. \eqref{TwoPtThetaFermion} with the spectral representation \eqref{specDecScalar}, we obtain\footnote{In fact, it is easier  to derive this formula from the form factor \eqref{FFThetaFermion}. } 
\beq
b_{\Theta  \Delta_n}^2=\frac{2^{3+8m+8n} m^2 \Gamma \left(n+\frac{3}{2}\right) \Gamma(m+n+1)^4\Gamma\left(2m+n+\frac{3}{2}\right) }{\pi^3 n! \Gamma\left(2m+n+1\right) \Gamma\left(4m +4n+3\right)}~.
\eeq

One can check that this expression satisfies the sum rule \eqref{sumRuleAdS2FromTT} for $C_T=\frac{1}{4\pi^2}$. On the other hand, the sum rules \eqref{s1better} and \eqref{s3better} do not converge, as it is easy to check. This is consistent with our expectations, since $\D_\mc{V}=1$---see subsection \ref{subsec:taubSpectral}.

In fact, one can readily see that the asymptotics \eqref{tauberian} of the spectral density is respected, including the prefactor. To do this, the value of the parameter $\lambda$ in eq. \eqref{Sdeformed} can be extracted from the short distance limit of eq. \eqref{TwoPtThetaFermion}:
\beq
	\langle \Theta(X_1) \Theta(X_2)\rangle \approx \frac{m^2}{4 \pi^2 x^2}~,
\eeq
where $\xi=-\frac{1-\sqrt{1+x^2}}{2}$ using the coordinates defined in eq. \eqref{AdScoordx}. Comparing this expression with \eqref{Ashort}, we learn that 
\beq\label{lambdaFermion}
	\lambda=\frac{m}{2\pi}~.
\eeq

We then consider the three-point function $\langle \hat{\psi}(P_1) \hat{\psi}(P_2)  \Theta(X)\rangle$, where the boundary fermion is denoted with a hat to avoid confusion with its bulk counterpart. We would like $\hat{\psi}$ to be unit normalized. This prompts the following definition:
\beq
\hat{\psi}(x)=\frac{1}{2^{m+1}\sqrt{C}}\lim_{z \rightarrow 0} 
z^{-\Delta_\psi}\psi(x,z)~.
\eeq
We compute in this way the bulk-to-boundary propagators from eqs. \eqref{Bulk2Bulk:1} and \eqref{Bulk2Bulk:4}: 
\begin{align}
\langle \psi(x_1,z_1) \hat{\psi}(x_2) \rangle &=  2 \sqrt{C}\,
\frac{\sqrt{z_1}}{(x_1-x_2)+iz_1} \left(\frac{2z_1}{(x_1-x_2)^2+z_1^2}\right)^{m}~, 
\label{Bulk2Bdy:1} \\
\langle \overline{\psi}(x_1,z_1) \hat{\psi}(x_2) \rangle &= -2 \sqrt{C} \,
\frac{\sqrt{z_1}}{(x_1-x_2)-iz_1} \left(\frac{2z_1}{(x_1-x_2)^2+z_1^2}\right)^{m}~.
\label{Bulk2Bdy:2}
\end{align}
$\Delta_\psi=m+\frac{1}{2}$ is the dimension of the boundary operator.
We can then compute
\begin{equation}
\begin{split}
	\langle \hat{\psi}(P_1) \hat{\psi}(P_2)  \Theta(X)\rangle &= 2im \langle \hat{\psi}(P_1) \hat{\psi}(P_2)  \colon \overline{\psi}(X) \psi(X) \colon\rangle\\
	&=-\frac{m\,\Gamma(m+1)}{\sqrt{\pi}\,\Gamma\left(m+\frac{1}{2}\right)}\frac{1}{(-2P_1\cdot P_2)^{\frac{1}{2}+m}} \chi^{m+1}~.
	\label{FFThetaFermion}
\end{split}
\end{equation}
By comparing with equation \eqref{3ptFctAdS2Trace}, we get
\beq\label{bThetaCFermion}
b_{\Theta  \Delta_n} c_{\psi\psi\Delta_n}=\frac{(-1)^{n+1} 4^m m \Gamma(m+n+1)^2 \Gamma\left(2m+n+\frac{3}{2}\right)}{\pi n! \Gamma(2m+1)  \Gamma\left(2m +2n+\frac{3}{2}\right)}~,
\eeq
where recall that $\Delta_n= 2\Delta_\psi+2n+1$ and $n=0,1,2,\dots$. Similarly, 
\beq
b_{T  \Delta_n} c_{\psi\psi\Delta_n}=\frac{\Delta}{\Delta-2}\frac{(-1)^n 4^m m \Gamma(m+n+1)^2 \Gamma\left(2m+n+\frac{3}{2}\right)}{\pi n! \Gamma(2m+1) \Gamma\left(2m +2n+\frac{3}{2}\right)}~.
\label{bTCFermion}
\eeq

Eq. \eqref{bTCFermion} can be used to verify the sum rule \eqref{offDiagSumRule}, much as in the case of the free boson. As expected, convergence is seen for $\al>\D_\psi+1/2=m+1$. In particular, for values $\al=\D_\psi+(2k+1)/2$, with $k>0$ and integer, the sum rule has a finite number of terms. 

We can also take the mass to be negative, in the range $-1/2<m<0$. This allows to confirm the convergence of the sum rule for $\al<1$. Again, the identity associated to this improved convergence, eq. \eqref{DiscConstraint2}, is verified. And again, eq. \eqref{DiscConstraint2}, with the free fermion spectrum, also implies
\beq
\sum_{n=0}^\infty c_{\p\p\D_n}b_{T\D_n} F(\D_n) \cos \frac{\pi}{2}\D_n=0~.
\eeq
Therefore, the large $\chi$ asymptotic of the form factor of the traceless part of $T_{\mu\nu}$ must respect the bound \eqref{boundOnGspin2} also for $m<0$. 

In fact, like for the free boson, the product \eqref{bThetaCFermion} saturates the Cauchy-Schwarz bound \eqref{bcBound}, as it is easy to check. Correspondingly, the bound \eqref{asymptBoundCauchy} must be saturated as well, up to the coefficient. This can be verified in the same way as for the free boson, exploiting the $\chi \to -\infty$ limit of the three-point function \eqref{FFThetaFermion}---see the comments after eq. \eqref{CoeffbThScalar}. In this case, we can explicitly check the coefficient in eq. \eqref{asymptBoundCauchy}, using $\Delta_\mathcal{V}=1$ and $\lambda=\frac{m}{2\pi}$. We get the inequality
\beq
	\frac{1}{\sqrt{\pi}} \frac{\Gamma(\Delta_\psi+\frac{1}{2})}{\Gamma\left(\Delta_\psi \right)}  \leq \frac{1}{8\pi} \frac{(4\Delta_\psi+2)^{2\Delta_\psi+1}}{(4\Delta_\psi)^{2\Delta_\psi}} ~,
\eeq
which is indeed satisfied for any values of $\Delta_\psi>0$, and is saturated in the conformal case, $\Delta_\psi=\frac{1}{2}$.

\vspace{0.5cm}
\paragraph{Flat space limit.} The two-to-two S-matrix, the two-particle form factor and the spectral density in flat space are \cite{Mussardo:2010mgq}
\beq
S(s)=-1~,\qquad\qquad
\mathcal{F}_2^\Theta(s)=-m\sqrt{4m^2-s} ~,\qquad\qquad
\rho_\Theta(s) = \frac{m^2\sqrt{s-4m^2}}{4\pi \sqrt{s}}~,
\label{freefermionflat}
\eeq
where the square root is evaluated on the principal branch, and $s$ is above the cut. It is easy to see that eqs. \eqref{FSL:S}-\eqref{FSL:rho} perfectly reproduce these results as well.

\section{Outlook: a bootstrap problem in AdS$_2$}
\label{sec:problem}

Let us summarize what we obtained so far, focusing on the two dimensional case. Given a QFT in AdS$_2$, we derived two constraints on the boundary OPE of the stress tensor, eqs. \eqref{sumRuleAdS2FromTT} and \eqref{offDiagSumRule}, which we reproduce here: 
\begin{subequations}
\begin{align}
&\sum_{\D>0} \frac{24\, \Gamma(2\D)}{4^\D \Gamma(\D+2)^2} b_{T\D}^2=C_T~, \label{sumRulesFinal1}\\
&\sum_{\D>0} 
\kappa(\D,\al)
c_{\phi \phi \D} b_{T \D} = \D_\phi~, \label{sumRulesFinal2}
\end{align}
\label{sumRulesFinal}
\end{subequations}
where
\begin{multline}
\kappa(\D,\al)=
\frac{\pi^{\frac{1}{2}} \D\, \Gamma\left(\D+\frac{1}{2}\right)}{(\D+1) \Gamma\left(\frac{\D}{2}+1\right)^2} \\
- \frac{\pi^\frac{1}{2}}{2}\frac{\D-2}{\D} \frac{\Ga\left(\alpha-\frac{1}{2}\right) \Ga\left(\D+\frac{1}{2}\right) \Ga\left(\alpha\right)}{\Ga\left(\alpha+\frac{\D}{2}+\frac{1}{2}\right) \Ga\left(\alpha-\frac{\D}{2}+1\right) \Ga\left(\frac{\D}{2}\right)^2} \left[ \pFq{3}{2}{1,\al-\frac{1}{2},\al}{\al-\frac{\D}{2}+1,\al+\frac{\D}{2}+\frac{1}{2}}{1}\right.  \\
\left.+\frac{\alpha}{\al-1}
 \pFq{4}{3}{1,\al-1,\al+1;\al-\frac{1}{2}}{\al,\al-\frac{\D}{2}+1,\al+\frac{\D}{2}+\frac{1}{2}}{1} \right]~.
\label{kappaFinal}
\end{multline}
The first sum rule is positive, and constrains the OPE coefficients in terms of the central charge of the bulk CFT in the ultraviolet. The second sum rule is not positive, and involves the three-point coefficients $c_{\phi\phi \D}$, $\phi$ being a boundary primary operator. If our aim is to bound the space of QFTs in AdS, eq. \eqref{sumRulesFinal1} is hardly sufficient: one equation for infinitely many unknowns. On the other hand, any QFT in AdS obeys a well known set of crossing constraints, which apply to the four-point functions of boundary operators. As demonstrated by the large conformal bootstrap literature \cite{Rattazzi:2008pe,Poland:2018epd}, the OPE data is highly constrained by crossing symmetry. In the AdS context, though, it is not obvious how to input in the four-point function the information that a local bulk theory exists. This is where the second sum rule \eqref{sumRulesFinal2} comes in: it provides a constraint on the three-point couplings $c_{\phi\phi \D}$ which is only obeyed by conformal theories on the AdS boundary. 

Together, the two sum rules \eqref{sumRulesFinal} and the crossing equation for the four-point function of $\phi$ form a positive semi-definite constraint, which can be fed to SDPB \cite{Simmons-Duffin:2015qma}. In order to make it explicit, let us write down the crossing equation for the four-point. The conformal blocks in one dimension read
\beq
G_\D(z)=z^\D {}_2F_1(\D,\D;2\D;z)~,
\eeq
where the cross-ratio is related to the position of the four operators on a line as follows:
\beq
z= \frac{x_{12}x_{34}}{x_{13}x_{24}}~.
\eeq
It is possible to restrict it to $z\in (0,1)$. 
Then, crossing reads
\beq
\sum_{\D>0} c^2_{\p\p \D} F^{\D_\p}_\D(z)=1~, \qquad 
F^{\D_\p}_\D(z)= \frac{z^{-2\D_\phi}G_\D(z)-(1-z)^{-2\D_\phi}G_\D(1-z)}{(1-z)^{-2\D_\phi}-z^{-2\D_\phi}}~.
\label{crossing4pt}
\eeq 
The one-dimensional conformal bootstrap has been object of intense scrutiny, and besides the numerical bootstrap tools, a set of analytic functionals is available as well \cite{Mazac:2016qev,Mazac:2018mdx,Mazac:2018ycv}. 

We can now apply three independent linear functionals to eqs. \eqref{sumRulesFinal} and \eqref{crossing4pt}. The space of functionals acting on  eq. \eqref{sumRulesFinal1} is one-dimensional: we can only multiply the equation  by a constant  $\La_1$. On the other hand, for eq. \eqref{sumRulesFinal2} we can use a linear functional $\La_2$ that acts on functions of $\alpha$.
The functionals for the crossing equation \eqref{crossing4pt}, instead, act on the space of functions of one-variable, as usual:
\beq
\La_3\left(F^{\D_\p}_\D\right) \in \mathbb{R}~.
\eeq
We can organize the sum of the resulting equations as follows:
\beq
\sum_{\D>0} 
\begin{pmatrix}
b_{T\D} & c_{\p\p \D}
\end{pmatrix}
\vec{\La} \cdot \vec{V}_\D^{\D_\p}
\begin{pmatrix}
b_{T\D} \\ c_{\p\p \D}
\end{pmatrix}
=
\La_1 C_T + \La_2 (\D_\p) + \La_3(1)~,
\label{TotalCrossing}
\eeq
where $\vec{\La}= (\La_1,\La_2,\La_3)$ and
\beq
\vec{V}_\D^{\D_\p} = 
\begin{pmatrix}
\begin{pmatrix}
\frac{24\, \Gamma(2\D)}{4^\D \Gamma(\D+2)^2}  & 0 \\
0  & 0
\end{pmatrix} \\
\begin{pmatrix}
0  & 
\frac{\kappa(\D,\al)}{2} \\
\frac{\kappa(\D,\al)}{2}  & 0
\end{pmatrix} \\
\begin{pmatrix}
0  & 0 \\
0  & F^{\D_\p}_\D(z)
\end{pmatrix}
\end{pmatrix}~.
\eeq
As usual, one can look for bounds on the CFT data by finding contradictions to eq. \eqref{TotalCrossing}. A contradiction is found if, for a certain choice of spectrum, external dimension $\D_\phi$ and central charge $C_T$, functionals can be found such that
\beq
\vec{\La} \cdot \vec{V}_\D^{\D_\p} \succeq 0~, \qquad 
\La_1 C_T + \La_2 \D_\p + \La_3(1) \leq 0~,
\eeq
with at least one strict inequality.

There are several interesting QFTs that can be studied using this setup. For example, we can studied relevant deformations of CFT minimal models that flow to massive phases or to other minimal models. Some of these theories were studied in AdS$_2$ using Hamiltonian truncation \cite{Hogervorst:2021spa}.
It would be interesting to compare these two approaches. The hope is that the bootstrap approach will allow us to go further into the strongly coupled phase where Hamiltonian truncation fails. Another important theory to study is the Sine-Gordon model. It would be very interesting to supplement the bootstrap study in \cite{Antunes:2021abs} with our new sum rules.

The flat space limit of this bootstrap setup is the mixed bootstrap studied in \cite{Karateev:2019ymz,Correia:2022dyp}.
In flat space, it is not obvious how to systematically add more constraints to the bootstrap setup because the analytic structure of $2\to (n\ge 3)$ scattering amplitudes is too complicated (there is an infinite number of Landau singularities). 
In AdS, it is trivial (at least conceptually) to include more four-point functions of other boundary operators. 
It would be interesting to apply this approach to the Ising Field Theory in the regime where there is only one stable particle in flat space.

It would also be interesting to study asymptotically free QFTs like the $O(N)$ models (for $N\ge 3$) in AdS$_2$ \cite{Carmi:2018qzm}. 
On the one hand, this would be a useful test of the bootstrap approach because these theories are integrable in flat space and therefore there is a lot of data to compare with.
On the other hand, they are toy models for gauge theories in 4D \cite{Aharony:2012jf} which are the holy grail for any non-perturbative QFT method.


\section*{Acknowledgements} 
The authors would like to thank Miguel Paulos for useful discussions and Manuel Loparco for pointing out a typo in the first version of this paper. MM would like to thank the organizers of the ``Bootstrapping nature'' conference held at the Galileo Galilei Institute in October 2022, where some important ideas for this project were first learnt. TS would like to thank the Swiss Study Foundation, the Geissbühler Foundation and the Werner Siemens Foundation for their support during the completion of this project.
 JP is supported by the Simons Foundation grant 488649 (Simons Collaboration on the Nonperturbative Bootstrap) and the Swiss National Science Foundation through the project
200020\_197160 and through the National Centre of Competence in Research SwissMAP. MM is supported by the SNSF Ambizione grant PZ00P2\_193472.

\appendix

\section{Three constraints from conservation}
\label{app:consConstr}

We explicitly write the three components of the vector $E=(E_1,E_2,E_3)$, defined in equation \eqref{consConstrEq}, which vanishes due to conservation of the stress tensor:

\begin{subequations}
\begin{multline}
E_1 =
\left(\zeta ^2-1\right)^2 \zeta ^3 h_1'
+(d+2) \left(\zeta ^2-1\right) \zeta ^3 h_2'
-4 \left(\zeta    ^2-1\right) \zeta ^2 h_3'
+(d+1) \zeta ^3 h_4'
+2 \zeta  h_5'\\
+(d+4) \left(\zeta ^2-1\right) \zeta ^4 h_1
+(d (d+3)+4) \zeta ^4    h_2
-4 \left((d+1) \zeta ^3+\zeta \right)h_3 
-2 \left(d \zeta    ^2+2\right)h_5 
 ~,
\end{multline}
\begin{multline}
E_2=
\left(\zeta ^2-1\right)^2 \zeta ^3 h_1'
+2 \left(\zeta   ^2-1\right) \zeta ^3 h_2'
-4 \left(\zeta ^2-1\right) \zeta^2 h_3'
+\zeta ^3 h_4'
+2 \zeta  h_5'\\
+(d+4) \left(\zeta ^2-1\right) \zeta ^4 h_1
+(d+4) \zeta ^4 h_2
-\left(2 (d+2) \zeta ^3+4 \zeta   \right)h_3 
-4 h_5
   ~,
\end{multline}   
\begin{multline}
E_3=   
\left(\zeta ^2-1\right)^2 \zeta ^3 h_1'
+2 \left(\zeta   ^2-1\right) \zeta ^3 h_2'
 -\left(\zeta ^2-1\right) \zeta ^2 \left(d   \zeta ^2+4\right) h_3'
+\zeta ^3 h_4' 
+\zeta  \left(d \zeta ^2+2\right) h_5'\\
+(d+4)   \left(\zeta ^2-1\right) \zeta ^4 h_1
+2 (d+2) \zeta ^4 h_2
-\left(d   (d+3) \zeta ^5+2 (d+2) \zeta ^3+4 \zeta \right)h_3 
- \left( d \zeta   ^2+4\right)h_5~.
  \end{multline}
\end{subequations}

\section{Sum rules for the stress tensor in flat space}
\label{app:flatStress}

In this appendix, we apply the strategy of section \ref{sec:twopStress} to flat space. This is the same setup discussed in detail in \cite{Karateev:2020axc}. We re-derive  known sum rules, including the $c$-theorem in $2D$ \cite{Zamolodchikov:1986gt}. Interestingly, we also find an additional linear constraint on the two-point function of the stress tensor in two dimensions. 

{\bf Notation alert:} In this appendix, we use $d$ to denote spacetime dimension following the standard convention of contemporary CFT literature. Notice that in the main text, we used $d+1$ for the spacetime dimension of AdS. Of course this needs to be taken into account when comparing the flat space limit of formulas from section \ref{sec:twopStress} with formulas in this appendix.

\subsection{The two-point function of the stress tensor}
\label{subsec:twopStressFlat}

In $d$ dimensions, by virtue of rotational, translational and parity invariance, the connected two point function can be written as
\begin{equation}\label{2ptFctFlat}
\begin{split}
    \mathbb{T}^{(\mu \nu),(\lambda \sigma)} = \langle0| T^{\mu\nu}(x)T^{\lambda \sigma}(0)| 0\rangle_\textup{connected}=\sum_{i=1}^5 \frac{1}{x^{2d}}h_i(x^2)\mathbb{T}_i^{(\mu \nu),(\lambda \sigma)}~,
\end{split}
\end{equation}
where we have defined the tensors
\begin{equation}\label{TensorStructuresFlatGeneralDim}
    \begin{split}
        \mathbb{T}_1^{(\mu \nu),(\lambda \sigma)}&=\frac{x^\mu x^\nu x^{\lambda } x^{\sigma}}{x^4}\\
        \mathbb{T}_2^{(\mu \nu),(\lambda \sigma)}&= \frac{x^\mu x^\nu \delta^{\lambda \sigma} + x^{\lambda } x^{\sigma} \delta^{\mu \nu }}{x^2}\\
        \mathbb{T}_3^{(\mu \nu),(\lambda \sigma)}&= \frac{x^{\mu } x^{\lambda } \delta^{\nu \sigma} + x^{\nu } x^{\sigma} \delta^{\mu \lambda }+x^{\nu } x^{\lambda } \delta^{\mu \sigma}+ x^{\mu } x^{\sigma} \delta^{\lambda \nu }}{x^2}\\
        \mathbb{T}_4^{(\mu \nu),(\lambda \sigma)}&= \delta^{\mu \nu } \delta^{\lambda \sigma}\\
        \mathbb{T}_5^{(\mu \nu),(\lambda \sigma)}&= \delta^{\mu \lambda } \delta^{\nu \sigma} + \delta^{\mu \sigma} \delta^{\nu \lambda }~.
    \end{split}
\end{equation}
In the following, we sometimes denote the square distance by
\beq
w=x^2~.
\eeq
The two point function of the trace of the stress tensors is 
\begin{equation}\label{FlatSpace2ptFctTrace}
    A(w)= \delta_{\mu \nu }\delta_{\lambda \sigma} \mathbb{T}^{(\mu \nu),(\lambda \sigma)}=\frac{1}{w^d}\left(h_1(w)+2 d\, h_2(w)+ 4h_3(w) +d^2 h_4(w)+ 2d\, h_5(w)\right)~.
\end{equation}
As in AdS, conservation yields three equations, which are independent for $d>1$, which can be collected in a vector:
\beq
E=x^{2d}\left( x^{\mu} \delta^{\lambda \sigma}, \frac{x^{\mu } x^{\lambda } x^{\sigma}}{x^2},\delta^{\mu \lambda }  x^{\sigma}\right)
\frac{\pa}{\pa x^{\nu}}\mathbb{T}^{(\mu \nu),(\lambda \sigma)}~.
\label{consVecFlat}
\eeq
As in AdS, we consider QFTs obtained by deforming a CFT with a relevant operator
\beq
S=S_\textup{CFT}+\lambda \int \mathcal{V}~, \qquad \D_\mathcal{V}<d~.
\label{SdeformedFlat}
\eeq
In the UV limit $w\to 0$, the two-point function then approximates the conformal invariant form
\begin{equation}\label{hiUVFlatSpace}
    h_1(w) \approx 4C_T~, \quad h_2(w) \sim o(w^0)~, \quad h_3(w)\approx -C_T~, \quad h_4(w)\approx -\frac{C_T}{d}~,\quad h_5(w)\approx \frac{C_T}{2}~.
\end{equation}
If the flow is massless, then the same limit is reached in the IR as well, with a different value for $C_T$. On the contrary, the $h_i$'s decay exponentially at large distances in a massive QFT. The two-point function of the trace has the short distance limit already written in eq. \eqref{Ashort}.

\subsection{A sum rule in any dimension}
\label{subsec:sumRulesHdFlat}

In flat spacetime, the procedure described in the beginning of section \ref{subsec:hdsumrules}
leads to only one candidate $C$-function:
\beq
 C(w)=-\frac{1}{w^{d/2}}\big(h_1(w)+ (d+1) h_2(w) + 4\, h_3(w) +d\, h_4(w) + 2 h_5(w)\big)~.
 \label{CHdFlat}
\eeq
The differential constraint reads
\beq
\dot{C}(w)= -\frac{1}{2}w^{d/2-1} A(w)~.
\label{diffHdFlat}
\eeq
Therefore,
\beq
C(0)-C(\infty) = \frac{1}{2}\int_0^\infty  dw\,w^{d/2-1} A(w) = \frac{1}{S_d} 
\int d^dx \braket{\Theta(x)\Theta(0)} ~.
\label{AtoIntTT}
\eeq
One way to proceed is to study the right hand side of the sum rule using the identity
\beq
\Theta(x) = \partial_\mu j_s^\mu~, \qquad j_s^\mu = x_\nu T^{\mu\nu}~,
\eeq
and  write
\beq
 \lim_{\substack{\ep \to 0 \\R\to \infty }} \int_{\eps<|x|<R}\!d^dx
\, \partial_\mu \braket{j_s^\mu(x) \Theta(0)} 
=   \lim_{\substack{\ep \to 0 \\R\to \infty }} 
\left(\int_{\Sigma(R)}- \int_{\Sigma(\ep)}\right) d\Sigma_\mu \braket{j_s^\mu(x) \Theta(0)} ~.
\eeq
In the last expression, the flux is computed over spheres $\Sigma$ with radii $R$ and $\ep$. We can now use the UV and IR approximations for the correlator to compute the integrals. This means that the IR contribution vanishes, while in the UV $j_s^\mu$ becomes the scaling current of the CFT, and $\Theta$ can be replaced via eq. \eqref{ThetaUV}. The flux of the current gives the charge, and so \cite{Delfino:1996nf} 
\beq
\int d^dx \braket{\Theta(x)\Theta(0)}  =   \D_\mathcal{V}  \braket{\Theta}~.
\label{sumBad}
\eeq

\subsection{The sum rules and a linear constraint in $2d$}
\label{subsec:sumRules2dFlat}

In two dimensions, the tensor structures \eqref{TensorStructuresFlatGeneralDim} are linearly dependent:
\begin{equation}
    2 \mathbb{T}_2^{\mu\nu \lambda\sigma}- \mathbb{T}_3^{\mu\nu \lambda\sigma}-2 \mathbb{T}_4^{\mu\nu \lambda\sigma}+ \mathbb{T}_5^{\mu\nu \lambda\sigma}=0~.
    \label{TLinearDep2dFlat}
\end{equation}
Hence, the functions $h_i$ are again defined up to a shift by an arbitrary function, analogous to the one in eq. \eqref{hred2d}. The story proceeds much as in the AdS case. One can write an ansatz for a $C$-function in terms of four shift invariant structures and ask that the $w-$derivative is proportional to the trace $A(w)$, once conservation \eqref{consVecFlat} is imposed. The resulting options are 
\begin{subequations}
\begin{align}
C_1(w) &=  -\frac{1}{w}\left(h_1(w)+3 h_2(w) +4 h_3(w) +2  h_4(w) +2 h_5(w)\right)~,\label{CBad} \\
C_2(w) &= -h_1(w) - 4 h_2(w) - 5 h_3(w) - 3 h_4(w) - 3 h_5(w)~,\label{CGood} \\
C_3(w) &= \frac{4}{w^2}\left(h_2(w) - h_3(w) +  h_4(w) - h_5(w)\right)~. \label{CUgly} 
\end{align}
\label{sumRules2dFlat}
\end{subequations}
The three functions obey the following equations:
\begin{subequations}
\begin{align}
\dot{C}_1(w) &= -\frac{1}{2}A(w)~, \label{diffBad}  \\
\dot{C}_2(w) &= -\frac{3}{2} w A(w)~,  \label{diffGood}\\
\dot{C}_3(w) &= 0~, \label{diffUgly} 
\end{align}
\label{diff2dFlat}
\end{subequations}
where a dot denotes a derivative with respect to $w$. Integrating the first two equations, we get sum rules involving the flat space spectral density. 
The first sum rule gives \eqref{sumBad} in two dimensions.
Eq. \eqref{diffGood} yields a sum rule for the central charges:
\beq
C^{UV}_T-C^{IR}_T = \frac{3}{2}\int\! dw\, w A(w)
=\frac{3}{\pi} \int d^2 x \,x^2
\langle \Theta(x)\Theta(0)\rangle~.
\label{sumGood}
\eeq
Since in a local unitary theory $A(w)>0$, eq. \eqref{sumGood} establishes a $c$-theorem \cite{Zamolodchikov:1986gt}.

Let us now move on to the third $C$-function \eqref{CUgly}. Eq. \eqref{diffUgly} tells us that $C_3$ is constant. We can therefore compute it in the far IR, where it clearly vanishes. We conclude that
\beq
h_2(w) - h_3(w) +  h_4(w) - h_5(w)=0~.
\label{UglyConstraint}
\eeq
This is a remarkable constraint for the two-point function of the stress tensor in any two dimensional QFT, which was not noticed before to the best of our knowledge. Together with the shift symmetry \eqref{TLinearDep2dFlat}, it implies that the two-point function only contains three independent functions. One can easily check the validity of eq. \eqref{UglyConstraint}, for instance, in a free massive theory, where it follows from an identity involving sum of products of Bessel functions. 

Taking the flat space limit of the AdS sum rules given in  \eqref{CAdS2} we obtain precisely \eqref{sumRules2dFlat}.

\subsection{The UV limit from Conformal Perturbation Theory}
\label{app:CPTflatspace}

\subsubsection{Two spacetime dimensions}

Eq. \eqref{sumBad} mandates the UV limit of $C_1(w)$ in eq. \eqref{CBad}. It is interesting to see that this agrees with a direct computation in conformal perturbation theory. 
The leading term in the short distance expansion of the $h_i$'s is given by the UV CFT limit \eqref{hiUVFlatSpace}. One can easily check that this does not contribute to $C_1(w)$ (nor $C_3(w)$)  in eq. \eqref{sumRules2dFlat}. In fact, to get a finite limit $C_1(0)$, one needs subleading terms linear in $w$ in the short distance expansion.
In conformal perturbation theory, we are instructed \cite{Amoretti:2017aze, Mussardo:2020rxh} to use the off-critical OPE to compute the short distance limit of the two-point function:
\beq
\braket{T^{\mu\nu}(x)T^{\la\si}(0)} \approx \frac{1}{w^2} 
\left(c_{TT \mathbb{1}}^{\mu\nu\la\si}(x) \braket{\mathbb{1}}+c_{TT \mathcal{O}}^{\mu\nu\la\si}(x) \braket{\mathcal{O}}+\dots\right)~,
\label{TTfromOPEflat}
\eeq
where $\mathcal{O}$ is a QFT operator with mass dimension $\D_\mathcal{O}$.\footnote{Each QFT operator $\mathcal{O}$ is a linear combination of CFT operators with dimension smaller or equal to $\D_\mathcal{O}$.}
The OPE coefficients $c$ are analytic in the coupling $\lambda$ and computable in a perturbative expansion. Their index structure is captured by the dimensionless structures \eqref{TensorStructuresFlatGeneralDim}, hence we will schematically denote them as functions of $w$ in the following. Expectation values of the operators, on the other hand, are non-perturbative: their dependence on the coupling is fixed by dimensional analysis. Our interest only lies in the $w$ dependence at small $w$, which can also be deduced from dimensional analysis and some input from perturbation theory. It is convenient to treat the traceless part the stress tensor $\tilde{T}$ and the trace $\Theta$  separately. Let us start from $c_{\tilde{T}\tilde{T}\mathbb{1}}$. At zeroth order in $\lambda$, we find the CFT answer, which we already know does not contribute to $C_1$ (nor $C_3$) in \eqref{sumRules2dFlat}. Since the fusion of $\tilde{T}$ with itself only contains the identity Virasoro module, the leading correction is proportional to $\lambda^2$. Using that
\beq
[\lambda] = (\textup{mass})^{2-\D_\mathcal{V}}~,
\eeq
we get, schematically,
\beq
c_{\tilde{T}\tilde{T}\mathbb{1}}(w) \sim (\textup{CFT value})+ \lambda^2 w^{2-\D_\mathcal{V}}
\eeq
at small $w$. A similar analysis can be applied to the other components of the two-point function, recalling that the trace can be expressed via eq. \eqref{ThetaUV} in terms of $\mathcal{V}$. One finds
\begin{align}
c_{\tilde{T}\Theta\mathbb{1}}(w) &\sim \lambda^2w^{2-\D_\mathcal{V}} \\
c_{\Theta\Theta\mathbb{1}}(w) &\sim \lambda^2w^{2-\D_\mathcal{V}}~. 
\end{align}
Similarly, for a generic operator $\mathcal{O} \neq \mathcal{V}$, keeping into account that
\beq
[\braket{\mathcal{O}}]=(\textup{mass})^{\D_\mathcal{O}}~,
\eeq
one finds
\beq
c_{\tilde{T}\tilde{T}\mathcal{O}}(w) \sim c_{\tilde{T}\Theta\mathcal{O}}(w) \sim 
c_{\Theta\Theta\mathcal{O}}(w)\sim \lambda^2 w^{\D_\mathcal{O}/2+2-\D_\mathcal{V}}~.
\eeq
The previous OPE coefficients are more suppressed in case $c_{\mathcal{V}\mathcal{V}\mathcal{O}}=0$ at the UV fixed point. Finally, the exchange of the perturbing operator $\mathcal{V}$---or equivalently of the trace $\Theta$---in the OPE leads to a correction
\begin{align}
&c_{\tilde{T}\tilde{T}\mathcal{V}}(w) \sim c_{\tilde{T}\Theta\mathcal{V}}(w) \sim \lambda w~, \label{leadContrFlat} \\
&c_{\Theta\Theta\mathcal{V}}(w) \sim \lambda^2 w^{2-\D_\mathcal{V}/2}~.
\end{align}
If $\D_\mathcal{V}<1$, eq. \eqref{leadContrFlat} gives the dominant term at small $w$, which indeed leads to a finite limit for $C_1(w)$ \eqref{CBad}, proportional to $\braket{\Theta}$. Finally, notice that the fixed point OPE coefficients which determine $c_{\tilde{T}\tilde{T}\mathcal{V}}(w)$ and $c_{\tilde{T}\Theta\mathcal{V}}(w)$ are theory independent, in agreement with eq. \eqref{sumBad}.

\subsubsection{Higher spacetime dimensions}

One can also check directly that $C(w)$ in eq. \eqref{CHdFlat} has the correct small $w$ limit, using conformal perturbation theory in a similar fashion as in two dimensions. The computation is analogous to the two-dimensional one, except for the contributions to the two-point function of the traceless part of the stress tensor. Indeed, when estimating the coefficients $c_{\tilde{T}\tilde{T} \textup{anything}}$, we took advantage of the fact that the stress tensor in a $2d$ CFT is a descendant of the identity. This is not the case in higher dimensions, and so the following coefficients might be leading, depending on the scaling dimensions involved:
\begin{equation}
c_{\tilde{T}\tilde{T} \mathbb{1}}(w) \sim (\textup{CFT value})+
\lambda\, w^{\frac{d-\D_\mathcal{V}}{2}}+\dots~, 
\qquad
c_{\tilde{T}\tilde{T} \mathcal{O}}(w) \sim w^{\frac{\D_\mathcal{O}}{2}}+\lambda\, 
w^{\frac{d-\D_\mathcal{V}+\D_\mathcal{O}}{2}}+\dots~,
\label{offendingHdFlat}
\end{equation}
where $\mathcal{O}$ is any operator, including $\mathcal{V}$. This is incompatible with eqs. (\ref{CHdFlat},\ref{AtoIntTT},\ref{sumBad}), which require $C(0)$ to be finite.
However, looking closely at the offending terms, one discovers that none of them contribute to $C(w)$. Indeed, the first correction to $c_{\tilde{T}\tilde{T} \mathbb{1}}$ is proportional to\footnote{In fact, the integral is infrared divergent, but such divergences cancel out in the full OPE \eqref{TTfromOPEflat}, so we are instructed to ignore them \cite{Mussardo:2020rxh}.}
\beq
 \langle \Tilde{T}^{\mu\nu}(x) \Tilde{T}^{\lambda \sigma}(0)\int \!d^dy\, \mathcal{V}(y)\rangle_{\textup{CFT}}~.
\eeq
This correlator is conserved and traceless in the $(\mu\nu)$ indices and $x$ coordinate: the delta function terms in the Ward identities are proportional to (a derivative of) $\braket{\Tilde{T}^{\lambda \sigma}(0)\mathcal{V}(x)}_\textup{CFT}$, which vanishes. Therefore, the associated $h_i$'s in eq. \eqref{2ptFctFlat} are all proportional to a pure power, and their proportionality constants are constrained by tracelessness and conservation. One can check that, when replacing them in eq. \eqref{CHdFlat}, $C$ vanishes. We conclude that the term of order $\lambda$ in $c_{\tilde{T}\tilde{T} \mathbb{1}}$ does not contribute to the $C$-function. A similar story holds for the order $\la^0$ term in $c_{\tilde{T}\tilde{T} \mathcal{O}}$. Vice versa, the fate of the order $\la$ correction crucially depends on whether $\mathcal{O}=\mathcal{V}$ or not. Indeed, in this case, taking the trace or the divergence of the stress tensor leads to a result proportional to $\braket{\Tilde{T}^{\lambda \sigma}(0)\mathcal{V}(x)\mathcal{O}(\infty)}_\textup{CFT}$, which only vanishes if $\mathcal{O}\neq\mathcal{V}$. Hence, the correction of order $\lambda$ in $c_{\tilde{T}\tilde{T} \mathcal{V}}$ is not conserved nor traceless, and contributes to $C$ with a term of the right order $w^{d/2}$ and proportional to $\braket{\Theta}$. A contribution of the same order arises from $c_{\tilde{T}\Theta \mathcal{V}}$, as in eq. \eqref{leadContrFlat}. These two terms, again, fix the leading behavior of eq. \eqref{CHdFlat} at small $w$ to be a constant, compatibly with eq. \eqref{AtoIntTT}.

\section{Details on the sum rules in AdS}
\label{app:details}

\subsection{The $C$-functions in AdS$_2$}

\begin{subequations}
\begin{multline}
C_1(\xi) = -64 \xi^3\left(\xi+1 \right)^3 h_1(\xi)-48\xi^2 \left(\xi+1 \right)^2 h_2(\xi)
-\frac{64 \xi^2\left(\xi+1\right)^2}{2 \xi +1}h_3(\xi)-8 \xi  (\xi +1)
   h_4(\xi) \\
   -\frac{8 \xi  (\xi +1) }{(2 \xi +1)^2} h_5(\xi)~.
   \label{C1AdS2}
\end{multline}
\begin{multline}
C_2(\xi)=-256\, \xi ^3 (\xi +1)^3 \log (\xi +1) h_1(\xi)
-64 \xi ^2 (\xi +1) \big[\xi +3 (\xi +1) \log (\xi +1)\big]h_2(\xi)  \\
-\frac{64}{2 \xi +1} \xi ^2 (\xi +1) \big[\xi  (2 \xi
   +1)+4 (\xi +1) \log (\xi +1) \big] h_3(\xi)
   -16 \xi  \big[\xi +2 (\xi +1) \log (\xi +1)\big] h_4(\xi)\\
   -\frac{16}{(2\xi +1)^2} \xi  
   \big[\xi  (2 \xi +1)+2 (\xi +1) \log (\xi +1)\big] h_5(\xi)~,
   \label{C2AdS2}
\end{multline}
\begin{multline}
C_3(\xi) = -64\, \xi ^3 (\xi +1)^3 \log (1+1/\xi)  \,h_1(\xi)
+16\, \xi  (\xi +1)  \left(1+2\xi -3\, \xi  (1+\xi) \log (1+1/\xi) \right) \,h_2(\xi)  \\
-\frac{16 \xi  (\xi +1) \left(1+4 \xi +6 \xi^2+4
   \xi^3+ 4\xi(1+\xi)  \log (1+1/\xi) \right)}{2 \xi +1} \,h_3(\xi)\\
   +\left(4+8\xi-8\xi (\xi +1)  \log (1+1/\xi ) \right) \,h_4(\xi) 
  -\frac{4+8 \xi (2+3\xi+2\xi^2)  +8 \xi (1+ \xi ) \log (1+1/\xi)}{(2 \xi +1)^2} \,h_5(\xi)
     \label{C3AdS2}
\end{multline}
\label{CAdS2}
\end{subequations}

The $C$-functions have a more compact expression in terms of the parametrization of the stress tensor two-point function defined in subsection \ref{subsec:twopBlocksStress}---see eq. \eqref{blocks2PtAdS2}. Explicitly, let us define the following functions:
\begin{subequations}
\begin{align}
\braket{T_{ww}(w_1,\bar{w}_1)T_{ww}(w_2,\bar{w}_2)}
&= \left(\frac{1}{z_1z_2} 
 \frac{\bar{w}_1-\bar{w}_2}{w_1-w_2}\right)^2 F(\xi)~, \\
\braket{T_{ww}(w_1,\bar{w}_1)\Th(w_2,\bar{w}_2)} &= 
 \left(\frac{1}{z_1} \frac{\bar{w}_1-\bar{w}_2}{w_1-\bar{w}_2}\right)^{2} G(\xi)~, \\
 \braket{\Th(w_1,\bar{w}_1)\Th(w_2,\bar{w}_2)} &= H(\xi)~,\\
  \braket{T_{ww}(w_1,\bar{w}_1)T_{\bar{w}\bar{w}}(w_2,\bar{w}_2)}
&= \left( \frac{1}{z_1z_2} 
 \frac{(w_1-w_2)(\bar{w}_1-\bar{w}_2)}{(w_1-\bar{w}_2)^2}\right)^2 I(\xi)~.
\end{align}
\label{TTcomplex}
\end{subequations}
Recall that $w=x+\ii z$ and  $\bar{w}=x-\ii z$ are coordinates on the upper half-plane $z>0$, with metric given in eq. \eqref{PoincarePatch}. The remaining components of the stress tensor two-point function can be found by using parity, which swaps $w \leftrightarrow - \bar{w}$. Eq. \eqref{TTcomplex} confirms that there are four independent functions in two dimensions. They can be expressed in terms of the $h$ basis of eq. \eqref{2ptFct} as follows:
\begin{subequations}
\begin{align}
F(\cross) &= \frac{1}{16} \left(\cross ^2-1\right)^2 h_1(\cross )-\frac{(\cross -1) (\cross +1)^2 }{4 \cross
   }h_3(\cross ) +\frac{(\cross +1)^2}{8 \cross ^2}h_5(\cross )~, \\
   G(\cross) &= -\frac{1}{4} \left(\cross ^2-1\right) (\cross -1)^2 h_1(\cross )-\frac{1}{2} (\cross -1)^2 h_2(\cross
   )+\left(\cross +\frac{1}{\cross }-2\right) h_3(\cross )+\frac{(\cross -1)^2 }{2 \cross ^2}h_5(\cross )~, \\
   H(\cross) &= \left(\cross ^2-1\right)^2 h_1(\cross )+4 \left(\cross ^2-1\right) h_2(\cross )+\left(\frac{4}{\cross }-4
   \cross \right) h_3(\cross )+4 h_4(\cross )+2 \left(\frac{1}{\cross ^2}+1\right) h_5(\cross )~, \\
 I(\cross) &= \frac{1}{16} (\cross -1)^4 h_1(\cross )+\frac{(\cross -1)^4 }{4 \cross  (\cross
   +1)}h_3(\cross )+\frac{(\cross -1)^4 }{8 \cross ^2 (\cross +1)^2}h_5(\cross )~.
\end{align}
\label{FGHIofh}
\end{subequations}
In terms of $F,\,G,\,H$ and $I$, the $C$-functions \eqref{CAdS2} read
\begin{subequations}
\begin{align}
   C_1(\xi) &= 8 \xi ^2 G(\xi )-2 \xi  (\xi +1) H(\xi )~.
\label{C1AdS2ofT}\\
C_2(\xi) &= 64 \xi ^2 F(\xi )+32 \xi ^2 \log (\xi +1)
   G(\xi )-4 \xi  (\xi +2 (\xi +1) \log (\xi +1)) H(\xi )~,   \label{C2AdS2ofT}\\
      C_3(\xi) &= 16 \xi^2 \left( F(\xi)-I(\xi) \right)+ 8 \xi ^2 \log   (1+1/\xi ) G(\xi )
      + \left(1+2\xi-2 \xi(1+\xi)  \log (1+1/\xi )\right) H(\xi ) ~,
   \label{C3AdS2ofT}
\end{align}
\label{CAdS2ofT}
\end{subequations}
Here we abused notation and simply replaced $F(\zeta)$ by $F(\xi)$ and similarly for the other functions.

An advantage of the parametrization \eqref{TTcomplex} is that the large distance limit is transparent. Indeed, comparing eq. \eqref{TTcomplex} with eq. \eqref{blocks2PtAdS2}, we see that $F,\,G,\,H$ and $I$ all behave like $\xi^{-\D_\textup{gap}}$ at large $\xi$, $\D_\textup{gap}$ being the scaling dimension of the leading operator (above the identity) in the boundary OPE of the stress tensor.
Explicitly, the spectral representation of the four functions is
\begin{subequations}
\begin{align}
F(\xi) &= \frac{1}{16} \sum_{\D>0}  b_{T\D}^2  (4\xi)^{-\D} {}_2 F_1 \left(\D-2,\D+2;2\D;-\frac{1}{\xi}\right)~, \\
G(\xi) &= -\frac{1}{4}\sum_{\D>0}  b_{T\D} b_{\Theta\D} (4\xi)^{-\D} {}_2 F_1 \left(\D-2,\D;2\D;-\frac{1}{\xi}\right)~,  \\
H(\xi) &= \sum_{\D>0}    b_{\Theta\D}^2  (4\xi)^{-\D} {}_2 F_1 \left(\D,\D;2\D;-\frac{1}{\xi}\right)~,\\
I(\xi) &=\frac{1}{16} \sum_{\D>0}  b_{T\D}^2   (4\xi)^{-\D} {}_2 F_1 \left(\D-2,\D-2;2\D;-\frac{1}{\xi}\right)~.
\end{align} 
\end{subequations}
Recalling the relation \eqref{bOPECons}, we conclude that
\begin{subequations}
\begin{align}
C_1(\xi) &\approx b_{T\D_\textup{gap}}^2\frac{\D_\textup{gap}-2}{4\D_\textup{gap}^2}(4\xi)^{2-\D_\textup{gap}}  ~, \\
C_2(\xi) &\approx b_{T\D_\textup{gap}}^2\frac{\D_\textup{gap}-2}{\D_\textup{gap}^2}(4\xi)^{2-\D_\textup{gap}} \log\xi~, \\
C_3(\xi) &\approx b_{T\D_\textup{gap}}^2\frac{4(\D_\textup{gap}-2)^2}{3\D_\textup{gap}^2(1+\D_\textup{gap} )}(4\xi)^{-1-\D_\textup{gap}}~,
\qquad \qquad \xi \to \infty~.
\end{align}
\label{CAdS2IR}
\end{subequations}
These limits match the infrared behavior of the integrals on the r.h.s. of the sum rules derived in subsection \ref{subsec:2dsumrules}.

%

\subsection{The short distance expansion and the sum rules in AdS$_2$}
\label{SD-AdS2}

Based on the discussion in section \ref{app:CPTflatspace}, we can write a power series ansatz with undetermined coefficients for the functions $F$, $G$, $H$ and $I$. Then we impose the 3 equations that follow from conservation of the stress tensor. This leads to mane linear relations between the coefficients. In this ansatz, we include integer powers of $\xi$ from special operators like the identity, the stress tensor or the perturbing relevant operator $\mathcal{V} \sim \Theta$, and non-integer powers of $\xi$ from other local operators with generic scaling dimensions.
It is clear that the conservation equations will not related the coefficients of integer powers of $\xi$ to non-integer powers of $\xi$. Therefore, we can treat each set of terms separately.
Let us start with a non-integer power series ansatz of the form
\begin{subequations}
\begin{align}
F(\xi) &= \xi^{-2-\alpha} \sum_{n=0} f_n \xi^n~, \\
   G(\xi) &= \xi^{-2-\alpha} \sum_{n=0} g_n \xi^n~, \\
   H(\xi) &= \xi^{-2-\alpha} \sum_{n=0} h_n \xi^n~, \\
 I(\xi) &= \xi^{-2-\alpha} \sum_{n=0} i_n \xi^n~.
\end{align} 
\end{subequations}
where $\alpha$ is not integer. 
It is not hard to show, by brute force computation of the linear relations that follow from conservation, that this ansatz gives a vanishing contribution to all C-functions \eqref{CAdS2ofT}, at least if $\alpha<1$ as we expect from the discussion in section \ref{app:CPTflatspace} and the condition $\Delta_\mathcal{V}<1$.

Consider now an ansatz with integer powers and logarithms,
\begin{subequations}
\begin{align}
F(\xi) &= \xi^{-2} \sum_{n=0} (f_n + \tilde{f}_n \log\xi ) \xi^n~, \\
   G(\xi) &= \xi^{-2} \sum_{n=0} (g_n + \tilde{g}_n \log\xi ) \xi^n~, \\
   H(\xi) &= \xi^{-2} \sum_{n=0} (h_n + \tilde{h}_n \log\xi ) \xi^n~, \\
 I(\xi) &= \xi^{-2} \sum_{n=0} (i_n + \tilde{i}_n \log\xi ) \xi^n~.
\end{align} 
\end{subequations}
By imposing conservation one can find many relations between the undetermined coefficients.
In particular one finds that $h_0=h_1=\tilde{h}_0=\tilde{h}_1=0$. 
This leads to
\begin{subequations}
\begin{align}
F(\xi) &= \frac{C_T}{64\xi^{2}} + O(\xi^{-1}) ~, \\
   G(\xi) &= \frac{\Delta_\mathcal{V} \braket{\Theta}}{16\pi \xi^{2}} + O(\xi^{-1})~, \\
   H(\xi) &=  h_2+\tilde{h}_2 \log \xi+O(\xi)~, \\
 I(\xi) &= \frac{C_T-4C_3(0) }{64\xi^{2}}-\frac{\Delta_\mathcal{V} \braket{\Theta}  }{32\pi \xi^{2}} \log\xi +\frac{h_2+\tilde{h}_2 \log \xi}{16\xi^2}+ O(\xi^{-1})~,
\end{align} 
\label{eq:sdFGHI}
\end{subequations}
where we expressed the leading coefficients $f_0,g_0,i_0$ in terms of the 3 sum rules.\footnote{Curiously, if we set $H(\xi)=0$, then the unique solution of the conservation equations is given by
\begin{subequations}
\begin{align}
F(\xi) &= \frac{C_T}{64\xi^{2}} -\frac{\Delta_\mathcal{V} \braket{\Theta} }{32\pi \xi^{2}}\log(1+\xi)~, \\
   G(\xi) &= \frac{\Delta_\mathcal{V} \braket{\Theta}}{16\pi \xi^{2}} ~, \\
   H(\xi) &=  0~, \\
 I(\xi) &= \frac{C_T-4C_3(0)}{64\xi^{2}}-\frac{\Delta_\mathcal{V} \braket{\Theta} }{32\pi \xi^{2}}\log\xi ~,
\end{align} 
\end{subequations}
which leads to constant C-functions $C_1(\xi)= \frac{\Delta_\mathcal{V} \braket{\Theta} }{2\pi}$,
 $C_2(\xi)=C_T$ and  $C_3(\xi)=C_3(0)$.}
 




\section{The spectral representation of spinning operators in AdS$_{d+1}$}
\label{subsec:twopBlocks} 

In this appendix, we discuss the spectral blocks for two-point functions of symmetric traceless bulk operators in AdS$_{d+1}$. Our main observation is that the technology developed in \cite{Lauria:2018klo} to compute conformal blocks in defect CFT provides a general closed form solution to this problem. We first present the general recipe and then consider the special case of spin-2 external operators.

The strategy is a standard tool in the kinematics of CFT \cite{Costa:2011dw} and QFT in AdS \cite{Costa:2014kfa}.
Firstly, notice that bulk operators in symmetric traceless representations of $SO(d)$ only exchange symmetric traceless tensors in their boundary OPE. This is easily seen, for instance, by writing down the leading term in the OPE in Poincaré coordinates.

Then, one starts from a \emph{seed} block, which in this case can be defined as the exchange of a boundary primary of spin $\ell$ and arbitrary dimension $\D$ in the correlator of two bulk operators with the same spin. The intuitive reason for this definition is that the same block cannot be exchanged if we reduce the rank of either of the two bulk operators. For this reason, seed blocks satisfy conservation equations, and compute bulk-to-bulk propagators of local spinning fields in AdS \cite{Costa:2014kfa}.
To obtain the block of spin $\ell$ exchanged by a correlator of bulk operators with larger spin, one acts on the seed block via certain differential operators defined in \cite{Lauria:2018klo}.

In \cite{Lauria:2018klo}, a recipe was given to compute in closed form all the seed blocks for the two-point function of traceless symmetric bulk operators. Here, we provide a map of their boundary CFT results to AdS.

Let us consider traceless symmetric external operators $\mathcal{O}_1$ with spin $\ell_1$ and $\mathcal{O}_2$ with spin $\ell_2$ placed at positions $X_1$ and $X_2$ with polarization vectors $W_1$ and $W_2$ respectively. The seed block corresponding to the two point function $\langle \mathcal{O}_1 \mathcal{O}_2 \rangle$ with spin $\ell_1=\ell_2=\ell$ exchanging an operator of dimension $\Delta$ and spin-$\ell$ is given by \cite{Lauria:2018klo}
\begin{equation}\label{SeedBlock}
    G_{\Delta \ell}(X_1,X_2,W_1,W_2)=\frac{P^{d+2}_{-\Delta,\ell}(X_1,W_1,X_2,W_2)}{(X_1\cdot X_1)^{-\frac{\Delta}{2}}(X_2\cdot X_2)^{-\frac{\Delta}{2}}}~.
\end{equation}
The function $P^{d+2}_{-\Delta,\ell}$ is the analytic continuation for $m\rightarrow -\Delta$ of the polynomial $P^{d+2}_{m,\ell}(X_1,W_1,X_2,W_2)$, which is found by fully contracting with the $X_i$ and the $W_i$ a projector $\pi_{m,l}^{SO(d+2)}$. $\pi_{m,l}^{SO(d+2)}$ projects tensors with $m+\ell$ indices onto a representation of $SO(d+2)$ whose Young tableau has two rows of length $m$ and $\ell$ respectively. Such projectors are computed in \cite{Costa:2016xah}. Concretely, the polynomials $P^{d+2}_{m,\ell}(X_1,W_1,X_2,W_2)$ are given in terms of Gegenbauer polynomials $C_m^{\frac{d}{2}}(\cross)$, and their analytic continuation is obtained via the prescription: 
\begin{equation}\label{AnalyticContinuationGegenbauer}
    C_m^{\frac{d}{2}}(\cross)\rightarrow \frac{2^m(\frac{d}{2})_m}{\Gamma(m+1)} \cross^m\prescript{}{2}{}F_1\left(\frac{1-m}{2},-\frac{m}{2}; -m-\frac{d}{2}+1;\frac{1}{\cross^2}\right),
\end{equation}
where $\cross=X_1\cdot X_2$.

The last ingredient is given by the differential operators which, when applied to a block, increase by one unit the spin of one of the bulk operators. These are simply given by
\begin{equation}\label{SpinningOperators}
     D_1= (W_1\cdot \partial_{X_1})~, \quad
   	 D_2= (W_2\cdot \partial_{X_2})~.
\end{equation}

Let us illustrate the method via the case of spin-2 traceless symmetric external operators, which corresponds to the bulk two-point function of the traceless symmetric part of the stress tensor. Spin-2 external operators can exchange scalar, spin-1 and spin-2 operators. We thus need to consider the seed block for those three cases and use spinning operators on the scalar and spin-1 seed blocks. We first compute the spin-2 seed block. The polynomial $P^{d+2}_{m,2}(X_1,W_1,X_2,W_2)$ is given by \cite{Costa:2016xah}
\begin{equation}\label{PolynSeedSpin2}
\begin{split}
    P_{m,2}^{d+2}(X_1,W_1,X_2,W_2)&=\pi_{m,2}^{SO(d+2)}(X_1,W_1,X_2,W_2)\\&=- \frac{c_2}{d} (X_1 \cdot X_1)^{\frac{m}{2}} (X_2\cdot X_2)^{\frac{m}{2}}\Big( f_1(\cross) \mathcal{Q}_1+  f_2(\cross) \mathcal{Q}_2 + f_3(\cross) \mathcal{Q}_3\Big)~,
\end{split}
\end{equation}
where we have introduced the following structures: 
\begin{equation}
\begin{split}
    \mathcal{Q}_1&= (W_1\cdot W_2)^2 + \frac{1}{\cross^2}(W_2\cdot X_1)^2(W_1\cdot X_2)^2-\frac{2}{\cross}(W_1\cdot W_2)(W_2\cdot X_1)(W_1\cdot X_2)~,\\
    \mathcal{Q}_2&=-\frac{1}{\cross^2}(W_2\cdot X_1)^2(W_1\cdot X_2)^2+\frac{1}{\cross}(W_1\cdot W_2)(W_2\cdot X_1)(W_1\cdot X_2)~,\\
    \mathcal{Q}_3&=\frac{1}{x^2}(W_2\cdot X_1)^2(W_1\cdot X_2)^2~,
\end{split}
\end{equation}
and the functions
\begin{equation}
\begin{split}
    f_1(\cross)&=(d+1)(d+2)(1-d \cross^2)\mathcal{C}_m^{(2)}(\cross)-2d(d+2)\cross(\cross^2-1)\mathcal{C}_m^{(3)}(\cross)-d(\cross^2-1)^2\mathcal{C}_m^{(4)}(\cross)~,\\
    f_2(\cross)&=2(d+1)(d+2)\mathcal{C}_m^{(2)}(\cross)+ 2(d+1)(d+2)\cross\mathcal{C}_m^{(3)}(\cross)+2d(\cross^2-1)\mathcal{C}_m^{(4)}(\cross)~,\\
    f_3(\cross)&=(d+1)(d+2)\mathcal{C}_m^{(2)}(\cross)+ 2(d+2)\cross\mathcal{C}_m^{(3)}(\cross)+(\cross^2-d)\mathcal{C}_m^{(4)}(\cross)~.\\
\end{split}
\end{equation}
In the previous equation, we have denoted $\mathcal{C}_m^{(n)}(\cross)\equiv \partial_\cross^n C_m^{\frac{d}{2}}(\cross)$. After using the prescription \eqref{AnalyticContinuationGegenbauer}, we find the desired spin-2 seed block by replacing \eqref{PolynSeedSpin2} in \eqref{SeedBlock}.
Let us then compute the seed block for spin-1 external operators, which is found by analytically continuing the polynomial 
\begin{equation}
    P_{m,1}^{d+2}(X_1,W_1,X_2,W_2)=\pi_{m,1}^{SO(d+2)}(X_1,W_1,X_2,W_2)= c_1\Big( d\, \mathcal{C}_m^{(1)} \mathcal{Q}_4 +\mathcal{C}_m^{(2)} \mathcal{Q}_5  \Big)~.
\end{equation}
Here, the tensor structures are given by
\begin{equation}
    \begin{split}
        \mathcal{Q}_4&=(W_1\cdot X_2)(W_2 \cdot X_1)-\cross (W_1 \cdot W_2)~,\\
        \mathcal{Q}_5&=\cross(W_1\cdot X_2)(W_2 \cdot X_1) - (\cross^2-1) (W_1 \cdot W_2)~.
    \end{split}
\end{equation}
Replacing the analytical continuation of this polynomial in \eqref{SeedBlock}, we find the spin-1 seed block. Applying both spinning operators \eqref{SpinningOperators} on the spin-1 seed block yields the desired block corresponding to spin-2 external operators exchanging a spin-1 operator:
\begin{equation}
D_1 D_2\frac{P_{-\Delta,1}^{d+2}(X_1,W_1,X_2,W_2)}{(X_1\cdot X_1)^{-\frac{\Delta}{2}}(X_2 \cdot X_2)^{-\frac{\Delta}{2}}}~.
\end{equation} 
Finally, the scalar seed block is given by
\begin{equation}
    G_{\Delta,0}(X_1,W_1,X_2,W_2)= c_0 \left(-\cross+\sqrt{\cross^2-1}\right)^{\Delta} \prescript{}{2}{}F_1\left(\frac{d}{2},\Delta;\Delta-\frac{d}{2}+1;\left(-\cross+\sqrt{\cross^2-1}\right)^2\right)~.
\end{equation}
We can apply on it both the differential operators \eqref{SpinningOperators} twice to find the block corresponding to spin-2 external operators exchanging a scalar operator, i.e. $D_1^2 D_2^2 G_{\Delta,0}$.

One can check that the blocks computed in this way satisfy the Casimir equations for the two-point function of spin-2 external operators and have the correct OPE limit.

\section{Ward identities in the presence of boundary operators}
\label{app:ward}

This appendix is dedicated to proving (a generalization of) the Ward identity \eqref{WardIdAdS}. Specifically, we will explain how to derive Ward identities for the stress tensor which act non-trivially on boundary operators, by assuming (axiomatically) only the ordinary contact terms involving bulk operators. The central point is that a charge is defined by integrating along a closed surface in the bulk, so we must break the contour open at the boundary in order to obtain a surface which encloses a boundary operator.

The derivation works equally well for boundary CFTs (\emph{i.e.} conformal field theories in AdS), and, \emph{mutatis mutandis}, for general manifolds with a boundary. It has the merit of showing that no ad-hoc axioms are required besides the ones appropriate for infinite space.

In AdS, the divergence of the metric at the boundary entails an additional subtlety: the flux of the stress tensor might not vanish for a general isometry-preserving boundary condition. This issue and its resolution is well known in the case of a free scalar \cite{Breitenlohner:1982jf,Klebanov:1999tb}. Here, we show in general how to construct the appropriate stress tensor, defined so that its integral on any slice of AdS is finite and computes the conserved charges.

\subsection{Boundary Ward identities from bulk Ward identities}
\label{subsec:ward}

Consider the conserved vector operator associated to a Killing vector (or, for a conformal field theory, a conformal Killing vector) $\xi^\mu$:
\beq
u^\nu(\xi)=\xi_\mu T^{\mu\nu}~.
\label{uofxi}
\eeq
As usual, we use $u$ to define a topological operator associated to a closed surface $\mathcal{B}$:
\beq
Q_{\xi}\!\left[\mathcal{B}\right] = \int_\mathcal{B} \star u(\xi)~.
\label{QTopDef}
\eeq
The bulk Ward identities consist in the following equation:
\beq
Q_{\xi}\!\left[\mathcal{B}\right]\Phi(x)-\braket{Q_{\xi}\!\left[\mathcal{B}\right]}\Phi(x)=\delta_\xi \Phi(x)~,
\label{BulkWard}
\eeq
where $\Phi(x)$ is any bulk operator enclosed by $\mathcal{B}$, and $\delta_\xi$ denotes the action of the isometry on it. Let us first insert the topological operator  $Q_\xi$ in a correlation function involving bulk operators only:
\beq
\braket{Q_{\xi}\!\left[\mathcal{B}\right] X(x_1,\dots x_n)}-\braket{Q_{\xi}\!\left[\mathcal{B}\right]}\braket{ X(x_1,\dots x_n) }~,
\label{QX}
\eeq
where $X$ denotes a product of bulk operators. We choose $\mathcal{B}$ to enclose all the operators.  Using eq. \eqref{BulkWard} and invariance of the theory under the isometries, we conclude that \eqref{QX} vanishes:
\beq
\braket{Q_{\xi}\!\left[\mathcal{B}\right] X(x_1,\dots x_n)}-\braket{Q_{\xi}\!\left[\mathcal{B}\right]}\braket{ X(x_1,\dots x_n) } = \braket{\delta_\xi X(x_1,\dots x_n)} = 0~.
\label{variationX}
\eeq
If we deform $\mathcal{B}$ towards the boundary of AdS, we see that invariance of the theory under the isometries imposes constraints on the boundary OPE of the stress tensor. In Poincaré coordinates, a Killing vector which preserves the boundary satisfies
\beq
 n_\mu \xi^\mu \overset{z\to 0}{\longrightarrow}\ \textup{finite}~,
\label{KillingBC}
\eeq
where $n^\mu$ is a unit vector orthogonal to the boundary, \emph{i.e.} $n_\mu \sim 1/z$. Then, the flux of $u(\xi)$ vanishes when computed on the boundary of AdS, if
\beq
\sqrt{\hat{g}}\, n^\mu n^\nu \left(T_{\mu\nu}- \braket{T_{\mu\nu}}\right) \overset{z\to 0}{\longrightarrow} 0~, \qquad
\sqrt{\hat{g}}\, n^\nu \left( T_{a\nu} -\braket{T_{a\nu}}\right)\overset{z\to 0}{\longrightarrow} 0~,
\label{CardyLike}
\eeq
where $a$ denotes a direction parallel to the boundary, and $\sqrt{\hat{g}}$ is the volume element of a surface placed at fixed (small) $z$, \emph{i.e.} $\sqrt{\hat{g}} \sim z^{-d}$~.  Let us first assume that this is the case: we shall come back to these conditions in subsection \ref{subsec:improved}.

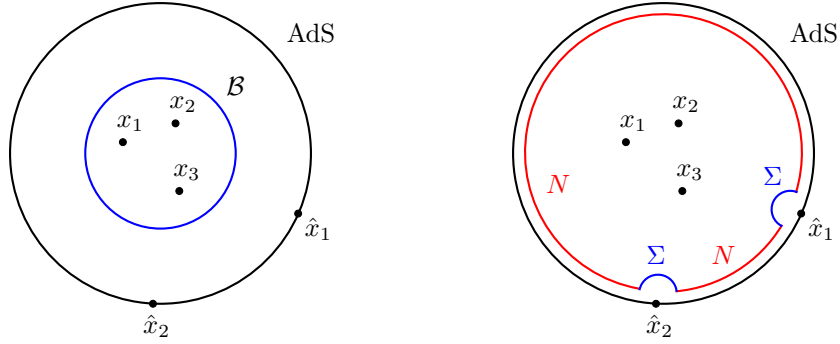
\begin{figure}[]
\centering
\begin{tikzpicture}[scale=1]
\draw[thick] (0,0) circle (2);
\draw[thick, blue] (0,0) circle (1);
\filldraw[black] (1.83,-0.8) circle (1.3pt);
\node[black] at (1,0.9) {$\mathcal{B}$};
\node [black] at (2.1,-1) {$\hat{x}_1$};
\filldraw[black] (-0.1,-2) circle (1.3pt) ;
\node [black] at (-0.05,-2.3) {$\hat{x}_2$};
\filldraw[black] (0.2,0.4) circle (1.3pt);
\node [black] at (0.3,0.65) {$x_2$};
\filldraw[black] (-0.5,0.15) circle (1.3pt);
\node [black] at (-0.4,0.4) {$x_1$};
\filldraw[black] (0.25,-0.5) circle (1.3pt);
\node [black] at (0.35,-0.25) {$x_3$};
\node[black] at (2,1.6) {AdS};
\end{tikzpicture}
\hspace{2cm}
\begin{tikzpicture}[scale=1]
\draw[thick] (0,0) circle (2);
\filldraw[black] (1.83,-0.8) circle (1.3pt);
\draw[thick,red] (-0.32,-1.8) arc (260:-17:1.84);
\draw[thick,red] (0.16,-1.835) arc (-85:-31:1.84);
\draw[thick,blue] (-0.32,-1.82) arc (170:5:0.25);
\draw[thick,blue] (1.58,-0.97) arc (245:67:0.248);
\node [black,blue] at (1.45,-0.3) {$\Sigma$};
\node [black,blue] at (-0.1,-1.35) {$\Sigma$};
\node [black,red] at (-1.4,-0.4) {$N$};
\node [black,red] at (0.8,-1.35) {$N$};
\node [black] at (2.1,-1) {$\hat{x}_1$};
\filldraw[black] (-0.1,-2) circle (1.3pt) ;
\node [black] at (-0.05,-2.3) {$\hat{x}_2$};
\filldraw[black] (0.2,0.4) circle (1.3pt);
\node [black] at (0.3,0.65) {$x_2$};
\filldraw[black] (-0.5,0.15) circle (1.3pt);
\node [black] at (-0.4,0.4) {$x_1$};
\filldraw[black] (0.25,-0.5) circle (1.3pt);
\node [black] at (0.35,-0.25) {$x_3$};
\node[black] at (2,1.6) {AdS};
\end{tikzpicture}
\caption{\emph{Left panel:} A choice of surface $\mathcal{B}$ such that the topological operator $Q_\xi[\mathcal{B}]$ generates the variation of the bulk operators via eq. \eqref{BulkWard}. \emph{Right panel:} Continuous deformation of $\mathcal{B}$ into the union of purely boundary contributions ($N$) and open surfaces ending on the boundary ($\Sigma$), such that $Q_\xi[\Sigma]$ generates the variation of the boundary operators via eq. \eqref{BoundaryWard}.}
\label{fig:charge}
\end{figure}

We now insert $Q_\xi$ in a correlation function involving both bulk and boundary operators, and again choose $\mathcal{B}$ to enclose the bulk operators: an example is given in the left panel of figure \ref{fig:charge}. This time, the isometries dictate
\begin{multline}
\braket{Q_{\xi}\!\left[\mathcal{B}\right] X(x_1,\dots x_n) \wh{X}(\hat{x}_1,\dots \hat{x}_n)}-\braket{Q_{\xi}\!\left[\mathcal{B}\right]}\braket{ X(x_1,\dots x_n) \wh{X}(\hat{x}_1,\dots \hat{x}_n)} \\
=\braket{\delta_\xi X(x_1,\dots x_n) \wh{X}(\hat{x}_1,\dots \hat{x}_n)}
=-\braket{ X(x_1,\dots x_n) \delta_\xi\wh{X}(\hat{x}_1,\dots \hat{x}_n)}~.
\label{variationXXhat}
\end{multline}
%
We deform $\mathcal{B}$ towards the boundary of AdS, as shown in the right panel of figure \ref{fig:charge}. Under the conditions \eqref{CardyLike}, the contribution from the surfaces marked by $N$ in the figure can be dropped. One is left with the fluxes across the surfaces marked by $\Sigma$, and comparing with eq. \eqref{variationXXhat} it is easy to derive the following Ward identity:
\beq
Q_{\xi}\!\left[\Sigma\right]O(\hat{x})-\braket{Q_{\xi}\!\left[\Sigma\right]}O(\hat{x})=
- \delta_\xi  O(\hat{x})~,
\label{BoundaryWard}
\eeq
for any boundary operator $O$ enclosed by the surface $\Sigma$. Notice that the minus sign on the r.h.s. is compensated by the orientation of the flux integral \eqref{QTopDef}, so that the Ward identities for bulk and boundary operators---eqs. \eqref{BulkWard} and \eqref{BoundaryWard}---look identical, once the `outward' flux is computed in both cases. Eq. \eqref{BoundaryWard} coincides with eq. \eqref{WardIdAdS}. 


\subsection{The improved stress tensor for fine-tuned boundary conditions}
\label{subsec:improved}

Let us now go back to the conditions \eqref{CardyLike}, see when they are satisfied and what happens if they are not. The boundary OPE of the stress tensor only contains scalars and traceless symmetric tensor primaries \cite{Lauria:2018klo}. Then, it is not hard to see that the asymptotics \eqref{CardyLike} are respected if the OPE satisfies
\begin{subequations}
\begin{align}
&\D>d~, \quad\textup{scalars,} \label{NoImprScalar}\\
&\D>d+1~,  \quad\textup{vectors,} \label{NoImprVec}
\end{align}
\label{spectrumNoImprovement}
\end{subequations}
for any local (not necessarily primary) operator besides the identity.\footnote{Notice that a spin two tensor of dimension $\D$ would contribute to the right equation in \eqref{CardyLike} as
\beq
T_{a n} \sim z^{\D-1}\partial^b t_{ab}~.
\eeq 
For $\D=d$, one would have a finite contribution, but unitarity implies that in this case $t_{ab}$ is a primary and it is conserved, so it actually does not contribute to this OPE. \label{foot:bt}} 

Let us first discuss boundary CFTs. There, the trace of the stress tensor vanishes, and conservation of the traceless part imposes that the only scalar and vector primaries appearing in the OPE have dimension $\D=d+1$ (the displacement and the energy flux operators respectively). Because of the unitarity bounds on higher spin operators, the boundary OPEs of $T_{zz}$ and $T_{az}$ are non singular in flat space.\footnote{The OPE in AdS is related to flat space by a factor $z^{d-1}$ coming from eq. \eqref{AdStoFlat}.} Hence, eq. \eqref{NoImprScalar} is always satisfied, while eq. \eqref{NoImprVec} is more subtle: a vector of dimension precisely equal to $d+1$ would not be invariant under AdS isometries, since eq. \eqref{variationX} would fail to be satisfied. Notice that instead the purported contribution from the descendant of a spin two tensor of dimension $d$ vanishes---see footnote \ref{foot:bt}. This leads to Cardy condition, which in flat space simply states that $T_{az}$ vanishes at the boundary.

For a quantum field theory in AdS, The discussion on eq. \eqref{NoImprVec} is identical. Indeed, vectors do not contribute to the trace of the stress tensor, and so are constrained by conservation in the same way as for a boundary CFT. Cardy condition, \emph{i.e.} the absence of energy flux across the boundary of AdS, equally follows from the request of invariance under isometries. On the other hand, an additional interesting complication arises when considering the scalar contribution to the OPE. Indeed, the dimension of scalar primary operators is no longer constrained, and conservation only leads to a relation between the OPE coefficients involving the trace and the traceless part of $T_{\mu\nu}$---eq. \eqref{bOPECons} for $d=1$. 

What happens if there is a relevant boundary operator, \emph{i.e.} $\D<d$? In this case, eq. \eqref{variationX} is still satisfied, thanks to a well-known remarkable kinematical property: the normal component $n^\mu u_\mu(\xi)$---eq. \eqref{uofxi}---is a total derivative on the boundary of AdS. The following is an instructive way of proving this fact. Consider the following tensor \cite{Breitenlohner:1982jf}:
\beq
\Delta T_{\mu\nu} = \left(g_{\mu\nu} \Box - \nabla_\mu \nabla_\nu+ R_{\mu\nu} \right) \Phi(x)~,
\label{DeltaT}
\eeq
where $\Phi$ is a scalar bulk operator and $R_{\mu\nu}$ is the Ricci tensor. In any curved manifold, $\Delta T_{\mu\nu}$ is covariantly conserved, like $T_{\mu\nu}$. Moreover, in AdS $T_{\mu\nu}$ and $\Delta T_{\mu\nu}$ transform in the same way under the isometries. Since these two requirements completely fix the scalar contribution to the boundary OPE, the scalar conformal blocks are identical for the two tensors, in any correlator.  For instance, one can easily check this statement for the  $d=1$ form factor conformal blocks \eqref{block3p0} and \eqref{blocks3p2}.

Now, the improvement tensor $\Delta T$ is \emph{identically} conserved, which means that the associated vector operator,
\beq
v_\nu(\xi) = \xi^\mu \Delta T_{\mu\nu}~,
\label{vImprovement}
\eeq
is not only conserved, but it's flux is a boundary term. In the language of differential forms, the Hodge dual of $v$ is exact:
\beq
\star v(\xi)  = d\alpha~, \qquad 
\alpha_{\mu_1,\dots \mu_{d-1}} = 
\frac{\sqrt{g}}{(d-1)!}\ep_{\rho \sigma \mu_1 \dots \mu_{d-1}} 
\left(\xi^\rho \nabla^\si \Phi+\frac{1}{2}\big(\nabla^\rho \xi^\si\big)\Phi\right)~.
\eeq
When pulled back on a $d$-dimensional surface $\mathcal{B}$, parametrized by coordinates $y^a$, with $e^\rho{}_a = \partial x^\rho / \partial y^a$, the previous equation reads
\beq
n_\mu v^\mu(\xi) = \nabla^a \left[ n_\rho e_{\sigma a} 
\Bigr( \xi^\rho \nabla^\si \Phi - \xi^\si \nabla^\rho \Phi +
\big(\nabla^\rho \xi^\si  \Phi \big)  \Bigr)
\right]~.
\eeq
Adapted to AdS, this proves the statement that the flux of $u(\xi)$ vanishes when integrated on the entire boundary, even in the presence of relevant scalar primaries.

What is the consequence on the Ward identity for boundary operators, eq. \eqref{BoundaryWard}? This time, the $N$ surfaces in figure \ref{fig:charge} leave behind a boundary contribution for each scalar primary of dimension $\D<d$ in the OPE of the stress tensor. We do not need to compute these terms explicitly. Firstly, they are purely divergent terms, because of the structure of the OPE. Secondly, as discussed, the full integral over the closed surface $\mathcal{B}$ is finite: therefore, these boundary terms precisely cancel the near-boundary divergences of the flux computed along $\Sigma$. One is left again with eq. \eqref{BoundaryWard}, where the operator $Q_\xi[\Sigma]$ is replaced by the finite part of the corresponding integral.

As a final comment, notice that $Q_\xi[\Sigma]$, when $\xi$ is a translation in global coordinates, is but the global time Hamiltonian. This makes it clear that the extra boundary contribution to it, coming from the relevant boundary operators, corresponds to fine tuning their coupling in order to ensure scale invariance of the boundary theory, in the regularization scheme where integrals are cut off at coordinate distance $z=\epsilon$. In the theory of a free scalar, in the range $-d^2/4<m^2<1-d^2/4$ and with the choice of singular boundary conditions, fine tuning the operator $\Phi^2$ is as simple as integrating  the action by parts and dropping the boundary term \cite{Klebanov:1999tb}. In this special case, the discussion in this appendix explicitly relates the procedure of \cite{Klebanov:1999tb} with the one adopted by Breitenlohner and Freedman \cite{Breitenlohner:1982jf}. The latter authors precisely used the improvement \eqref{DeltaT}, with the replacement $\Phi \to \Phi^2$, to define a stress tensor which would be regular at the boundary of AdS. Here we showed that this procedure is equivalent to fine tuning the relevant operator, and that it works in any interacting QFT in AdS, at least when there is a single relevant boundary operator: in particular, one can always choose the trace of the stress tensor $\Theta$ as the operator $\Phi$ in eq. \eqref{DeltaT}, and tune the coefficient $\gamma$ in $T_{\mu\nu}+\gamma\Delta T_{\mu\nu}$ to obtain a non-singular stress tensor.\footnote{The relation between the scalar conformal block of $T_{\mu\nu}$ and of $\Delta T_{\mu\nu}$ has a $\D$ dependent factor, as it can be checked for instance when $d=1$, therefore using the trace of the stress tensor to define the improvement does not allow to fine tune multiple relevant operators.}


  

\section{A bound on the growth of $c_{\phi\phi \D} b_{\Theta \D}$}
\label{app:OPEasymptotics}

This appendix is dedicated to deriving the bound \eqref{CauchySchwarzSection4} on the form factor of $\Theta$ and the integral bound \eqref{bcBound} on the growth of the absolute value of the product $|c_{\phi\phi\D}b_{T \D}|$, as $\D \to \infty$. The two main ingredients for this derivation are the Cauchy-Schwartz inequality applied to the form factor of the stress tensor, and the Tauberian theorems which are by now standard items in the conformal field theory toolbox \cite{Pappadopulo:2012jk,Qiao:2017xif}.

The Cauchy-Schwartz inequality can be applied to the form factor of either $\Theta$ or the traceless part of the stress tensor. Here, we discuss the form factor of the trace, while the spin two case is discussed in subsection \ref{subsec:spin2LBConv}. We divide the proof in two steps: in the next subsection, we use the Cauchy-Schwartz inequality to bound the growth of a certain modified form factor as $z \to 1$. In subsection \ref{app:tauberianFF}, we use a Tauberian theorem to obtain the desired estimate on the growth of the OPE coefficients.

\subsection{A bound from the Cauchy-Schwarz inequality}
\label{app:Cauchy-Schwarz}

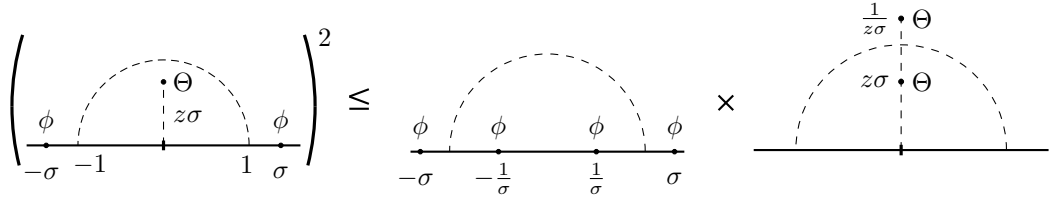
\begin{figure}
    \centering
\begin{tikzpicture}[scale=0.65]
\begin{scope}
    \clip (-2.5,0) rectangle (2.5,2.5);
    \draw[densely dashed] (0,0) circle (1.75);
\end{scope}
    \draw[very thick,black] (0,-0.1) -- (0,0.1);
    \draw[thick] (-2.8,0) -- (2.8,0);
    \node[black] at (-1.5,-0.4) {$-1$};
    \node[black] at (1.7,-0.4) {$1$};
    \filldraw[black] (-2.4,0) circle (1.3pt);
    \node[black] at (-2.4,0.5) {$\phi$};
    \node[black] at (-2.5,-0.55) {$-\sigma$};
    \filldraw[black] (2.4,0) circle (1.3pt);
    \node[black] at (2.4,0.5) {$\phi$};
    \node[black] at (2.4,-0.55) {$\sigma$};
    \draw[dashed] (0,0) -- (0,1.3);
    \node[black] at (0.4,1.3) {$\Theta$};
    \filldraw[black] (0,1.3) circle (1.3pt);
    \node[black] at (0.5,0.6) {$z\sigma$};
    \draw[black,very thick] (3.1,0.8) arc (0:-14:5);
    \draw[black,very thick] (3.1,0.8) arc (0:17:5);
    \draw[black,very thick] (-3.1,0.8) arc (180:194:5);
    \draw[black,very thick] (-3.1,0.8) arc (180:163:5);
    \node at (3.3,2.2) {$2$};
    \draw[black, thick] (3.8,1) -- (4.2,1.15);
    \draw[black, thick] (3.8,1) -- (4.2,0.85);
    \draw[black, thick] (3.8,0.75) -- (4.2,0.75);
    \node at (0,-1) {};
\end{tikzpicture}
\hspace{0.01cm}
\begin{tikzpicture}[scale=0.65]
\begin{scope}
    \clip (-2.5,0) rectangle (2.5,2.5);
    \draw[dashed] (0,0) circle (2);
\end{scope}
    \draw[thick] (-2.8,0) -- (2.8,0);
    \filldraw[black] (-2.6,0) circle (1.3pt);
    \node[black] at (-2.6,0.5) {$\phi$};
    \node[black] at (-2.7,-0.55) {$-\sigma$};
    \filldraw[black] (2.6,0) circle (1.3pt);
    \node[black] at (2.6,0.5) {$\phi$};
    \node[black] at (2.6,-0.55) {$\sigma$};
    \filldraw[black] (-1,0) circle (1.3pt);
    \node[black] at (-1,0.5) {$\phi$};
    \node[black] at (-1.1,-0.55) {$-\frac{1}{\sigma}$};
    \filldraw[black] (1,0) circle (1.3pt);
    \node[black] at (1,0.5) {$\phi$};
    \node[black] at (1,-0.55) {$\frac{1}{\sigma}$};

   \draw[black, thick] (3.5,0.85) -- (3.8,1.15);
   \draw[black, thick] (3.8,0.85) -- (3.5,1.15);
\end{tikzpicture}
\hspace{0.01cm}
\begin{tikzpicture}[scale=0.7]
\begin{scope}
    \clip (-2.5,0) rectangle (2.5,2.5);
    \draw[dashed] (0,0) circle (2);
\end{scope}
    \draw[very thick,black] (0,-0.1) -- (0,0.1);
    \draw[thick] (-2.8,0) -- (2.8,0);
    \draw[dashed] (0,0) -- (0,2.5);
    \node[black] at (0.4,1.3) {$\Theta$};
    \filldraw[black] (0,1.3) circle (1.3pt);
    \node[black] at (-0.5,1.3) {$z\sigma$};
    \node[black] at (0.4,2.5) {$\Theta$};
    \filldraw[black] (0,2.5) circle (1.3pt);
    \node[black] at (-0.5,2.5) {$\frac{1}{z\sigma}$};
    \node at (0,-0.84) {};
\end{tikzpicture}
\caption{A depiction of the Cauchy-Schwarz inequality of eq. \eqref{CSfirst}, where the correlators are evaluated in AdS in Poincaré coordinates. The boundary operators are inserted at $x=\pm\sigma$ and $x=\pm1/\sigma$, where $\sigma>1$ and otherwise arbitrary. The dotted hemi-circle is the quantization surface.}
\label{fig:CauchySchwarzFF}
\end{figure}

Consider first the correlator depicted in the left panel of figure \ref{fig:CauchySchwarzFF}. By quantizing the theory on the unit hemi-circle, we can write it as an overlap, and apply the Cauchy-Schwarz inequality to it:
\begin{multline}
\big|\braket{\Theta(\sigma z)\phi(\sigma)\phi(-\sigma)} \big|=
\sigma^{-4\Delta_\phi}\big| \braket{\phi(1/\sigma)\phi(-1/\sigma)|\Theta(\sigma z)} \big| \\
\leq \sigma^{-4\Delta_\phi}\sqrt{\braket{\phi(1/\sigma)\phi(-1/\sigma)|\phi(1/\sigma)\phi(-1/\sigma)}\braket{\Theta(\sigma z)|\Theta(\sigma z)}} \\
= \sigma^{-2\Delta_\phi} \sqrt{\braket{\phi(\sigma)\phi(-\sigma)\phi(1/\sigma)\phi(-1/\sigma)}\braket{\Theta(1/(\sigma z))\Theta(\sigma z)}}~.
\label{CSfirst}
\end{multline}
Notice that we only denoted the value of the coordinate orthogonal to the boundary in the argument of the bulk operator. When translating between correlation functions and state overlaps, we used that Hermitian conjugation is implemented by an inversion \cite{Pappadopulo:2012jk}, and we assumed for ease of notation that the operators are real:
\begin{align}
\bra{\phi(x)} &=(x^2)^{-\Delta_\phi}\bra{0} \phi(1/x)~, \\
\bra{\Theta(z)} &=\bra{0} \Theta(1/z)~.
\end{align}
The parametrization of the position of the operators in term of $\sigma$ has been chosen so that, up to an isometry, the configuration on the left in figure \ref{fig:CauchySchwarzFF} coincides with the one in figure \ref{fig:interface}. Eq. \eqref{CSfirst} requires
\beq
\sigma z < 1 < \sigma~,
\label{lambdaCond}
\eeq
As it is clear from the figure. 

The bound \eqref{CSfirst} is weak: as $\sigma \to 1$, the right hand side diverges, while the left hand side is regular. As we will see, this is an effect of the oscillations in the OPE data appearing in the form factor, which generate cancellations and tame the singularity. What we are after is a bound on their absolute value, and it turns out that the Cauchy-Schwarz inequality can be saturated in that case. In fact, we can construct an overlap of states which is fully determined by the absolute values $|c_{\phi\phi\D}b_{\Theta \D}|$, and it is still bounded by the r.h.s. of eq. \eqref{CSfirst}. 

Let us begin by comparing the conformal block decomposition of the l.h.s. with the decomposition in a complete set of states on the hemi-circle. On the one hand, using eq. \eqref{FFblockZ} for the conformal block in $z$ coordinates, we get \eqref{3ptFctAdS2Trace}
\beq
\braket{\Theta(\sigma z)\phi(\sigma)\phi(-\sigma)} =
(2\sigma)^{-2\D_\phi} \sum_{\D} c_{\p\p\D}b_{\Theta \D} 
(2z)^\D {}_2F_1 \left(\frac{1}{2},\Delta; \Delta+\frac{1}{2};-z^2 \right)
=\sigma^{-2\D_\phi} \sum_{\D_\al} g_\al z^{\D_\al}~, 
\label{OPEzFF}
\eeq
where the label $\D_\al$ runs over all the powers of $z$, without distinction between primaries and descendants. 
On the other hand, denoting the states with scaling dimension $\Delta_\al$ as
\beq
\ket{\al,k_\al}~, 
\label{alphaStates}
\eeq
$k_\al$ denoting a finite multiplicity,\footnote{In AdS$_2$, this degeneracy is only present if there primary operators whose dimensions differ by an integer.} one also has
\beq
\braket{\Theta(\sigma z)\phi(\sigma)\phi(-\sigma)} = 
\sigma^{-2\D_\phi} \sum_{\alpha} z^{\Delta_\al}
\braket{0| \p(1) \p(-1) | \al,k_\al}\braket{\al, k_\al |\Theta(1) |0}~,
\label{RadQuantFF}
\eeq
where we used
\begin{align}
\braket{0| \p(\sigma) \p(-\sigma) | \al,k_\al} &= 
\sigma^{-2\D_\phi-\D_\al}\braket{0| \p(1) \p(-1) | \al,k_\al}~, \\
\braket{\al, k_\al |\Theta(\sigma z) |0} &=
(\sigma z)^{\D_\al} \braket{\al, k_\al |\Theta(1) |0}~.
\end{align}
Comparing eqs. \eqref{OPEzFF} and \eqref{RadQuantFF}, we conclude that
\beq
g_\al = \sum_{k_\al} \braket{0| \p(1) \p(-1) | \al,k_\al}\braket{\al, k_\al |\Theta(1) |0}~.
\eeq	
Let us first consider the non-degenerate case, where the label $k_\al$ takes only one value. Then, we build the correct overlap by replacing one of the two states, say, the one created by the boundary operators, as follows:
\beq
\ket{\phi(1/\sigma)\phi(-1/\sigma)} \quad \longrightarrow \quad
\ket{\widetilde{\phi(1/\sigma)\phi(-1/\sigma)}}=\sum_\al \textup{sign} g_\al\, \sigma^{2\D_\p-\D_\al}\braket{\al| \p(1) \p(-1) |0 } \ket{\al}~.
\label{tildeState}
\eeq
The overlap with the stress tensor gives the desired sum over absolute values:
\beq
\braket{\widetilde{\phi(1/\sigma)\phi(-1/\sigma)}|\Theta(\sigma z)}
= \sigma^{2\D_\p} \sum_\al |g_\al| z^{\D_\al}~.
\label{absoluteSum}
\eeq
Crucially, the norm of the original and the modified state in eq. \eqref{tildeState} coincide, so the sum \eqref{absoluteSum} is still bounded by the product of correlators on the r.h.s. of eq. \eqref{CSfirst}. Comparing with eq. \eqref{OPEzFF}, we obtain the following bound involving the OPE coefficients:
\beq
\sum_{\D} |c_{\p\p\D}b_{\Theta \D} |
(2z)^\D {}_2F_1 \left(\frac{1}{2},\Delta; \Delta+\frac{1}{2};z^2 \right)
\leq 
4^{\D_\phi} \sqrt{\braket{\phi(\sigma)\phi(-\sigma)\phi(1/\sigma)\phi(-1/\sigma)}\braket{\Theta(1/(\sigma z))\Theta(\sigma z)}}~.
\label{CSfinal}
\eeq
Notice the change of sign in the last argument of the hypergeometric function, with respect to eq. \eqref{OPEzFF}. These are the positive blocks dubbed $g^+_\D(z)$ in eq. \eqref{posBlock0}. In fact, eq. \eqref{CSfinal} is valid also in the presence of degeneracies. Indeed, the only danger comes from a primary being degenerate with a descendant. But these are orthogonal states in the list \eqref{alphaStates}, therefore the definition \eqref{tildeState} can be modified to separately render positive the contribution of each of the degenerate states, thus leading back to eq. \eqref{CSfinal}.

When $z \to 1$, eq. \eqref{lambdaCond} also implies $\sigma \to 1$. Then, the r.h.s. in eq. \eqref{CSfinal} has a power law singularity. On the l.h.s., the blocks only have a logarithmic singularity, hence a singularity which competes with the bound can only be generated by the infinite sum. Intuitively, this means that the growth of the OPE coefficients at large $\D$ is bounded by eq. \eqref{CSfinal}. In the next subsection we make this fact precise. For the moment, let us specify the bound  \eqref{CSfinal} in this limit.

Since the l.h.s of eq.\eqref{CSfinal}  is $\sigma$ independent, we get the strictest bound by minimizing the r.h.s. over $\sigma$, compatibly with the conditions \eqref{lambdaCond}. As both $\sigma$ and $z$ approach 1, the correlators on the r.h.s. behave as follows: 
\begin{align}
\braket{\phi(\sigma)\phi(-\sigma)\phi(1/\sigma)\phi(-1/\sigma)} 
&\overset{\sigma\to 1^+}{\sim} \frac{1}{[2(\sigma-1)]^{4\Delta_\phi}}~, \\
\braket{\Theta(1/(\sigma z))\Theta(\sigma z)} &\underset{z \to 1^-}{\overset{\sigma\to 1^+}{\sim}} \frac{(2-\Delta_{\mc{V}})^2\la^2}{[2(1-\sigma z)]^{2\Delta_{\mc{V}}}}~, 
\end{align}
where we used eq. \eqref{Ashort}. To find the strictest bound, we should maximize the function
\beq
(\sigma-1)^{4\D_\p}(1-\sigma z)^{2\D_\mc{V}}~,
\eeq
over $\sigma$ obeying \eqref{lambdaCond}, when $z\to 1$. 
Changing variables to $x=\sigma-1$ and $y=1-z$,
\beq
\frac{(\sigma-1)^{4\D_\p}(1-\sigma z)^{2\D_\mc{V}}}{(1-z)^{4\D_\p+2\D_\mc{V}}} = 
\left(\frac{x}{y}\right)^{4\D_\p} \left(1-\frac{x}{y}+x\right)^{2\D_\mc{V}}~.
\eeq
In view of eq. \eqref{lambdaCond}, $0\leq x/y\leq1$ as $y\to 0$. Defining $\nu=\lim_{y\to0} x/y$, we get
\beq
(\sigma-1)^{4\D_\p}(1-\sigma z)^{2\D_\mc{V}} \approx  \nu^{4\D_\p}(1-\nu)^{2\D_\mc{V}} (1-z)^{4\D_\p+2\D_\mc{V}}~.
\eeq 
Maximizing over $0\leq\nu\leq1$ we get
\beq
(\sigma-1)^{4\D_\p}(1-\sigma z)^{2\D_\mc{V}} \approx
\frac{(4\D_\p)^{4\D_\p}(2\D_\mc{V})^{2\D_\mc{V}}}{\left(4\D_\p+2\D_\mc{V}\right)^{4\D_\p+2\D_\mc{V}}} (1-z)^{4\D_\p+2\D_\mc{V}}~.
\eeq
Putting it all together, we get the following asymptotic bound:
\beq
\sum_{\D} |c_{\p\p\D}b_{\Theta \D} |
g^+_\D(z)
\overset{z\to 1}{\lessapprox}
\frac{\left(4\D_\p+2\D_\mc{V}\right)^{2\D_\p+\D_\mc{V}}}{(4\D_\p)^{2\D_\p}(2\D_\mc{V})^{\D_\mc{V}}} \frac{(2-\Delta_{\mc{V}})\la}{2^{\Delta_\mc{V}}}
\frac{1}{(1-z)^{2\D_\p+\D_\mc{V}}}~,
\label{asymptBoundCauchy}
\eeq
where recall that
\beq
g^+_\D(z)=(2z)^\D {}_2F_1 \left(\frac{1}{2},\Delta; \Delta+\frac{1}{2};z^2 \right)~.
\eeq
In the next subsection, we turn this inequality into an inequality for the integrated density of OPE coefficients.


\subsection{A bound on the OPE coefficients from a Tauberian theorem}
\label{app:tauberianFF}

We will now use an idea of \cite{Qiao:2017xif}---see also \cite{Fitzpatrick:2012yx,Komargodski:2012ek,Fitzpatrick:2014vua}. Firstly, notice that the bound \eqref{asymptBoundCauchy} still holds if the sum on the l.h.s. is restricted to operators of dimension greater than any finite threshold $\D_0$. Indeed, the $z\to 1$ asymptotics of a single block is logarithmic. Then, we can rewrite eq. \eqref{asymptBoundCauchy} using the uniform large $\D$ asymptotics \eqref{largeDuniformFF0}, which, translated to the function $g_\D^+(z)$, is the one derived in \cite{Qiao:2017xif} and is valid for $0<z<1$:
\beq
\sum_{\D} |c_{\p\p\D}b_{\Theta \D} | (2z)^\D \sqrt{\D}\, e^{\frac{1-z^2}{2}\D} K_0\!\left(\frac{1-z^2}{2}\D\right) \overset{z\to 1}{\lesssim}
\frac{1}{(1-z)^{2\D_\p+\D_\mc{V}}}~.
\eeq
Then, \cite{Qiao:2017xif} proved that the factor $z^\D \exp \D(1-z^2)/2$ can be dropped without altering the bound.\footnote{In their case, the bound is replaced by an asymptotic equality, but the argument works in the same way.} We obtain
\beq
\sum_{\D} |c_{\p\p\D}b_{\Theta \D} | 2^\D \sqrt{\D}\, K_0\left(\frac{\D}{Y}\right) \overset{Y\to \infty}{\lesssim}
Y^{2\D_\p+\D_\mc{V}}~, \qquad Y=\frac{2}{1-z^2}~.
\label{CauchySchwarzY}
\eeq
We can then use the trick explained in appendix D of \cite{Qiao:2017xif} to obtain a Tauberian theorem without exact prefactors, or in our case, the weaker one-sided bound. One simply takes advantage of the fact that $K_0(x)$ is monotonically decreasing and positive to write
\beq
\sum_{\D} |c_{\p\p\D}b_{\Theta \D} | 2^\D \sqrt{\D}\, K_0\left(\frac{\D}{Y}\right) \geq
K_0(1)\sum_{\D}^Y 2^\D \sqrt{\D} |c_{\p\p\D}b_{\Theta \D} | ~.
\label{MonotoTrick}
\eeq
Combined with eq. \eqref{CauchySchwarzY}, eq. \eqref{MonotoTrick} yields
\beq
\sum_{\D}^Y 2^\D \sqrt{\D} |c_{\p\p\D}b_{\Theta \D} | \overset{Y\to \infty}{\lesssim} Y^{2\D_\p+\D_\mc{V}}~,
\eeq
which is eq. \eqref{bcBound}.

\section{Closed forms of the local blocks}
\label{app:local_blocks}

In this appendix, we derive the closed form of the local blocks defined in eqs.  \eqref{localBlock0} and \eqref{localBlock2}.

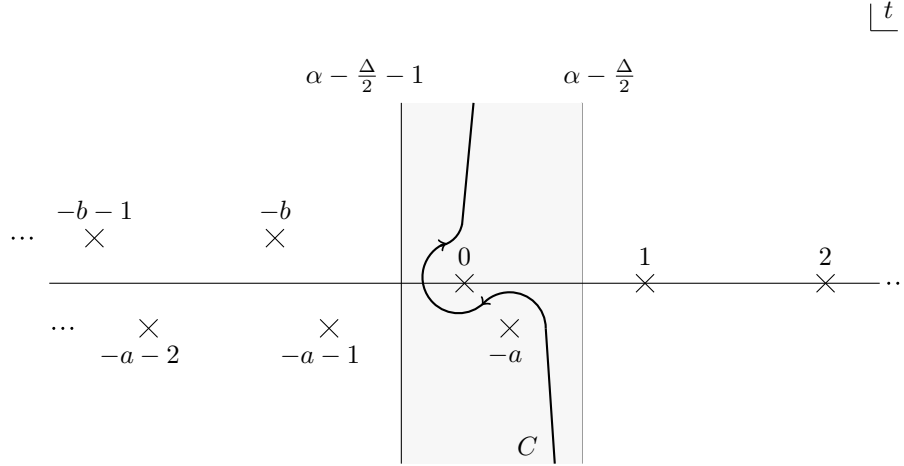
\begin{figure}[ht]
\centering
\begin{tikzpicture}[scale=1.2]
     \draw[thin] (-0.7,-2) -- (-0.7,2);
     \draw[thin] (1.3,-2) -- (1.3,2);
     \fill[black!3!white] (-0.695,-2) rectangle (1.295,2);
     \node at (1.5,2.3) {$\alpha-\frac{\Delta}{2}$};
     \node at (-1.1,2.3) {$\alpha-\frac{\Delta}{2}-1$};

     \draw (4.5,3.1) -- (4.5,2.8);
     \draw (4.5,2.8) -- (4.8,2.8);
     \node at (4.7,3.03) {$t$};

     \draw[thin] (-4.6,0) -- (4.6,0);

     \node at (0,0.3) {0};
     \draw[thin] (-0.1,-0.1) -- (0.1,0.1);
     \draw[thin] (-0.1,0.1) -- (0.1,-0.1);

     \node at (2,0.3) {1};
     \draw[thin] (1.9,-0.1) -- (2.1,0.1);
     \draw[thin] (1.9,0.1) -- (2.1,-0.1);

     \node at (4,0.3) {2};
     \draw[thin] (3.9,-0.1) -- (4.1,0.1);
     \draw[thin] (3.9,0.1) -- (4.1,-0.1);

     \filldraw[black] (4.7,0) circle (0.3pt) ;
     \filldraw[black] (4.8,0) circle (0.3pt) ;
     \filldraw[black] (4.9,0) circle (0.3pt) ;
     
     \node at (0.45,-0.8) {$-a$};
     \draw[thin] (0.4,-0.6) -- (0.6,-0.4);
     \draw[thin] (0.4,-0.4) -- (0.6,-0.6);

     \node at (-1.6,-0.8) {$-a-1$};
     \draw[thin] (-1.6,-0.6) -- (-1.4,-0.4);
     \draw[thin] (-1.6,-0.4) -- (-1.4,-0.6);

     \node at (-3.6,-0.8) {$-a-2$};
     \draw[thin] (-3.6,-0.6) -- (-3.4,-0.4);
     \draw[thin] (-3.6,-0.4) -- (-3.4,-0.6);

     \filldraw[black] (-4.35,-0.5) circle (0.3pt) ;
     \filldraw[black] (-4.45,-0.5) circle (0.3pt) ;
     \filldraw[black] (-4.55,-0.5) circle (0.3pt) ;

     \node at (-2.1,0.8) {$-b$};
     \draw[thin] (-2.2,0.4) -- (-2,0.6);
     \draw[thin] (-2.2,0.6) -- (-2,0.4);

     \node at (-4.1,0.8) {$-b-1$};
     \draw[thin] (-4.2,0.4) -- (-4,0.6);
     \draw[thin] (-4.2,0.6) -- (-4,0.4);

     \filldraw[black] (-4.8,0.5) circle (0.3pt) ;
     \filldraw[black] (-4.9,0.5) circle (0.3pt) ;
     \filldraw[black] (-5,0.5) circle (0.3pt) ;

     \draw[thick] (1,-2) -- (0.9,-0.5); 
     \draw[thick,->] (0.9,-0.5) arc (0:140:0.4);
     \draw[thick,->] (0.217,-0.215) arc (-45:-250:0.4);
     \draw[thick] (-0.27,0.415) arc (-80:-10:0.3);
     \draw[thick] (-0.027,0.65) -- (0.1,2); 
     \node at (0.7,-1.8) {$C$};
\end{tikzpicture}
\caption{Contour of integration in eq. \eqref{HyperMellin}. The poles in the integrand are marked by crosses. They are all on the real axis, but some of them are displaced for artistic reasons. The width of the grey band, $\al-\D/2-1<\Re t<\al-\D/2$ is explained in the main text.}
\label{fig:2F1MellinContour}
\end{figure}

It is convenient to perform the integrals in eqs. \eqref{localBlock0} and \eqref{localBlock2} after a Mellin transform. The Mellin space representation of the hypergeometric function is 
\beq
	 {}_2 F_1 (a,b;c;z)=\frac{\Gamma(c)}{2\pi i \Gamma(a)\Gamma(b)} \int_{-i\infty, C}^{i\infty} \frac{\Gamma(a+t)\Gamma(b+t)\Gamma(-t)}{\Gamma(c+t)}(-z)^t dt~,
	 \label{HyperMellin}
\eeq
where $a,\,b,\,c$ denote either of the arguments of the two ${}_2F_1's$ appearing in the definitions of $G_\D(\chi)$ and $H_\D(\chi)$. We pick the contour $C$ as depicted in figure \ref{fig:2F1MellinContour}, for reasons to become clear shortly. A generic local block can then be written as follows: 
\beq\label{LocalBlockGeneric}
	L(\chi)= \frac{1}{2\pi i} \int_{-i\infty,C}^{i \infty} \Tilde{L}(t) \chi^{\frac{\D}{2}+t} dt = \chi^\alpha \sin\left[\frac{\pi}{2}(\Delta-2\alpha)\right]\int_{-\infty}^0 \frac{d\chi'}{\pi} \frac{1}{\chi'-\chi} \lvert \chi' \rvert^{\frac{\Delta}{2}-\alpha} {}_2 F_1 \left(a,b;c;\chi'\right)~.
\eeq
Replacing eq. \eqref{HyperMellin} and performing the $\chi'$ integration gives
\beq
	-\int_0^\infty \frac{d\lvert\chi' \rvert}{\pi} \frac{1}{\chi+\lvert \chi' \rvert} \lvert \chi' \rvert^{\frac{\Delta}{2}-\alpha+t} = \frac{\chi^{\frac{\Delta}{2}-\alpha+t}}{\sin\left[\pi \left( \frac{\Delta}{2}-\alpha +t\right)\right]}~.
	\label{tIntForL}
\eeq
Convergence of this integral at infinity requires 
\beq
\Re t<\al-\frac{\D}{2}~.
\eeq
On the other hand, eq. \eqref{HyperMellin} requires the contour to be on the right of the poles at $t=-a,\,-a-1 \dots$ and $t=-b,\,-b-1 \dots$, which means that some part of $C$ satisfies $\Re t>-\frac{\D}{2},$ for spin 0, and $\Re t>-\frac{\D}{2}+1,$ for spin 2.
We see that convergence at infinity requires $\al>0$ for $G_\D(\chi)$ and $\al>1$ for  $H_\D(\chi)$, a condition that can easily be derived from the original expressions \eqref{localBlock0} and \eqref{localBlock2}. 

Convergence of the integral \eqref{tIntForL} at $\chi'=0$ requires 
\beq
\Re t>\al-\frac{\D}{2}-1~,
\eeq
which must be compatible with the requirement from eq. \eqref{HyperMellin} that the pole at 0 is on the right of the contour. One gets $\al<\D/2+1$, which was also pointed out below eq. \eqref{localBlock2}. These considerations explain the choice of contour in figure \ref{fig:2F1MellinContour}.

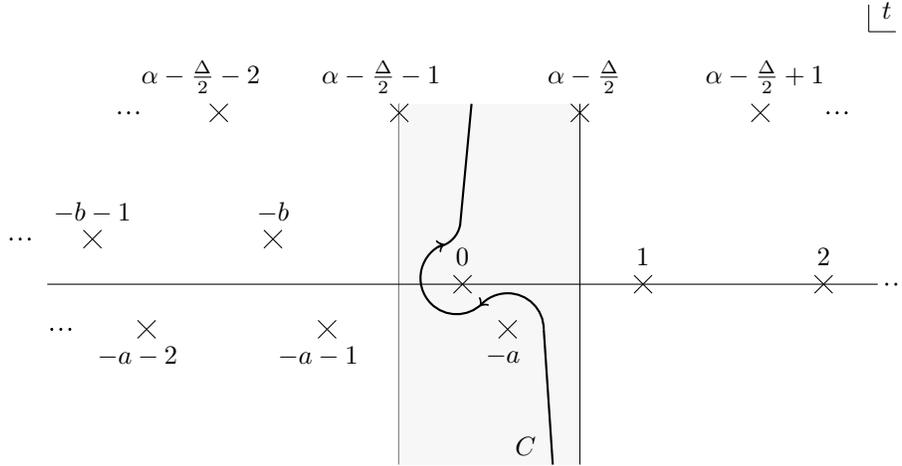
\begin{figure}[ht]
\centering
\begin{tikzpicture}[scale=1.2]
     \draw[thin] (-0.7,-2) -- (-0.7,2);
     \draw[thin] (1.3,-2) -- (1.3,2);
     \fill[black!3!white] (-0.695,-2) rectangle (1.295,2);

     \draw (4.5,3.1) -- (4.5,2.8);
     \draw (4.5,2.8) -- (4.8,2.8);
     \node at (4.7,3.03) {$t$};

     \draw[thin] (-4.6,0) -- (4.6,0);

     \node at (0,0.3) {0};
     \draw[thin] (-0.1,-0.1) -- (0.1,0.1);
     \draw[thin] (-0.1,0.1) -- (0.1,-0.1);

     \node at (2,0.3) {1};
     \draw[thin] (1.9,-0.1) -- (2.1,0.1);
     \draw[thin] (1.9,0.1) -- (2.1,-0.1);

     \node at (4,0.3) {2};
     \draw[thin] (3.9,-0.1) -- (4.1,0.1);
     \draw[thin] (3.9,0.1) -- (4.1,-0.1);

     \filldraw[black] (4.7,0) circle (0.3pt) ;
     \filldraw[black] (4.8,0) circle (0.3pt) ;
     \filldraw[black] (4.9,0) circle (0.3pt) ;
     
     \node at (0.45,-0.8) {$-a$};
     \draw[thin] (0.4,-0.6) -- (0.6,-0.4);
     \draw[thin] (0.4,-0.4) -- (0.6,-0.6);

     \node at (-1.6,-0.8) {$-a-1$};
     \draw[thin] (-1.6,-0.6) -- (-1.4,-0.4);
     \draw[thin] (-1.6,-0.4) -- (-1.4,-0.6);

     \node at (-3.6,-0.8) {$-a-2$};
     \draw[thin] (-3.6,-0.6) -- (-3.4,-0.4);
     \draw[thin] (-3.6,-0.4) -- (-3.4,-0.6);

     \filldraw[black] (-4.35,-0.5) circle (0.3pt) ;
     \filldraw[black] (-4.45,-0.5) circle (0.3pt) ;
     \filldraw[black] (-4.55,-0.5) circle (0.3pt) ;
     
     \node at (-2.1,0.8) {$-b$};
     \draw[thin] (-2.2,0.4) -- (-2,0.6);
     \draw[thin] (-2.2,0.6) -- (-2,0.4);

     \node at (-4.1,0.8) {$-b-1$};
     \draw[thin] (-4.2,0.4) -- (-4,0.6);
     \draw[thin] (-4.2,0.6) -- (-4,0.4);

     \filldraw[black] (-4.8,0.5) circle (0.3pt) ;
     \filldraw[black] (-4.9,0.5) circle (0.3pt) ;
     \filldraw[black] (-5,0.5) circle (0.3pt) ;

     \node at (-2.9,2.3) {$\alpha -\frac{\Delta}{2}-2$};
     \draw[thin] (-2.8,1.8) -- (-2.6,2);
     \draw[thin] (-2.8,2) -- (-2.6,1.8);

     \node at (-0.9,2.3) {$\alpha -\frac{\Delta}{2}-1$};
     \draw[thin] (-0.8,1.8) -- (-0.6,2);
     \draw[thin] (-0.8,2) -- (-0.6,1.8);

     \node at (1.35,2.3) {$\alpha-\frac{\Delta}{2}$};
     \draw[thin] (1.2,1.8) -- (1.4,2);
     \draw[thin] (1.2,2) -- (1.4,1.8);

     \node at (3.35,2.3) {$\alpha-\frac{\Delta}{2}+1$};
     \draw[thin] (3.2,1.8) -- (3.4,2);
     \draw[thin] (3.2,2) -- (3.4,1.8);
     
     \filldraw[black] (4.05,1.9) circle (0.3pt) ;
     \filldraw[black] (4.15,1.9) circle (0.3pt) ;
     \filldraw[black] (4.25,1.9) circle (0.3pt) ;

     \filldraw[black] (-3.6,1.9) circle (0.3pt) ;
     \filldraw[black] (-3.7,1.9) circle (0.3pt) ;
     \filldraw[black] (-3.8,1.9) circle (0.3pt) ;
     
     \draw[thick] (1,-2) -- (0.9,-0.5); 
     \draw[thick,->] (0.9,-0.5) arc (0:140:0.4);
     \draw[thick,->] (0.217,-0.215) arc (-45:-250:0.4);
     \draw[thick] (-0.27,0.415) arc (-80:-10:0.3);
     \draw[thick] (-0.027,0.65) -- (0.1,2); 
     \node at (0.7,-1.8) {$C$};
     
\end{tikzpicture}
\caption{Contour of integration in eq. \eqref{LocalBlockGeneric}, together with the three infinite families of poles discussed in the main text. Again, the crosses are vertically displaced to improve readability of the picture, but all the poles lie on the real axis.}
\label{fig:LocalBlockContour}
\end{figure}

We get
\beq\label{MellinLocalBlock}
	\Tilde{L}(t)=C(a,b,c)  \frac{\Gamma(a+t)\Gamma(b+t)\Gamma(-t)}{\Gamma(c+t)\sin\left[\pi\left( \frac{\Delta}{2}-\alpha +t\right) \right]}~,
\eeq
where
\beq
	C(a,b,c)= \frac{\Gamma(c)}{\Gamma(a) \Gamma(b)} \sin\left[\frac{\pi}{2} (\Delta -2\alpha) \right]~.
\eeq
This function contains additional poles at $\alpha-\frac{\Delta}{2}+n$, $n\in \mathbb{Z}$ which we need to take into consideration when integrating over $t$, as depicted in figure \ref{fig:LocalBlockContour}. For $0<\Re\chi<1$, we close the contour to the right, which leaves us with contributions from two sets of poles: 
\beq\label{polesLocalBlocks}
	t_n^{(1)}= n~, \qquad t_n^{(2)}= \alpha - \frac{\Delta}{2}+n~,
\eeq
where now $n\in \mathbb{N}$. The contribution from the first set of poles $t_n^{(1)}$ gives back a hypergeometric function:
\beq
	l(\chi)=\chi^{\frac{\Delta}{2}} {}_2 F_1 (a,b;c;\chi)~,
\eeq
as the sine factor simplifies to
\beq
	\sin\left[\pi\left( \frac{\Delta}{2}-\alpha +t_n^{(2)}\right) \right] = (-1)^n \sin\left[\pi\left( \frac{\Delta}{2}-\alpha\right) \right]~.
\eeq
The residues of the second set of poles $t_n^{(2)}$ are Res$_{t_n^{(2)}}=\frac{(-1)^n}{\pi}$. Putting it all together, we find: 
\beq
	L(\chi)=l(\chi) - \chi^\alpha \sum_{n=0}^\infty \frac{\Gamma\left(a+t_n^{(2)}\right)}{\Gamma(a)}\frac{\Gamma\left(b+t_n^{(2)}\right)}{\Gamma(b)}\frac{\Gamma(c)}{\Gamma\left(c+t_n^{(2)}\right)}\frac{1}{\Gamma\left(1+t_n^{(2)}\right)} \chi^n~.
\eeq
Specifying the unknowns for the spin 0 case, for which $a=b=\frac{\Delta}{2}$ and $c=\Delta+\frac{1}{2}$, we find the local block
\beq
	G_\Delta^\alpha(\chi)= g_\Delta(\chi) - \chi^\alpha \frac{\Gamma(\alpha)^2 \Gamma\left(\Delta +\frac{1}{2}\right)}{\Gamma\left(\frac{\Delta}{2}\right)^2\Gamma\left(\alpha-\frac{\Delta}{2}+1\right) \Gamma\left(\alpha+\frac{\Delta}{2}+\frac{1}{2}\right)} \pFq{3}{2}{1,\alpha,\alpha}{\alpha -\frac{\Delta}{2}+1,\alpha +\frac{\Delta}{2}+\frac{1}{2}}{\chi}~,
	\label{LocalBlockClosed0}
\eeq
where $g_\Delta(\chi)$ is the conformal block defined in \eqref{block3p0}.
For the spin 2 case, for which $a=\frac{\Delta}{2}+1$, $b=\frac{\Delta}{2}-1$ and $c=\Delta+\frac{1}{2}$, we find the local block
\beq
	H_\Delta^\alpha (\chi)= h_\Delta(\chi) -  \chi^\alpha \frac{\Gamma(\alpha+1)\Gamma(\alpha-1) \Gamma\left(\Delta +\frac{1}{2}\right)}{\Gamma\left(\frac{\Delta}{2}+1\right)\Gamma\left(\frac{\Delta}{2}-1\right)\Gamma\left(\alpha-\frac{\Delta}{2}+1\right) \Gamma\left(\alpha+\frac{\Delta}{2}+1\right)} \pFq{3}{2}{1,\alpha+1,\alpha-1}{\alpha -\frac{\Delta}{2}+1,\alpha +\frac{\Delta}{2}+1}{\chi}~,
		\label{LocalBlockClosed2}
\eeq
where $h_\Delta(\chi)$ is the conformal block defined in \eqref{blocks3p2}.

The structure of the result makes it clear that the closed form expressions are equivalent to performing the defining integrals \eqref{localBlock0} and \eqref{localBlock2} by pulling the $\chi'$-contour to the right, as shown in figure \ref{fig:contourLocalBlock}. Thus, one obtains the sum of the original conformal block and a contribution from the discontinuity along $\chi'>1$.

\section{Hypergeometric florilegium}
\label{app:hyper_formulas}

In this work, a few different hypergeometric functions appear, depending on the choice of cross ratio which is more convenient for a given task.
In fact, all the conformal blocks in AdS$_2$ which are used in this paper can be obtained from the $sl(2,\mathbb{R})$ blocks appearing in the four-point function of local operators in a two dimensional CFT. This follows from the method of images in boundary CFT, as discussed in detail in subsection \ref{subsec:twopBlocksStress}. In this appendix, we collect all the relations which ensue.  

Then, in subsection \ref{app:asymptBlocksFF}, we derive various useful asymptotics for the form factor conformal and local blocks. Besides eqs. \eqref{LocBlockLargeD0App}-\eqref{largeDal02app}, which are of specific interest for this work, the reader might find the asymptotics \eqref{largeDuniformFF2} useful: it corresponds to a large $\D$ approximation for the exchange of a spin 2 primary in a $2d$ CFT, and it is uniform in the value of the cross ratio. It was obtained by extending work done in \cite{Qiao:2017xif}.

Finally, in subsection \ref{subsec:TaubRQ} we report the Tauberian theorem proven in \cite{Qiao:2017xif}.

\subsection{Useful relations among the conformal blocks}

All the changes of coordinates in this subsection are easily obtained via the method of images. 

The spectral block in eq. \eqref{scalarSpecBlocks}, when $d=1$, can be mapped to the scalar four-point block via the following change of coordinates:
\beq
u=\frac{1}{\xi+1}~.
\eeq
Then:
\beq
f_\D(\xi(u)) = \left(\frac{u}{4}\right)^\D 
{}_2 F_1 \left(\D,\D;2\D;u\right)~.
\label{2ptblockTo1d}
\eeq

The form factors blocks, eqs. \eqref{block3p0} and \eqref{blocks3p2}, have a $\rho$-coordinate expression in terms of the variable $z$ defined in eq. \eqref{chiofz}, 
which we report here,
\begin{equation}
\chi = \left(\frac{2z}{1+z^2}\right)^2~.
\label{chiofzApp}
\end{equation}
One gets
\begin{align}
g_\D(\chi) &=  \chi^{\Delta/2} {}_2F_1 \left(\frac{\Delta}{2},\frac{\Delta}{2}; \Delta+\frac{1}{2};\chi \right) =
(2z)^\D {}_2F_1 \left(\frac{1}{2},\Delta; \Delta+\frac{1}{2};-z^2 \right)~, 
\label{FFblockZ} \\
h_{\D}(\chi) &=  \chi^{\Delta/2} {}_2F_1 \left(\frac{\Delta}{2}-1,\frac{\Delta}{2}+1; \Delta+\frac{1}{2};\chi \right) =
(2z)^\D(1+z^2)^2 {}_2F_1 \left(\frac{5}{2},\Delta+2; \Delta+\frac{1}{2};-z^2 \right)~.
\label{FFblock2Z}
\end{align}
We sometimes abuse notation and write $g_\D(z)$, $h_\D(z)$ for the rightmost expressions.

The relation between the form factor blocks and the $sl(2,\mathbb{R})$ blocks of the four-point function then is the same as the expression for the latter blocks in the $\rho$-coordinate of \cite{Hogervorst:2013sma}, up to a (crucial) replacement $z^2 \to -z^2$. The relevant maps are
\begin{align}
(4z)^\D {}_2F_1 \left(\frac{1}{2},\Delta; \Delta+\frac{1}{2};z^2 \right)
&= u^\D {}_2F_1 \left(\D,\D;2\D;u\right)~, 
\label{sl2blockwtoz} \\
(4z)^\D(1-z^2)^2 {}_2F_1 \left(\frac{5}{2},\Delta+2; \Delta+\frac{1}{2};z^2 \right)
&=  u^\D (1-u)\, {}_2F_1 \left(\D+2,\D;2\D;u\right)~, \qquad u=\frac{4z}{(1+z)^2}~.
\label{sl2blockwtoz2}
\end{align}
The arguments of the ${}_2F_1$ on the right hand side of eq. \eqref{sl2blockwtoz2} are easily understood using the method of images and comparing with eq. 10 in \cite{Osborn:2012vt}.
For reference, we explicitly report the map in terms of $g_\Delta(\chi),\,h_\Delta(\chi)$:
\begin{align}
(-4\chi)^{\Delta/2} {}_2F_1 \left(\frac{\Delta}{2},\frac{\Delta}{2}; \Delta+\frac{1}{2};\chi \right) &= u^\D {}_2F_1 \left(\D,\D;2\D;u\right)~, 
\label{sl2blockwtochi} \\
(-4\chi)^{\Delta/2} {}_2F_1 \left(\frac{\Delta}{2}-1,\frac{\Delta}{2}+1; \Delta+\frac{1}{2};\chi \right) &= u^\D (1-u)\, {}_2F_1 \left( \D+2,\D; 2\D; u\right)~, \qquad \chi=\frac{-u^2}{4(1-u)}~.
\label{sl2blockwtochi2}
\end{align}
where the cuts are oriented so that the r.h.s. is real and positive when $0<u<1$. 

\subsection{Asymptotic expressions for the conformal and the local blocks of the form factor}
\label{app:asymptBlocksFF}

The large $\D$ limit of the conformal blocks for the form factors, at fixed $\chi$, can easily be obtained from their expressions in term of the $z$ coordinate, eqs. \eqref{FFblockZ} and \eqref{FFblock2Z}:
\begin{subequations}
\begin{align}
g_\D(\chi(z)) &\overset{\D \to \infty}{\approx} \frac{ (2z)^\D}{\sqrt{1+z^2}}~, \\
h_\D(\chi(z)) &\overset{\D \to \infty}{\approx} \frac{ (2z)^\D}{\sqrt{1+z^2}}~.
\end{align}
\label{largeDFF}
\end{subequations}
Notice that the two asymptotic expressions differ only at subleading order. 

In order to find the local blocks at large $\D$, it is useful to obtain an expression for the conformal blocks valid in the limit of large $\D$ and fixed  $\D(z^2+1)$. By using an integral representation and a saddle point approximation one obtains
\begin{subequations}
\begin{align}
g_\D(\chi(z)) &\overset{\D \to \infty}{\approx} 
(2\ii)^\D \sqrt{\frac{\D}{\pi}}\, K_0\left(\frac{z^2+1}{2}\D\right)~, 
\label{largeDfixedProd0}\\
h_\D(\chi(z)) &\overset{\D \to \infty}{\approx} 
(2\ii)^\D \sqrt{\frac{\D}{\pi}}\, K_2\left(\frac{z^2+1}{2}\D\right)~, \qquad \textup{fixed  } \ \frac{z^2+1}{2}\D~,
\end{align}
\label{largeDfixedProdFF}
\end{subequations}
where we decided to evaluate the factor $z^\D$ above its branch cut.

There is an approximation which interpolates between eqs. \eqref{largeDFF} and \eqref{largeDfixedProdFF}. This formula encodes the large $\D$ limit, uniformly in $z$. It was found in \cite{Qiao:2017xif} for the spin 0 $sl(2,\mathbb{R})$ block, and can be easily translated to the form factor block by use of eqs. \eqref{sl2blockwtoz} and \eqref{FFblockZ}:
\beq
g_\D(\chi(z)) \overset{\D \to \infty}{\approx} (2z)^\D \sqrt{\frac{\D}{\pi}} e^{\frac{z^2+1}{2}\D} K_0\left(\frac{z^2+1}{2}\D\right)~, \qquad  -1<z^2<0~.
\label{largeDuniformFF0}
\eeq
Here the power law prefactor should be evaluated above the cut. The analog of eq. \eqref{largeDuniformFF0} for $h_\D(\chi(z))$ can be derived with similar technology as in \cite{Qiao:2017xif}, by exploiting the map \eqref{sl2blockwtoz2}:
\beq
h_\D(\chi(z)) \overset{\D \to \infty}{\approx} (2z)^\D \sqrt{\frac{\D}{\pi}} e^{\frac{z^2+1}{2}\D} K_2\left(\frac{z^2+1}{2}\D\right)~, \qquad  -1<z^2<0~.
\label{largeDuniformFF2}
\eeq

The previous expressions allow to approximate the local blocks for large $\D$, fixed $\chi$, and large or fixed $\al$.
Let us discuss in detail the spin 0 case, spin 2 being similar. Starting with fixed $\alpha$, we change variable via eq. \eqref{chiofzApp} in the integral expression for the local block, eq. \eqref{localBlock0}, and obtain
\beq
G_\D^\al(\chi) =
\chi^\al \sin\left[\frac{\pi}{2}(\D-2\al)\right] \int_{-1}^0 d(z^2) 
\frac{4(z^2-1)}{\pi(z^2+1)^3}\left(\frac{4z^2}{(1+z^2)^2}-\chi\right)^{-1}
\left|\frac{4z^2}{(1+z^2)^2}\right|^{-\al} \big|g_\D(\chi(z))\big|~.
\label{LocBlockofz0}
\eeq
On the other hand, one sees from eq. \eqref{largeDuniformFF0} that the integrand is exponentially suppressed at large $\D$, unless $z^2$ is close to $-1$. Here, it is important that $\al$ does not scale with $\D$. We are led to replace the conformal block with its large $\D$, fixed $y=\D(z^2+1)$, limit---eq. \eqref{largeDfixedProd0}:
\beq
G_\D^\al(\chi) \overset{\D\to\infty}{\approx} 
-\chi^\al \sin\left[\frac{\pi}{2}(\D-2\al)\right] 
\frac{2^{\D-2\al+1}}{\pi^{3/2}} \D^{\frac{1}{2}-2\al}
\int_0^{\infty}\! dy\, y^{2\al-1} K_0\left(\frac{y}{2}\right)~,
\eeq
which yields
\beq
G_\D^\al(\chi) \overset{\D\to\infty}{\approx} 
-\chi^\al \sin\left[\frac{\pi}{2}(\D-2\al)\right] 
\frac{2^{\D+2\al-1}}{\pi^{3/2}}  \Gamma(\al)^2 \D^{\frac{1}{2}-2\al}~.
\label{LocBlockLargeD0App}
\eeq
An analogous computation gives the following, for the spin 2 local block:
\beq
H_\D^\al(\chi) \overset{\D\to\infty}{\approx} 
-\chi^\al \sin\left[\frac{\pi}{2}(\D-2\al)\right] 
\frac{2^{\D+2\al-1}}{\pi^{3/2}}  \Gamma(\al+1)\Gamma(\al-1)
 \D^{\frac{1}{2}-2\al}~.
\label{LocBlockLargeD2App}
\eeq

If $\al$ scales with $\D$, the limit of eq. \eqref{LocBlockofz0} is dictated by a different saddle. In section \ref{sec:flat}, we need the following limit:
\beq
\al=\D_\p+q~,\quad \D,\,\D_\p \to \infty~, \quad \frac{\D}{\D_\p}=\textup{fixed} >2~.
\label{limitFlat}
\eeq
In this case, the integral \eqref{LocBlockofz0} localizes at the value
\beq
z^2=\frac{1-\frac{\D}{2\D_\p}}{1+\frac{\D}{2\D_\p}}~.
\eeq
Therefore, we can replace the large $\D$ fixed $z$ limit of the conformal block---eq. \eqref{largeDFF}---and perform the saddle point approximation. The result is
\begin{multline}
G_\D^{\D_\p+q}(\chi) \overset{\eqref{limitFlat}}{\approx} H_\D^{\D_\p+q}(\chi) \overset{\eqref{limitFlat}}{\approx} \\
 \sin \left[\frac{\pi}{2} \left(2\D_\p-\D+2q\right)\right]
\frac{\chi^{\D_\p+q}}{\chi+\left(\frac{\D}{2\D_\p}\right)^2-1} 
\frac{2^{\D-\frac{1}{2}}\D}{\sqrt{\pi} \D_\p^{3/2}} 
\left(\frac{\Delta }{2\Delta_\phi }-1\right)^{\Delta/2-\Delta_\phi+\frac{1}{2}-q} 
 \left(\frac{\Delta}{2\Delta_\phi   }+1\right)^{-\Delta/2 - \Delta_\phi
   +1-q }~,
\label{largeDal02app}
\end{multline}
where we already included the fact that, as it is easy to check, the spin 2 local block has the same asymptotic behavior.

\subsection{A Tauberian theorem}
\label{subsec:TaubRQ}

The Tauberian theorem proved in \cite{Qiao:2017xif} 
states that, given a positive spectral density $p(\D)$, if 
\beq
\int_0^\infty\! d\D \,p(\D) w^\D {}_2F_1 \left(\D,\D;2\D;w\right) \overset{w\to 1}{\approx} (1-w)^{-\ga/2}~,\qquad \gamma>0\,,
\label{TaubHypoth}
\eeq
then
\beq
\int_0^Y \! d\D\, 4^\D \sqrt{\D}\, p(\D) \overset{Y\to \infty}{\approx} 
\frac{4\sqrt{\pi}}{\ga \Gamma\left(\frac{\gamma}{2}\right)^2} \,Y^\ga~.
\label{TaubRQ}
\eeq

\bibliography{./auxi/bibliography}
\bibliographystyle{./auxi/JHEP}


\end{document}